\documentclass[prd,preprint,superscriptaddress,nofootinbib,longbibliography,aps]{revtex4-1}
\usepackage[utf8]{inputenc}
\usepackage{graphicx}
\usepackage{longtable}
\usepackage{amsmath}
\usepackage{color}
\usepackage{amssymb}
\usepackage{subfigure}
\usepackage{microtype}
\usepackage[linktoc=all]{hyperref}
\usepackage{cleveref}
\usepackage{stackengine}
\usepackage{color}
\usepackage{bm}
\usepackage{braket}
\usepackage{slashed}
\usepackage{tikz}
\usepackage{adjustbox}
\usepackage{comment}
\newtheorem{theorem}{Theorem}
\def\mpl{M_{\rm Pl}}
\newcommand{\rd}{\mathrm{d}}

\newcommand{\beq}{\begin{equation}}
	\newcommand{\eeq}{\end{equation}}

\begin{document}
	
	\preprint{Imperial/TP/2021/MC/02}
	
	\title{Scattering Amplitudes for Binary Systems beyond GR}
	
	\author{Mariana Carrillo Gonz\'alez}
	\email{m.carrillo-gonzalez@imperial.ac.uk}
	\author{Claudia de Rham}
	\email{c.de-rham@imperial.ac.uk}
	\author{Andrew J. Tolley}
	\email{a.tolley@imperial.ac.uk}
	\affiliation{Theoretical Physics, Blackett Laboratory, Imperial College, London, SW7 2AZ, U.K}

	\date{\today}
	
	\begin{abstract}
	\noindent Amplitude methods have proven to be a promising technique to perform Post-Minkowskian calculations used as inputs to construct gravitational waveforms. In this paper, we show how these methods can be extended beyond the standard calculations in General Relativity with a minimal coupling to matter. As proof of principle, we consider spinless particles conformally coupled to a gravitational helicity-0 mode. We clarify the subtleties in the matching procedure that lead to the potential for conformally coupled matter. We show that in the probe particle limit, we can reproduce well known results for the field profile. With the scattering amplitudes at hand, we compute the conservative potential and scattering angle for the binary system. We find that the result is a non trivial expansion that involves not only the coupling strengths, but also a non trivial dependence on the energy/momentum of the scattered particles.
	\end{abstract}
	
	\maketitle
	
	\tableofcontents
\section{Introduction}

The new opportunities that gravitational wave observations by the LIGO and Virgo collaborations \cite{Abbott:2016blz,TheLIGOScientific:2017qsa} present have led to a quest for higher precision theoretical computations. For example, for the gravitational waves generated by the merging of two compact objects, theoretical computations allow us to build waveform templates required for detection and analysing the parameter space. These waveforms can be obtained using the effective one-body (EOB) formalism \cite{Buonanno:1998gg,Buonanno:2000ef} supplemented by numerical relativity \cite{Campanelli:2005dd,Baker:2005vv,Pretorius:2005gq} and self-force \cite{Mino:1996nk,Quinn:1996am} computations in the strong field regime and analytic computations during the inspiral phase. Since perturbative gravity calculations are valid during the inspiral phase, one can pursue fully analytic results. One prominent method consists of doing a Post-Newtonian (PN) expansion which is valid for a virialized system with small velocities so that $Gm/r\sim v^2\ll c^2$. See \cite{Blanchet:2013haa,Schafer:2018kuf,Barack:2018yvs,Barack:2018yly,Porto:2016pyg, Levi:2018nxp} for comprehensive reviews of different techniques used to perform PN computations. Away from the small velocity limit, one can use a Post-Minkowskian (PM) expansion which consists of taking the special relativity limit,  $Gm/r\ll c^2$. Computations in a PM expansion are especially well suited when using scattering amplitude techniques. Recently, the connection between quantum scattering amplitudes and classical observables has been investigated in detail, and extensive progress has been made in the past few years \cite{Cheung:2018wkq,Bern:2019nnu,Bern:2019crd,Bjerrum-Bohr:2018xdl,Ciafaloni:2018uwe,Bjerrum-Bohr:2019kec,Cachazo:2017jef,Cristofoli:2019neg,Damgaard:2019lfh,Cristofoli:2020uzm,Kosower:2018adc,Maybee:2019jus,KoemansCollado:2019ggb,Mougiakakos:2020laz,Parra-Martinez:2020dzs,Bern:2020buy,Bern:2021dqo,Herrmann:2021tct,DiVecchia:2020ymx,Kalin:2019rwq,Kalin:2019inp,Bjerrum-Bohr:2021vuf,Cheung:2020gyp,Damour:2020tta,Kalin:2020lmz,DiVecchia:2021bdo,Liu:2021zxr,DiVecchia:2021ndb,Cho:2021mqw,Bjerrum-Bohr:2021din,Dlapa:2021npj,Cristofoli:2021vyo,Bautista:2021wfy,Kosmopoulos:2021zoq,delaCruz:2020bbn,delaCruz:2021gjp}. Most of the progress focuses on the case of  the Einstein-Hilbert action describing the gravitational dynamics with a minimal coupling to matter. However, in order to test gravity, it is important to not only develop techniques that can give high precision theoretical computations for General Relativity (GR), but also for well motivated deviations of it. Within this context, interesting results have been developed for various effective field theories of gravity,
\cite{Brandhuber:2019qpg,AccettulliHuber:2019jqo,Emond:2019crr,AccettulliHuber:2020oou,AccettulliHuber:2020dal}.
Following within this logic, in this paper we shall be interested in understanding whether these scattering amplitudes techniques can be applied to complementary  types of `re-organized' effective field theories, and determine the precise scalings and other specific considerations that need to be taken into account in such cases.

Consistent local and Lorentz-invariant modified gravity theories generically involve extra degrees of freedom. In some cases, these extra degrees of freedom can give rise to the accelerated expansion observed today, although in what follows we shall remain agnostic on the precise role played by the modification of gravity on cosmological scales. Generically, these extra degrees of freedom can present themselves as new polarizations of the graviton as is for instance the case in theories that either involve higher dimensions or higher-spin states.
Irrespectively of their precise origin, from a gravitational four-dimensional perspective, these extra degrees of freedom can most of the time be associated with additional helicity-0, -1 or -2 modes. The effect of the helicity-1 modes can often be associated with a generalized Proca model (see for instance \cite{Deffayet:2010zh,Heisenberg:2014rta, Tasinato:2014eka, Allys:2015sht,Hull:2015uwa, Allys:2016jaq,Heisenberg:2016eld,Jimenez:2016isa,DeFelice:2016yws,Allys:2017map,deRham:2020yet}), although since all such Lorentz-invariant effective field theories also carry a helicity-0 mode, to start with we shall focus on the effect on a helicity-0 degrees of freedom.
Taking an effective field theory point of view, one can construct the most general theories that involve a helicity-0 mode besides the standard helicity-2 ones that appear in GR. In a particular scaling limit, the effects of the helicity-0 mode can be identified to those of a scalar degree of freedom with particular sets of non-linear interactions. An arbitrary scalar degree of freedom generically mediates a fifth force which is extremely tightly constrained by existing tests of gravity on Earth and in the Solar system \cite{Will:2014kxa,Bertotti:2003rm}. In considering models beyond GR, the fifth force constraints are typically evaded thanks to a screening mechanism and in what follows we shall focus on models enjoying a Vainshtein screening  \cite{Vainshtein:1972sx,Babichev:2013usa}. It is worth noting that when considered from a purely scalar-tensor theory point of view, screening mechanisms are typically features that have to be added by hand within the framework so as to ensure consistency with known constraints of gravity, however when the scalar degree of freedom originated from a helicity-0 mode of the graviton, many models of modified gravity like DGP \cite{Dvali:2000hr}, cascading gravity \cite{deRham:2007xp,deRham:2007rw,deRham:2009wb} and massive gravity \cite{deRham:2010ik,deRham:2010kj} have such screening mechanisms built in \cite{Deffayet:2001uk,deRham:2014zqa}.
This screening mechanism relies on classical non-linearities becoming strong, thus the screened region cannot be accessed via a weak coupling expansion, unless resummation is possible \cite{deRham:2014wfa}. It is also worth noting that the validity of the screening mechanism has only been shown for a small number of cases with large amount of symmetries in the static and weak field regimes \cite{Joyce:2014kja} and very few time-dependent cases \cite{deRham:2012fw,Chu:2012kz,deRham:2012fg,Dar:2018dra,Brax:2017wcj,Kuntz:2019plo,deAguiar:2020urb,Bezares:2021yek,Brax:2020ujo,Renevey:2021tcz,Dima:2021pwx,Bezares:2021dma}.

Related to the question of gravitational radiation, one could expect the presence of a scalar degree of freedom, or a helicity-0 mode for the graviton, to lead to monopole radiation which would entirely dominate the radiation process. This question has been investigated analytically in depth in \cite{deRham:2012fw}, where it was shown that in the scalar-tensor models we shall consider, those endowed with a Vainshtein mechanism,  monopole and dipole radiation into the scalar mode are suppressed by energy-momentum conservation, leaving the quadrupole as the main source of scalar radiation. While technically present, in these models scalar quadrupole radiation is significantly suppressed as compared to the additional standard GR tensor quadrupole radiation that is also present. When properly endowed with an active Vainshtein mechanism, the models we will study here, will hence lead to very limited observational signatures in standard isolated binary pulsar systems. Such results were confirmed numerically in \cite{Dar:2018dra}, while subtleties with higher multipoles may be occurring for some  theories \cite{deRham:2012fg}.

In this paper, we will focus on the decoupling limit in which the helicity-2 and helicity-0 modes do not interact with each other. This limit is obtained by taking $\mpl\rightarrow\infty$ and keeping the scale $\Lambda$ fixed, where $\Lambda$ measures the strength of the helicity-0 mode self-interactions. In the specific case of hard \cite{deRham:2010ik,deRham:2010kj} or soft  \cite{Dvali:2000hr,deRham:2007xp,deRham:2007rw,deRham:2009wb} massive gravity theories, $\Lambda = (m_g^2 \mpl)^{1/3}$ where $m_g$ is the scale of the graviton mass. Since we require a large hierarchy $m_g \ll \mpl$ it is also true that $\Lambda \ll \mpl$, and it is because of this hierarchy that the helicity-0 nonlinearities dominate over those of the helicity-2 mode, justifying the applicability of the decoupling limit over a wide range of scales $m_g \ll  E \sim \Lambda\ll \mpl$. Thus, we will focus on the new dynamics arising from the helicity-0 mode.

Note that the standard gravitational tensor (helicity-2) mode interactions can be included straightforwardly in the scattering amplitudes computations, by further accounting for the $1/\mpl$ suppressed interactions, so all our results presented here are to be understood as being in addition to the standard GR contributions.  To be concrete, we will work with a specific subclass of the scalar-tensor theories that display screening, the Galileon, but we point out that these techniques are applicable more generally. Galileon theories \cite{Nicolis:2008in} have several special properties such as their invariance under the shift symmetry $\delta\pi=c+b_\mu x^\mu$ which in turn leads to the enhanced vanishing of the scattering amplitudes in the soft limit, $\mathcal{A}\sim p^2$. These theories can arise as limits of theories of massive gravity and from brane constructions in higher dimensions \cite{Dvali:2000hr,Luty:2003vm,Nicolis:2004qq,deRham:2007rw,Nicolis:2008in,deRham:2009rm,deRham:2010gu,deRham:2010eu,deRham:2010ik,deRham:2010kj}. In these theories, a generic feature in the decoupling limit is that matter has a Planck suppressed conformal coupling to the Galileon, but since matter fields are treated as external sources in this limit, this conformal coupling does not break the shift symmetry. When the scalar degree of freedom is understood to arise as the helicity-0 mode of a modified or extra-dimensional representations of gravity, then the infrared corrections that emerge beyond the decoupling limit ensure a consistent coupling with matter. If on the other hand, the scalar degree of freedom is considered to be taken a dynamics of its own, then covariant embeddings of this decoupling limit would involve an explicit breaking of the Galileon shift symmetries. However in some cases, those breaking can remain soft since $\Lambda\ll \mpl$. Here, we will consider spinless matter conformally coupled to a Galileon which only has cubic self-interactions. While generically one could expect that in a binary of compact objects, each object is inside each other's Vainshtein radius, this is highly dependent on the background on which the binary lives in. Depending on the background, the couplings of the binary system can get redressed leading to both objects being outside the other's redressed Vainshtein radius. In such situations, a weak coupling expansion is valid and the methods considered in this paper are applicable. We will discuss this in more detail in the final remarks section.

The aim of this paper is to demonstrate that scattering amplitude techniques can help us understand theories beyond GR and minimal couplings. The rest of the paper is organized as follows. In Section \ref{sec:ClassA}, we review how to extract classical physics from quantum scattering amplitudes. Then, we explain two equivalent methods to extract the potential for the binary system from the classical pieces of the scattering amplitudes and briefly review the GR case in Section \ref{sec:VfromA}. In Section \ref{sec:galileon}, we introduce the Cubic Galileon theory that we will be focusing on. We compute the potential between two scalar particles conformally coupled to the Galileon. In this case, we do not have a simple PM expansion, instead there is a double expansion due to the energy scale measuring the strength of the self-interactions of the Galileon EFT. We compute the scattering amplitudes up to two loops contributing at order $Gr_V^6$ and $G^2r_V^3$, where $G$ is the standard Gravitational Newton constant and $r_V$ is the Vainshtein radius, outside which the Galileon remains weakly coupled.  Following a similar prescription as for GR, we obtain the corresponding potential at those orders. We pay special attention to the matching procedure for conformally coupled particles and point out some important subtleties. Furthermore, we show that we can reproduce well-known results in the probe particle limit. In Section \ref{sec:scatAngle}, we compute important observables, the scattering angle and the phase shift. We point out that the series expansion is not a simple expansion in the Vainshtein radius, but it has a rather simple velocity dependence. This velocity dependence is expected in dynamical systems and its precise scaling could be relevant for understanding phenomenologies beyond purely static configurations. Finally we conclude with some final remarks in Section \ref{sec:final}.

\section{Classical Scatterings} \label{sec:ClassA}
In this section, we analyze classical scatterings of spinless particles. We will first review the kinematics of these scatterings and afterwards we will explain how to extract classical physics from quantum scattering amplitudes. We explain the regions of momenta that can contribute to the classical scatterings and how to perform the loop integration in these regimes.
\subsection{Kinematics}
We are interested in the elastic scattering of two massive scalars $\phi_a$ and $\phi_b$ through a massless mediator. The external incoming  four-momenta are denoted by $p$ and $p'$ for the scalars $a$ and $b$ respectively, the transferred momenta is $q$, and $k$ denotes a generic loop momenta. The outgoing momenta of the scalars $a$ and $b$ are thus $p+q$ and $p'-q$ respectively. We will work in the center of mass (CM) frame in which the incoming momenta are given by $p=(E_1({\bf p}),{\bf p})$ and $p'=(E_2({\bf p}),-{\bf p})$ with  $E^2_i({\bf p})={\bf p^2}+m_i^2$. Here, we will consider a classical scattering so that the momentum transferred, $q=(0,{\bf q})$, is much smaller than the mass of the scalar particles. This guarantees that the particle separation, measured by the impact parameter $| \bm{b}|\sim1/|\boldsymbol{q}|$, is larger than the de Broglie wavelength of each particle. In other words, we work in the limit of large angular momentum
\begin{equation}
	J\sim| \bm{p} \times \bm{b}|\sim \frac{|\boldsymbol{p}|}{|\boldsymbol{q}|} \gg 1 \ .
\end{equation}
Thus, the hierarchy of the scales involved in the classical limit is
$	\bm{q} \ll |\boldsymbol{p}|, m_1, m_2 \ .$ In certain cases we will further restrict to the non-relativistic limit in which $|\boldsymbol{p}|\ll m_1, m_2$.
\subsection{Classical Limit of Scattering Amplitudes}
In order to extract classical physics from quantum calculations, we simply restore the $\hbar$ factors and take the limit as $\hbar\to 0$. Here, we are interested in the classical limit of scattering amplitudes. It is well known that, when both massless and massive particles are involved, not only the tree-level, but also the loop graphs will contribute to the classical limit. The loop graphs that contribute in the classical limit require the interplay of at least one massive and one massless particle in the loops, so that the {\it textbook} counting of $\hbar$ is modified. These loops give rise to non-analyticity in momentum space due to interactions through a massless particle which contributes to the classical physics.  A common feature when looking at the classical limit of scattering amplitudes is the restoration of $\hbar$ in the transferred momentum as $q \to \hbar q$, which simply states that we want to write our results in terms of the wavenumber, which appears in the Fourier transformation to coordinate space. By using dimensional analysis, one can write the scaling of scattering amplitudes in the classical limit for a given theory as
\begin{equation}
	A^\text{clas.}=M^n g_1^{n_1}g_2^{n_2}\cdots g_m^{n_m} |q|^{\alpha_{n_1,\dots,n_m}} \ , \label{Vgeneral}
\end{equation}
where $g_i$ are the coupling constants of your theory and $M$ represents the mass scale of the external particles. If the power $\alpha_{n_1,\dots,n_m}$ is even, a factor of $\log(q^2)$ should multiply the overall expression so that the coordinate space result is not a contact term. This shows that we can count the powers of $q$ of a given Feynman graph to understand whether or not it will contribute to the classical limit. This procedure helps to greatly reduce the amount of computations required. Before proceeding with performing this counting, we should understand the different scaling of external, transferred, and loop momenta.  Appendix~\ref{ap:Classical} includes all the details related to the types of graphs that contribute in the classical limit of the cubic Galileon theory.

\subsection{Hard and Soft Regions}
For a generic momentum, we can consider its scaling to belong to two different regions commonly called hard and soft. The hard region describes momentum which scales as $k\sim m$. If we have a loop graph where the loop momenta is hard, the loop integral will be independent of $q$ at leading order. The subleading corrections will be powers of $q/M$, and will not give rise to the require non-analyticity to contribute in the classical limit. From now on, we only focus on soft momenta loops, where $k\sim q$. We will use the method of regions were we restrict the loop momenta to a specific region, perform the appropriate expansions in that limit, and integrate over the whole momentum domain~\cite{Beneke:1997zp,Smirnov:2004ym}. This allows us to consider the following power-counting rules:
\begin{equation}
	p\cdot q\sim m|q| \ , \quad q,k^\text{loop}\sim |q| \ , \quad \frac{1}{P^2+m^2}\sim\frac{1}{m|q|}, \quad \frac{1}{K^2}\sim\frac{1}{|q|^2}\ ,
\end{equation}
where $P$ is a linear combination of external and loop momenta and $K$ is a linear combination of transferred and loop momenta. Now, we are in position of understanding the scaling of a given Feynman graph. Unfortunately, this is not enough to obtain only the classical contributions. It has been shown that box topologies give rise to superclassical scalings, that is, a graph that scales like $1/\hbar$, but this scaling is only the leading order in an expansion in $q/m$, so higher orders can, in principle, give classical contributions. Furthermore, bubble topologies, which are quantum and scale like $\hbar$, can combine with boxes to give rise to a counting that looks like a classical scaling but it does not contribute in the classical limit.

Fortunately, this power counting is improved in the non-relativistic limit ($v\ll1$) where we can further split the soft region into the so-called quantum soft, $k\sim q(1,1)$; potential, $k\sim q(v,1)$;  and radiation modes, $k\sim q(v,v)$. In this paper we shall only be interested in the conservative dynamics, thus understanding the velocity scaling will help us extract the relevant contributions that are even in $v$. Given the scaling of each mode, we can see that the conservative part arises due to potential gravitons. This is the only source up to two-loops, but at higher loops the radiation modes can contribute to the conservative part through, for example, the so-called radiation reaction effects \cite{Damour:2020tta,DiVecchia:2020ymx,DiVecchia:2021bdo,DiVecchia:2021ndb}. Looking at the velocity scaling, helps us identify superclassical contributions, since they will diverge as $v\to 0$. Similarly, quantum times super-classical contributions will lead to scaleless energy integrals but those vanish in dimensional regularization since all loops that only include massless internal propagators vanish within that prescription. Therefore, the momentum transferred and velocity power countings allow us to identify the only graphs that can contribute in the classical limit.
\subsection{Loop Integrals}
We now turn to explain our methods for performing the loop integrals. At one loop, one can easily perform the full relativistic integration in the classical limit. To do so, we extract the classical scaling of the integrand, and relate the tensor integrals to scalar ones using the Veltman-Passarino Reduction \cite{Passarino:1978jh}. The two-loop graphs are more involved and we will consider instead an expansion in the non-relativistic limit which greatly simplifies the integrands. Note that the amplitudes for the helicity-0 graviton mode arise from higher-derivative operators which lead to large powers of momenta in the numerator of the integrand. Thus, expanding in the non-relativistic limit can largely simplify these integrands. We use non-relativistic integration methods as described in \cite{Bern:2019crd}. First, we perform the energy integrals which are simplified by realizing that integrals without matter poles do not have support in the potential region and thus can be set to zero. For the spatial momentum integrals, we re-express numerator factors in terms of the denominator ones so that all our integrals reduce to the form of the following master integral \cite{Smirnov:2004ym}
\begin{align}
	&\int \!\!\!\frac{\rd^{D-1} \boldsymbol{\ell}}{(2 \pi)^{D-1}} \frac{\boldsymbol{\ell}^{\mu_{1}} \boldsymbol{\ell}^{\mu_{2}} \cdots \boldsymbol{\ell}^{\mu_{n}}}{\left[\boldsymbol{\ell}^{2}\right]^{\alpha}\left[(\boldsymbol{\ell}+\boldsymbol{w})^{2}\right]^{\beta}}=\frac{(-1)^{n}(4 \pi)^{\frac{1-D}{2}}}{\left[\boldsymbol{w}^{2}\right]^{\alpha+\beta-\frac{D-1}{2}}} \sum_{m=0}^{\lfloor n / 2\rfloor} A(\alpha, \beta ; n, m)\left[\frac{\boldsymbol{w}^{2}}{2}\right]^{m}\!\!\!\!\!\!\left\{[\delta]^{m}[\boldsymbol{w}]^{n-2 m}\right\}^{\mu_{1} \mu_{2} \cdots \mu_{n}} \!\!  ,  \nonumber  \\
	&\text{with } \  A(\alpha, \beta ; n, m)=\frac{\Gamma\left(\alpha+\beta-m-\frac{D-1}{2}\right) \Gamma\left(n-m-\alpha+\frac{D-1}{2}\right) \Gamma\left(m-\beta+\frac{D-1}{2}\right)}{\Gamma(\alpha) \Gamma(\beta) \Gamma(n-\alpha-\beta+D-1)} \ , \label{eq:masterI}
\end{align}
where the curly brackets denote a fully symmetric tensor with $n-2m$ vector $\boldsymbol{w}$, each time with a respective power $\mu_i$ and $m$ appearances of the $D-1$ Kronecker delta.
For UV divergent integrals, we use dimensional regularization in $D=4-2\epsilon$ and the $\overline{\text{MS}}$ scheme.

\section{Potential from Amplitudes} \label{sec:VfromA}
There are two different, but equivalent methods that are commonly used to obtain a potential from scattering amplitudes which we summarize in the following. In both cases we obtain the potential in momentum space before  Fourier transforming back to coordinate space. All the required transformations follow from these two formulae
\begin{align}
	&	\int \frac{\rd^{3} \boldsymbol{q}}{(2 \pi)^{3}} e^{-i \boldsymbol{r} \cdot \boldsymbol{q}}|\boldsymbol{q}|^{n}=\frac{2^{n} \Gamma\left(\frac{1}{2}(3+n)\right)}{\pi^{3 / 2} \Gamma\left(-\frac{1}{2} n\right)}r^{-(n+3)} \ , \label{FT1} \\
	&	\int \frac{\rd^{3} \boldsymbol{q}}{(2 \pi)^{3}} e^{-i \boldsymbol{q} \cdot \boldsymbol{r}}|\boldsymbol{q}|^{2} \log |\boldsymbol{q}|^{2}=\frac{3}{\pi r^{5}} \ ,\label{FT2}
\end{align}
where $r=|\boldsymbol{r}|$ is the distance between the two scattered particles.
\subsection{Lippmann-Schwinger Equation and the Born Approximation}
In this subsection, we briefly review how to relate the classical potential and scattering amplitudes directly. We want to analyze the 2-2 scattering of scalars with large impact parameters to extract the classical gravitational potential between them. While the formulation is fully relativistic, here we express the results in terms of the three-momenta and energies of the particles since we work in the CM frame. For this purpose one can consider the Lippmann-Schwinger equation which relates the potential and scattering amplitude as
\begin{align}
	\widetilde{\mathcal{M}}_4({\bf{p}},{\bf p+ q})&=-{V}({\bf p},{\bf p+ q})+ \int_{\boldsymbol{k}} \frac{V({\bf p},{\bf k})\widetilde{\mathcal{M}}_4({\bf k},{\bf p +q})}{(E_p-E_k + i \epsilon)} \ ,
	\label{LS}
\end{align}
where $\widetilde{\mathcal{M}}_4$ is the scattering amplitude in the CM frame with non-relativistic normalization $\widetilde{\mathcal{M}}_4=\mathcal{M}/4 E_1({\bf p}) E_2({\bf p})$, $\mathcal{M}$ has the standard relativistic normalization, $E_p=E_1({\bf p})+E_2({\bf p})$, and $\int_{\boldsymbol{k}}=\int \frac{\rd^3{\bf k}}{(2\pi)^3}$. The Fourier space potential and the scattering amplitude are defined with respect to the relativistic scattering states $\ket{p}$ and $\ket{p+q}$ as
\begin{align}
V(p, q) &\equiv\left\langle p\left|\widehat V\right| p+q\right\rangle \ , \label{VScatSates}\\
\mathcal{M}\left(p, p^{\prime}\right)&\equiv\lim _{\epsilon \rightarrow 0}\left\langle p\left|\widehat{T}\left(E_{p}+i \epsilon\right)\right| p+q\right\rangle \ ,
\end{align}
where $\widehat{T}$ is the transfer-matrix and $V$ is the potential of the binary system. We will see later on that care should be taken when interpreting the potential given the chosen scattering states.

The Lippmann-Schwinger equation can be solved recursively. At leading order, we can find the potential simply from the Fourier transform of the scattering amplitude as
\begin{equation}
	V({\bf{p}},r)=-\frac{1}{4 E_1({\bf p}) E_2({\bf p})} \int \frac{\rd^3{\bf q}}{(2\pi)^3}e^{- i {\bf{q}}\cdot{ \bf{r}}} \mathcal{M}_4({\bf p},{\bf p+q}) \ .
	\label{Born}
\end{equation}
If we want to go beyond leading order in the coupling strength, we have to consider a recursive relation that arises from the non-leading terms in the Born approximation. In this case, we can write the potential at order $g^n$, with $g$ the coupling strength, as
\begin{align}
	V({\bf{p}},{\bf p+ q})\big|_{\mathcal{O}(g^n)}&=-\frac{1
	}{4 E_1({\bf p}) E_2({\bf p})} \mathcal{M}_4({\bf p},{\bf p+ q}) \Bigg|_{\mathcal{O}(g^n)} \nonumber \\
	&- \frac{1}{4 E_1({\bf p}) E_2({\bf p})}  \int_{\boldsymbol{k}} \frac{\mathcal{M}_4({\bf p},{\bf k})\mathcal{M}_4({\bf k},{\bf p +q})}{4 E_1({\bf k}) E_2({\bf k})(E_p-E_k + i \epsilon)} \Bigg|_{\mathcal{O}(g^n)} +\cdots  \ .
	\label{NLOBorn}
\end{align}
This is commonly refer to as {\it Born subtraction}. In the context of General Relativity computations, this was first introduced in \cite{Iwasaki:1971vb}.
\subsection{EFT Matching}
Consider an EFT for the scalar particles $\phi_a(\boldsymbol{p})$ in the CM frame
\begin{align}
\mathcal{L}^\text{EFT}=&\int_{\boldsymbol{p}} \sum_{a=1,2} \phi_{a}^{\dagger}(-\boldsymbol{p})\left(i \partial_{t}-\sqrt{\boldsymbol{p}^{2}+m_{i}^{2}}\right) \phi_{a}(\boldsymbol{p})\nonumber \\
-&\int_{\boldsymbol{p}, \boldsymbol{p}^{\prime}} \phi_{1}^{\dagger}\left(\boldsymbol{p}\right) \phi_{2}^{\dagger}\left(-\boldsymbol{p}\right) {V}\left(\boldsymbol{p}, \boldsymbol{p}^{\prime}\right) \phi_{1}(\boldsymbol{p}^{\prime}) \phi_{2}(-\boldsymbol{p}^{\prime}) \ , \label{nrS}
\end{align}
where ${V}\left( \boldsymbol{p},\boldsymbol{p}^{\prime}\right)$ can be thought of as a Wilson coefficient for the contact interaction; since it is an off-shell coupling, it is not invariant under field redefinitions. The transferred momentum, whose conjugate variable is $\boldsymbol{r}$ (the distance between the two particles), is given by $\boldsymbol{q}=\boldsymbol{p}^{\prime}-\boldsymbol{p}$. Note that this choice corresponds to the isotropic gauge where terms involving $\boldsymbol{p}\cdot\boldsymbol{r}$ do not appear. Different gauges can be obtained by a coordinate transformation of the canonical variables $({\bf{r}},{\bf{p}})$ which preserves the Poisson brackets. Similarly, field redefinitions change the gauge.\\

This EFT can be thought of as arising from a theory where the mediator, via which the scalars $\phi_i$ interact, has been integrated out.
This procedure ought to be understood as an integration at the level of the path integral but in practise, we can derive the EFT by simply matching the amplitudes derived from the full theory with the on-shell EFT amplitudes, accounting for the non-relativistic normalization. More precisely, the matching we use is,
\begin{equation}
\mathcal{M}^\text{EFT}=\frac{\mathcal{M}}{4 E_1 E_2} \ .
\end{equation}
Given this matching, we obtain ${V}\left(\boldsymbol{p}, \boldsymbol{p}^{\prime}\right)$ which can be interpreted as the conservative potential between the massive scalars. Note that starting at next-to-leading order (NLO) order one has to include loop amplitudes in the EFT. This leads to an expression for the potential which in fact is given by the Born subtraction in Eq.\eqref{NLOBorn}. This EFT matching approach has been employed in \cite{Neill:2013wsa,Cheung:2018wkq,Bern:2019nnu,Bern:2019crd,Bern:2020buy,Cheung:2020gyp,Bern:2021dqo,Parra-Martinez:2020dzs} for GR computations. Furthermore, we can construct the Hamiltonian and restrict it to the 2-particle subspace of the massive scalars. The resulting Hamiltonian is the first-quantized Hamiltonian for 2 classical massive particles. This fact will be useful to compare to known results in the probe particle limit. The matching for minimally coupled particles is straightforward, but we will see that the matching in the presence of a conformal coupling is more delicate.
\subsection{Brief Review of the Gravitational Case} \label{sec:GR}
In this subsection we briefly review the construction of the conservative potential between two massive particles interacting gravitationally in the GR case. Let us consider a set of real scalar fields $\phi_i$ with mass $m_i$ which are only gravitationally coupled,
\begin{equation}
	S=\int \rd^4x \left(\frac{1}{16 \pi G}\sqrt{-g}~ R -\frac{1}{2}\sum_i \left( (\partial\phi_i)^2+m_i^2\phi_i^2 \right) \right) \, .	
\end{equation}
Since we are interesting in extracting the classical limit, we will put  the factors of $\hbar$ back in our expressions. This can be done by rescaling
\begin{equation}
	G\rightarrow\frac{G}{\hbar}, \quad q\rightarrow \hbar q ,
\end{equation}
so that the correct mass and length units are restored. We can analyze the regimes in which classical non-linearities become relevant and when quantum corrections are relevant. The parameters that measure these regions are well known and read
\begin{equation}
	\alpha_\text{cl non-lin}=\frac{h}{\mpl}\sim \frac{r_\text{Sch}}{r} \ , \quad \alpha_q=\frac{\partial^2}{\mpl^2}\sim \frac{1}{\mpl^2 r^2} \ .
\end{equation}
By bringing back the $\hbar$ factors, we can indeed see that $\alpha_\text{cl non-lin}\propto\hbar^0$, while $\alpha_\text{cl non-lin}\propto\hbar$, where we have assumed $r\sim1/q$. In the current computation, we are assuming that the classical non-linearities are small so that the perturbative calculation is valid. Nevertheless, one can hope to be able to resum the results to get access to the regime were the non-linearities are large, yet the quantum corrections are small. This regime is accessed as we get closer to the Schwarzschild radius, but still far from the Planck scale.

We now proceed with establishing which loop corrections contribute in the classical limit. Using dimensional analysis, we can understand the scaling of these corrections which we parametrize as
\begin{equation}
	V=V_\text{tree}\sum_{a, b =0}^{\infty} f_a(M,q,G)\left(\frac{p}{M}\right)^b \ ,
\end{equation}
where $M$ stands for a mass scale corresponding to the mass $m_i$ of the scalars $\phi_i$ or a combination of those masses. Here, $q=|\boldsymbol{q}|$ is the transferred momentum and $p=|\boldsymbol{p}|$ the CM momentum. We already know that the tree-level gravitational potential is given by $GM^2/q^2$, but we want to understand the subleading contributions coming from the loop corrections. By requiring that $f_a(M,q,G)$ scales as $\hbar^0$ and has units $(ML)^0$ we find that the only possibility is $f_a(M,q,G)=(MqG)^a$, so the classical gravitational potential is of the form
\begin{equation}
	V=\frac{GM^2}{q^2}\sum_{a, b =0}^{\infty} c_{a,b} (MqG)^a\left(\frac{p}{M}\right)^b \ .
\end{equation}
Note that this is an expansion in the Schwarzschild radius, $r_\text{Schw}(M)=2GM$. Now, we can apply the methods of Sections~\ref{sec:ClassA} and \ref{sec:VfromA} to obtain the classical potential. It is also interesting to note that one can further match to a point-particle EFT in the probe particle limit. By considering an ansatz for the metric perturbation that is consistent with the symmetries of the problem, one can reproduce the Schwarzschild metric from this potential~\cite{Neill:2013wsa}.
\section{Heliclity-0 Mode of the Graviton} \label{sec:galileon}
Having introduced all the relevant methods within the context of GR, we can  proceed with applying them to a specific example of theory beyond GR. As mentioned previously, any local and Lorentz invariant theory beyond GR will necessarily carry additional degrees of freedom, often hidden as additional gravitational polarizations. Here, it is understood that the polarization mode  $\pi$ we shall consider plays the role of  an additional gravitational mediator between the two scalars $\phi_i$, in addition to the standard helicity-2 gravitational ones. So even though in the limit we shall be working, there is no sign of the actual tensor gravitational mode, we are still dealing with a gravitational theory, albeit just focusing for now on its helicity-0 mode. This is manifest in many models beyond GR by taking a specific decoupling limit where the standard tensor mode of gravity can be treated separately from the helicity-0 mode of gravity. Beyond that decoupling limit, both modes will further mix non-trivially and will lead to further corrections although those are typically extremely suppressed.
For concreteness, we focus on models beyond GR for which the helicity-0 mode behaves as a cubic Galileon scalar field in the decoupling limit, for which the action takes the form
\begin{equation}
	S=\int \rd^4x \left(-\frac{1}{2}(\partial\pi)^2-\frac{1}{\Lambda^3} \square\pi (\partial\pi)^2 -\frac{1}{2}\sum_i \left( (\partial\phi_i)^2+m_i^2\phi_i^2 \right) +\mathcal{L}_{\pi\phi\phi} \right) \ ,	\label{Spiphi}
\end{equation}
where $\mathcal{L}_{\pi\phi\phi} $ contains the interaction between the Galileons and massive scalars. Here, we will consider an interaction of the form
\begin{equation}
	\mathcal{L}_{\pi\phi\phi}=\Omega(\pi)m_i^2\phi^2_i \ ,\quad \Omega(\pi)=\sum_{n} C_n g^n \pi^n/\mpl^n \ , \label{coupl}
\end{equation}
which gives rise to the Feynman rules for the matter-Galileon vertices
\begin{equation}
	V_{\phi_i^2\pi^n}=i \frac{2 \, C_n g^n  \, n! \,  m_i^2}{\mpl^n}\,. \label{vertexC}
\end{equation}
Other interactions involving derivatives of the scalars $\phi$ will ultimately reduce to the parametrization above since in the classical limit we neglect all higher order corrections in $q/m$ since they are of order $\hbar$. For example, an interaction term of the form $g(\partial\phi_i)^2\pi/\mpl$ will give rise to a vertex
$i \, 2 g\frac{p_1\cdot p'_1}{\mpl}$ where we assume $p_1$ is incoming and $p'_1=p_1+k$ is outgoing.
Here, $k$ stands for either a loop momentum, the transfer momentum or a linear combination of both. We are interested in scatterings of two scalars $\phi_1$ and $\phi_2$ with incoming momenta $p_1$ and $p_2$, and outgoing momenta  $p_1+q$ and $p_2-q$ respectively. If $k$ simply corresponds to the transferred momentum $q$, it is clear that $p_1\cdot p'_1 = -m^2_i+\mathcal{O}((q/m_i)^2)$, after using momentum conservation which tell us that $p_1\cdot q=-q^2/2$. On the other hand, if $k$ involves a loop momenta one has to realize that every higher order term in $k$ will give rise to an extra factor of $(q/m_i)^2$ after integration. Similarly, if $k$ is a combination of the transferred and loop momentum we will get $p_1\cdot p'_1 = -m^2_i+\mathcal{O}((q/m_i)^2)$. So in any case the vertex reduces to $- i \, 2g\frac{m^2_i}{\mpl}$ in the classical limit, that is, it scales just like the coupling in Eq.~\eqref{coupl}.
The same argument applies for other couplings that include derivatives of $\pi$, which would then involve higher powers of the momentum transfer $k$ which will again involve additional factors of $q/m_i$ after integration.
Given this, we can relate any other couplings involving derivatives to the parametrization introduced above.

We use the interaction in Eq.\eqref{coupl} to compute our results since this can be matched to more general couplings. At the end, we will be interested in analyzing the case of a conformal coupling whose interactions can be written in terms of the Wilsonian coefficients $C_n$. We proceed to analyze this coupling below.

\paragraph*{\bf Conformal coupling:}
Let us consider the conformal coupling $\tilde{g}_{\mu\nu}=A^2(\pi) g_{\mu\nu}$ for the massive scalars. In such a case, we have
\begin{equation}
	S_{\phi_i}=\int \rd^4x \sqrt{-g} \left(-\frac{1}{2}A^2(\pi)(\partial \phi_i)^2-\frac{1}{2}A^4(\pi)m^2\phi_{i}^2\right) \ ,
\end{equation}
with
\begin{equation}
A(\pi)=1+\sum_{n} D_n g^n \pi^n/\mpl^n \ . \label{eq:AconfCoupl}
\end{equation}
Although we introduced a metric here, we consider the decoupling limit where the contributions from the helicity-2 mode decouple from those of $\pi$ and end up being precisely the same as in GR. In practise, we can therefore take $g_{\mu\nu}=\eta_{\mu\nu}$, the Minkowski metric, with the understanding that we are focusing on the contributions from the helicity-0 mode $\pi$ that come in addition to the standard GR ones. Let us consider the field redefinition $\tilde{\phi}_i=A(\pi)\phi_i$, the action now reads
\begin{equation}
S_{\tilde{\phi}_i}= \int \rd^4x \sqrt{-g} \left(-\frac{1}{2}(\partial \tilde{\phi}_i)^2-\frac{1}{2} \tilde{\phi}_i^2   \frac{\Box  A(\pi)}{A(\pi) }-\frac{1}{2}A^2(\pi)m^2\tilde{\phi}_i^2 \right) \, . \label{confRedef}
\end{equation}
This field redefinition is required so that the state $\tilde{\phi}_i(x)\ket{0}$ has the standard canonical normalization of a conformally coupled particle
\begin{equation}
	\tilde{\phi}_i(x)\ket{0}=\int_{\boldsymbol{p}}\frac{1}{\sqrt{2E(\boldsymbol{p})}} a_{\boldsymbol{p}}^\dagger \ket{0} e^{-i{\bm{x}}\cdot{\bm{p}}} \ , \label{phistate}
\end{equation}
where $E(\boldsymbol{p})=\sqrt{{\bm{p}}^2+m_i^2A^2(\pi)}$. Different normalizations (field redefinitions) spoil the mapping to the conformally coupled point-particle which will be explored in the Section \ref{pp}. One can understand this from the EFT matching point of view by noting that the truncation of the Hamiltonian to the 2-particle subspace assumes the standard normalization, as in \eqref{phistate}, of the states $\tilde{\phi}_i(x)\ket{0}$. Similarly, in the Born approximation method, the potential is defined in Eq.~\eqref{VScatSates} with respect to scattering states $\ket{p}=a_{\boldsymbol{p}}^\dagger \ket{0}$, so we need to perform the field redefinition above to match between the first quantized point-particle and the truncated second-quantized scalar field potentials. In other words, we should consider the correct scattering states when extracting the classical physics.

Notice that in Eq.~\eqref{confRedef} the second term does not contribute in the classical limit since it is of order $(q/m)^2$, which can be seen by an argument analogous to that under Eq.~\eqref{vertexC}. Thus, the derivative couplings can be neglected and we have a simple coupling to the mass term \eqref{coupl} for which $\Omega(\pi) = - \frac{1}{2} (A(\pi)^2-1)$. Now, we can write the Wilson coefficients $C_n$ considered above in terms of the $D_n$ coefficients from the conformal coupling. For example, the first two are:
\begin{equation}
	C_1=-D_1 \, \quad C_2=-\frac{1}{2}\left(D_1^2+2D_2\right) \ ,
\end{equation}
which allow us to use the results from the general coupling in Eq.~\eqref{coupl} to obtain those of conformally coupled matter.

\subsection{Scaling of the classical Galileon potential}
In this section, we will analyze the scaling of the classical potential from dimensional analysis. First, we need to understand the factors of $\hbar$ that need to be restored in $\Lambda$ and $\mpl$. To do so, we can look at the scaling of the action in Eq.~\eqref{Spiphi} and use the fact that $[S]=ML$. From the kinetic term, we see that $\pi$ has units of $\sqrt{M/L}$. Then, from the cubic self-interaction and the interaction with the massive scalars we find that restoring $\hbar$ corresponds to the replacements
\begin{equation}
\Lambda^3\rightarrow \hbar^{5/2} \Lambda^3 \ , \quad  G\rightarrow  G / \hbar \ ,
\end{equation}
where we used $\mpl=1/\sqrt{8\pi G}$. Like in the gravitational case, we can look at the parameters that determine when the classical non-linearities and quantum corrections become important; for the Galileon these are
\begin{equation}
	\alpha_\text{cl non-lin}=\frac{\partial\partial\pi}{\Lambda^3} \ , \quad \alpha_q=\frac{\partial^2}{\Lambda^2} \ .
\end{equation}
Let us define the Vainshtein radius as
\begin{equation}
	r_{V_i}=\frac{1}{\Lambda}\left(\frac{|C_1| g \ m_i}{\mpl} \right)^{1/3}  \,,
\end{equation}
where $m_i$ corresponds to the mass $m_i$ of the scalars $\phi_i$. Note that in this non-static case, the Vainshtein radius defined above is not exactly the radius that separates screened and unscreened regions. We will see later on that this separation is expected to depend on the momentum of the binary system. Restoring the $\hbar$ factors in this expression is done via
\begin{equation}
	r_V\rightarrow\frac{r_V}{\hbar} \ .
\end{equation}

Again, by dimensional analysis we see that the classical potential in this case can be parametrized as
\begin{equation}
	V=V^\text{tree}\left(1+ \sum_{a, b} f_a(M,q,G,\Lambda)\left(\frac{p}{M}\right)^b\right) \ ,
\end{equation}
where we require that $f_a(M,q,\lambda,\Lambda)$ scales as $\hbar^0$, has dimensions of $(ML)^0$, and that only powers of $\Lambda^3$ appear on the denominator. This together with Eq.~\eqref{FT1} and Eq.~\eqref{FT2} tells us that the classical potential should be of the form
\begin{align}
	V=\begin{cases}
		\frac{8 \pi g^2 G M^2}{q^2}\left(\sum\limits_{a, n, c \,  \in  \, \mathbb{Z}^+} c_{a,c,n} \left(g^2 G M q\right)^{n}(r_V q)^{3 a}\left(\frac{p}{M}\right)^c\right)  & \quad n+3a \ \text{odd}  \ , \\
		\frac{8 \pi g^2 G M^2}{q^2}\left(\sum\limits_{a, n, c \,  \in  \, \mathbb{Z}^+} c_{a,c,n} \left(g^2 G M q\right)^{n}(r_V q)^{3 a}\log{q^2}\left(\frac{p}{M}\right)^c\right) & \quad  n+3a \ \text{even} \ .
	\end{cases}
\label{galV}
\end{align}
Note that the contributions from terms with $n=0$ correspond to a simple series in $r_V$. In Appendix~\ref{ap:Classical}, we explore in detail the Feynman graphs that can give rise to this classical potential.
\subsection{Scattering amplitudes}
Considering the coupling in Eq.~\eqref{coupl} and the insights from the previous sections and the Appendix~\ref{ap:Classical}, we proceed to compute the scattering amplitudes that will contribute to the classical potential. Note that we will ignore infrared divergent terms that cancel when computing the potential. These arise from box and cross-box diagrams. For example, at 1-loop this give rise to non-analytic structures of the form $\log{q^2}/q^2$, which have superclassical scaling.  Although we do not show them here explicitly, we have checked that the superclassical terms cancel in the computation of the potential as they should. We also ignore classical contact term contributions, that is, graphs with the correct scaling in $q$ to give a classical contribution, but whose Fourier transformation leads to a delta function in coordinate space.

The resulting scattering amplitudes correspond to a series expansion in both the Newton's constant and the Vainshtein radius. Note that higher order contributions in $G$ are largely suppressed compare to higher contributions in $r_V$. Thus, in following we compute the order $G$ contributions arising up to 2 loops. We also compute the order $G^2$ at one-loop since new features in the calculation of the potential will arise at this order. Nevertheless, we expect this contribution to be highly suppressed with respect to the order $Gr_V^6$.

\begin{figure}[!h]
	\includegraphics[width=\textwidth]{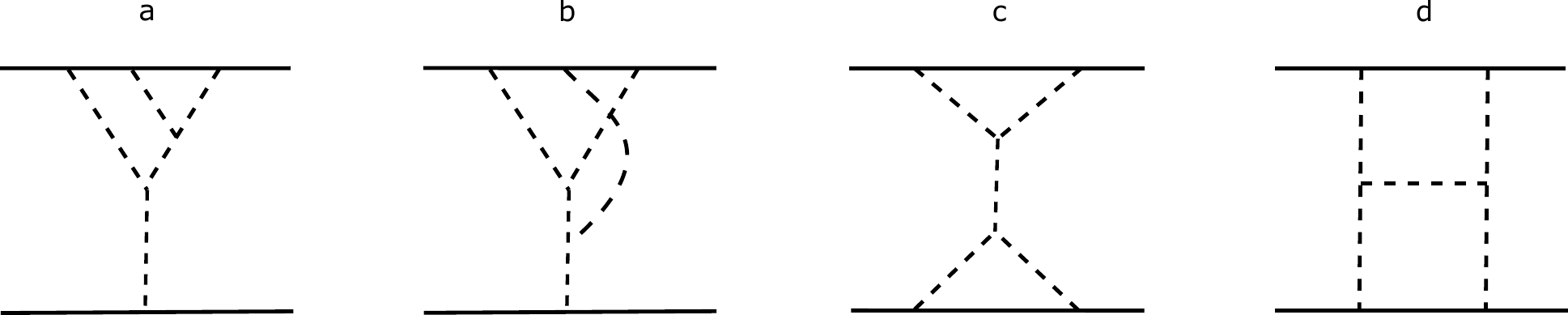}
	\caption{Graphs contributing at order $Gr_V^6$. Graphs a and b are the double triangle graphs, graph c is the triangle squared, and graph d is the H type. For each graph we include the inequivalent permutations of external legs. The number of different labelings of the external legs that are considered are 4 for graph a, 2 for graph b, 1 for graph c, and 2 for graph d.}
	\label{2loop}
\end{figure}

\paragraph{Tree level:}
The $t$-channel is the only classical contribution  to the $\phi_a\phi_b$ scattering at tree level and reads
\begin{equation}
	\mathcal{M}^{\mathcal{O}(G)}= \frac{32 \pi C_1^2g^2 G m_a^2 m_b^2}{q^2} \ . \label{Mtree}
\end{equation}
\paragraph{1-loop:}
At 1-loop order we find that the only classical contributions arise from triangle graphs. In our case, we have two classical contributions from triangle and inverted triangle graphs that read
\begin{align}
		\mathcal{M}^\text{1-loop $\mathcal{O}(Gr_V^3)$}&= -\frac{2 \pi C_1^2g^2 G m_a^2 m_b^2}{q^2} \left((r_{V_a} q)^3+(r_{V_b} q)^3\right) \text{sign}(C_1) \ , \label{1looprv}\\
	\mathcal{M}^\text{1-loop $\mathcal{O}(G^2)$}&= \frac{32 \pi^2 C_1^2C_2 g^4 G^2 m_a^2 m_b^2}{q} (m_a+m_b) \ . \label{1loopg4}
\end{align}
\begin{figure}[!h]
	\includegraphics[width=0.9 \textwidth]{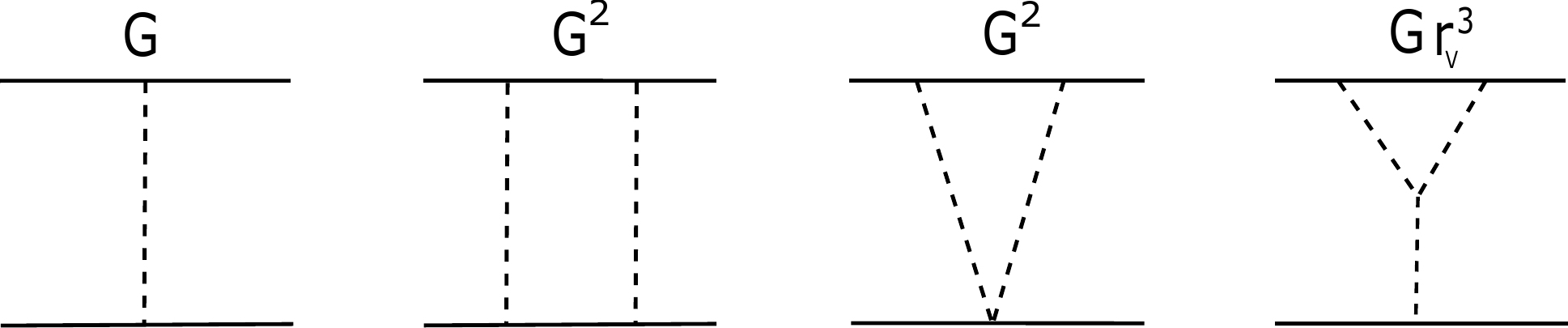}
	\caption{Graphs contributing at order $G$, $G^2$, and $Gr_V^3$. These correspond to tree and one loop level graphs. Note that we have checked that the box only has quantum contributions. We also include in our computation the graphs in which we exchange the external states $\phi_a\leftrightarrow\phi_b$.}
	\label{1loop}
\end{figure}
\paragraph{2-loops:}
At order $G$, we can divide the graphs contributing to the classical limit in 3 types: the H type, the double triangle and the triangle squared, see Fig.~\ref{2loop}. The H and cross H graphs give
\begin{align}
	\mathcal{M}_\text{cl}^\text{2-loop $\mathcal{O}(Gr_V^6)$  H}=&- \frac{C_1^2g^2 G m_a^3 m_b^3}{ E_a E_b q^2} \frac{(r_{V_a} q)^3(r_{V_b} q)^3}{80 \ \pi} \log(q^2)\times \nonumber \\
	&\left(1-\frac{25}{42}\frac{(3 E_a^2+4 E_a E_b+3 E_b^2)}{ E_a^2 E_b^2}|{\bf p}|^2+\mathcal{O}(|{\bf p}|^4)\right)\ . \label{2looprv6H}
\end{align}
For the double triangle type we have
\begin{equation}
 	\mathcal{M}_\text{cl}^{\text{2-loop $\mathcal{O}(Gr_V^6)$ } \triangle\triangle}= -\frac{8 C_1^2g^2 m_a^2 m_b^2}{\mpl^2 q^2} \frac{(r_{V_a} q)^6+(r_{V_b} q)^6}{210\pi} \log(q^2)\ .\label{2looprv6TT}
 \end{equation}
Meanwhile, the triangle squared type will not contribute to the classical potential. This is easy to see since the integration splits into two one-loop integrals each giving a $q^3$ term. The graph then scales like $q^4$ which corresponds to a contact term (as can be seen after  Fourier transform).

\subsection{Classical potential from amplitudes} \label{subsec:pot}
\subsubsection{Potential at order $G r_V^6$}
Using Eq.~\eqref{Born} we can obtain the leading order classical potential in coordinate space up to order $Gr_V^6$. Note that at order $G r_V^n$ there's no need to add Born subtraction terms, the potential is simply given by the scattering amplitude with non-relativistic normalization. From Eqs.~\eqref{Mtree}, \eqref{1looprv}, \eqref{2looprv6H}, and \eqref{2looprv6TT} we find
\begin{align}
	&V^{\mathcal{O}(G)}({\bf p},r)=-\frac{2 C_1^2 g^2 G  m_a^2 m_b^2}{E_a E_b r}\Bigg(1+ \left(\frac{ r_{V_a}^3+r_{V_b}^3}{4 \pi r^3}\right) \text{sign}(C_1)  \label{VGrv6} \\
	&+\frac{2}{7 \pi^2 r^6} \Bigg( r_{V_a}^6+r_{V_b}^6 +\frac{399m_am_b}{32E_aE_b}r_{V_a}^3r_{V_b}^3\left(1-\frac{25}{42}\frac{(3 E_a^2+4 E_a E_b+3 E_b^2)}{ E_a^2 E_b^2}|{\bf p}|^2+\mathcal{O}(|{\bf p}|^4)\right) \Bigg)\Bigg) \ , \nonumber 
\end{align}
where the energy factors depend on the CM momentum, but we do not show the dependence explicitly for simplicity. Note that these factors can be expanded in the non-relativistic limit to explicitly extract the order $p^2$ terms. In order to obtain a fully post-Minkowskian result, without expanding in the non-relativistic limit, one could try to resum the higher momentum contributions arising in the H-type graphs. We do not require to construct the potential to that precision here, so instead we expand Eq.~\eqref{VGrv6} in the non-relativistic limit and find the post-Newtonian result
\begin{align}
	V^{\mathcal{O}(G|{\bf p}|^2)}({\bf p},r)&=-\frac{2 C_1^2 g^2 G \nu m^2}{ r}\Bigg(\left(1-\left(\frac{1-2\nu}{m^2\nu^2}\right)\frac{|{\bf p}|^2}{2}\right) \nonumber  \\ 
	&+ \left(\frac{ r_{V_a}^3+r_{V_b}^3}{4 \pi r^3}\right) \text{sign}(C_1)\left(1-\left(\frac{1-2\nu}{m^2\nu^2}\right)\frac{|{\bf p}|^2}{2}\right)  \nonumber \\
	& +\frac{2}{7 \pi^2 r^6} \Bigg( r_{V_a}^6+r_{V_b}^6+\frac{399}{32}r_{V_a}^3r_{V_b}^3\left(1-\frac{2}{21\nu}\left(\frac{24}{\nu}-23\right)\frac{|{\bf p}|^2}{m^2}\right) \Bigg)\Bigg) \ , 
\end{align}
where we have defined the symmetric mass ratio $\nu$ and the total mass $m$ as
\begin{equation}
	\nu\equiv\frac{m_am_b}{(m_a+m_b)^2} \ , \quad m\equiv m_a+m_b \ . \label{eq:masses}
\end{equation}
We can see from the result above that this is not a simple expansion in the $r_{V_i}$, instead it depends on both $r_{V_i}$ and the CM momentum $|{\bf p}|$. In fact, this is expected in a dynamical system. Interestingly we see that in a fully dynamical system, the form of the potential does not simply follow the same scaling as one would infer from the static expansion \cite{deRham:2012fw}.
\subsubsection{Potential at order $G^2$}
At order $G^2$, we actually have to include the Born subtraction terms in order to obtain the correct potential. The result is a combination between the contribution from Eq.~\eqref{1loopg4} and the Born subtraction term
\begin{equation}
	V^\text{$\mathcal{O}(G^2)$}({\bf p},r)=-\frac{2 C_1^2g^2 G m_a^2 m_b^2}{ E_a({\bf p}) E_b({\bf p}) r} \left(\frac{2 C_2g^2 G (m_a+m_b)}{r}\right) + V^\text{$\mathcal{O}(G^2)$}_\text{Born subtr.}({\bf p},r)\ , \label{Vg2}
\end{equation}
where the Born subtraction is computed using Eq.\eqref{NLOBorn} with the leading order amplitude from Eq. \eqref{Mtree} and reads
\begin{equation}
	V^\text{$\mathcal{O}(G^2)$}_\text{Born subtr.}({\bf p},r)=-\frac{2 C_1^4g^4G^2 m_a^4 m_b^4 \left(E_a^2({\bf p})+E_b^2({\bf p})+E_a({\bf p}) E_b({\bf p})\right)}{E_a^3({\bf p}) E_b^3({\bf p}) \left( E_a({\bf p})+E_b({\bf p})\right)  r^2}\ .
\end{equation}

\subsection{Probe particle limit}\label{pp}
In this subsection we look at the probe particle limit: $m_b\ll m_a$.  From the amplitudes calculation for the conformally coupled scalar in Eq.\eqref{confRedef}, the potential in this limit is given by
\begin{align}
	V({\bf p},r)=-\frac{2 D_1^2 g^2 G m_a m_b}{ r} \Bigg(&\left(1-\frac{|{\bf p}|^2}{2 m_b^2}\right)\left(1-\frac{1}{4 \pi }\frac{r_{V_a}^3}{r^3}+\frac{2 r_{V_a}^6 }{7 \pi^2 r^6}\right)
	  \nonumber \\
	&+\left(-D_2-(D_1^2-D_2)\frac{|{\bf p}|^2}{ 2m_b^2}\right) \frac{2g^2 G  m_a}{ r}
	+\dots\Bigg) \ . \label{VfromACho}
\end{align}
We can see that the result greatly simplifies since we do not have a non-trivial expansion involving energy factors. Looking at this limit, we expect to be able to reproduce known results for the Galileon field profile.

By restricting the scalars EFT Hamiltonian to the 2-particle subspace, we obtain the first quantized Hamiltonian of two classical point-particles. In our case, since the scattering states correspond to  conformally coupled scalar, we would like to further match our results to an EFT of conformally coupled point-particles. This will be relevant in the probe particle limit where we can compute exact results with a different method. First, lets consider the following action $S=S_\text{Gal}+S_{\rm p.p.}$ where
\begin{align}
S_\text{Gal}&=\int \rd^4x \left(-\frac{1}{2}(\partial\pi)^2-\frac{1}{\Lambda^3} \square\pi (\partial\pi)^2 \right) \ ,  \label{Sgal} \\
S_{\rm p.p.}&=-\sum_{i=1}^2 m_i\int \rd^4x \int \rd \tau A(\pi) \sqrt{-g_{\mu\nu}\frac{\rd x^\mu}{\rd \tau}\frac{\rd x^\nu}{\rd \tau}} \delta^4(x-x_i(\tau)) \ . \label{Spp}
\end{align}
Here, we have included a conformal coupling of the form $S_\text{p.p.}[A^2(\pi) g;x^\mu(\tau)]$ in the point particle action where $A(\pi)$ is given by Eq.~\eqref{eq:AconfCoupl}. This gives rise to a coupling $A(\pi)T_\mu^\mu$ in the action where $T_\mu^\mu$ is the trace of the stress energy tensor defined with respect to the metric $g_{\mu\nu}$. In the probe particle limit, the field profile is simply generated by the heavy particle. Assuming spherical symmetry, we can write the solution for $\pi$ as a series expansion in $r_V$ when we are in the unscreened region,
\begin{equation}
	\pi(r)=\frac{\alpha_1 g m_a}{\mpl r}+\frac{\alpha_2 g m_a r_{V_a}^3}{ \mpl
		r^4}+\frac{\alpha_3 g m_a r_{V_a}^6}{\mpl r^7}+\cdots
\end{equation}
Now, we consider a second point particle with $m_b\ll m_a$ so that it does not affect the $\pi$ dynamics. This point-particle has the same conformal coupling with $A(\pi)$ as shown in Eq.~\eqref{Spp}. We proceed with computing its Hamiltonian in order to extract the potential $V({\bf p},r)$ and comparing it with that obtained from the amplitude's derivation. The Hamiltonian reads
\begin{equation}
	\mathcal{H}_b=\sqrt{|{\bf p}_b|^2+m_b^2 A^2(\pi)}= m_b A(\pi)+\frac{|{\bf p}_b|^2}{2m_b A(\pi)}+\mathcal{O}(|({\bf p}_b|/m_b)^4) \ , \quad {\bf p}_b=\frac{m_b A(\pi)}{\sqrt{-g_{\mu\nu}\frac{\rd x^\mu}{\rd t}\frac{\rd x^\nu}{\rd t}}}\frac{\rd {\bf{x}_b}}{\rd t} \ ,  \label{Hamilt}
\end{equation}
where  we have taken the non-relativistic limit in the second equality. Defining the potential by
\begin{equation}
	\frac{\rd {\bf{p}_b}}{\rd t}=-\boldsymbol{\nabla}V({\bf p}_b,r) \ , \label{PotFromH}
\end{equation}
and using $\mpl^{-1}=8\pi G$ we find
\begin{align}
	V({\bf p}_b,r)=\frac{8\pi \alpha_1  D_1 g^2 G m_a m_b}{ r} \Bigg(&\left(1-\frac{|{\bf p}_b|^2}{2 m_b^2}\right)\left(1+\frac{ \alpha_2 r_{V_a}^3}{\alpha_1 r^3}+\frac{\alpha_3 r_{V_a}^6 }{\alpha_1  r^6}\right) \nonumber \\
	&+ \left(D_2+(D_1^2-D_2)\frac{|{\bf p}_b|^2}{2 m_b^2}\right)\frac{8\pi \alpha_1g^2 G m_a}{D_1 r}
	+\dots\Bigg) \ .
	\label{generalVpp}
\end{align}
Note that this potential has been computed in the rest frame of particle $a$, so to compare it with the results derived from the amplitudes, we first need to transfer it to the CM frame. Since we are in the non-relativistic, probe particle limit, this is quite straightforward and we simply have to change ${\bf p}_b$ to the CM momentum ${\bf p}$. With this simple switch in place, we can now compare the potentials in Eq.~\eqref{generalVpp} with that derived using the amplitude's method in \eqref{VfromACho}. Performing the matching procedure we find that the coefficients are
\begin{equation}
	\alpha_1=-\frac{D_1}{4\pi} \ , \quad \alpha_2=\frac{D_1}{16\pi^2} \ , \quad  \alpha_3=-\frac{D_1}{14\pi^3} \ .
\end{equation}

On the other hand, the potential for a conformally coupled point-particle in the non-relativistic, probe particle limit is easily found by solving the classical equations of motion. The field profile generated by the heavy mass is (taking the stable ghost-free branch)
\begin{equation}
\pi'(r)=\frac{\Lambda^3 r}{8}\left(-1+\sqrt{1+\frac{4D_1 r_{V_a}^3}{\pi r^3}}\right) \ , \label{dGalileon}
\end{equation}
where we neglected the back-reaction from $\pi(0)$. Expanding Eq.~\eqref{dGalileon} for small $r_V/r$ gives
\begin{equation}
	\pi(r)=-\frac{D_1 g m_a}{4 \pi  \mpl r}+\frac{D_1 g m_a r_{V_a}^3}{16 \pi ^2 \mpl
		r^4}-\frac{D_1 g m_a r_V^6(m_a)}{14 \pi ^3 \mpl r^7}+\cdots \label{Galileon}
\end{equation}
The effects of higher order terms in the conformal factor $A(\pi)$ can be computed by solving the equations of motion perturbatively, and viewing the result above as the leading order term in a series expansion. At each order, including the backreaction from the Galileon corresponds to shifting the mass as $m_a\rightarrow m_a A'(\pi(0))/(D_1 g)$. Since this effect corresponds to strong coupling contributions from inside the Vainshtein radius, we neglect it for the current analysis. Comparing to the amplitudes result, we find perfect agreement. This shows that we are able to reproduce the galileon field profile from the scattering amplitudes computation, and also serves as a consistency check for our calculations.

\subsection{Galileon Potential in the Screened Region in the Probe Particle Limit} \label{screened}
The perturbative calculation performed above is valid outside the Vainshtein radius. Inside $r_V$, we are in the strong coupling regime and the potential is screened. While we cannot obtain the potential by standard scattering amplitudes calculations, we can obtain it in the probe particle limit from the exact result for the field profile. Note that in the unscreened region, the expansion parameters for the potential where
\begin{equation}
	\epsilon_G=g^2\frac{r_\text{Sch}}{r} \ , \quad \epsilon_{Vr}=\frac{r_V}{r} \ .
\end{equation}
On the other hand, in the strong coupling limit the expansion parameters are now
\begin{equation}
	\epsilon_{GV}=g^2\frac{r_\text{Sch}}{r_V} \ , \quad  \epsilon_{Vr}^{-1}=\frac{r}{r_V} \ .
\end{equation}
As previously, the Hamiltonian is given by Eq.~\eqref{Hamilt} and the potential is obtained from Eq.~\eqref{PotFromH}, but in this case we expand the Galileon field profile in the large $r_V/r$ limit. In this limit the Galileon field is
\begin{equation}
	\pi(r)=-\frac{\beta \, D_1 g m_a}{\mpl r_V}-\frac{D_1 g m_a \  \sqrt{r}}{2 \sqrt{\pi } \mpl r_V^{3/2} }-\frac{D_1 g m_a r^2}{16 \mpl
		r_V^3}+\cdots \, \label{GalileonS}
\end{equation}
where $\beta$ is a constant.
From this we can obtain the Hamiltonian $\mathcal{H}=\mathcal{H}_a+\mathcal{H}_b$, which in the CM frame reads
\begin{align}
\mathcal{H}_a=m_a+\frac{|{\bf p}|^2}{2 m_a^2}-\frac{D_1^2 g^2 m_a^2}{\mpl^2  r_{V_a}} \Bigg(&\beta
\left(1-\frac{|{\bf p}|^2}{2 m_a^2}\right)-\frac{ \beta^2 g^2 m_a}{\mpl^2  r_{V_a}} \left(3D_2-\left(3D_2-D_1^2\right)\frac{|{\bf p}|^2}{2 m_a^2}\right)\Bigg)+\cdots \ ,\\
\mathcal{H}_b=m_b+\frac{|{\bf p}|^2}{2 m_b^2}-\frac{D_1^2 g^2 m_b m_a}{\mpl^2  r_{V_a}} \Bigg(&\beta
\left(1-\frac{|{\bf p}|^2}{2 m_b^2}\right)-\frac{ \beta^2 g^2 m_a}{\mpl^2  r_{V_a}} \left(3D_2-\left(3D_2-D_1^2\right)\frac{|{\bf p}|^2}{2 m_b^2}\right)\nonumber \\
		&-\frac{1 }{2 \sqrt{\pi } }\left(\frac{r}{r_{V_a}}\right)^{1/2}\left(1-\frac{|{\bf p}|^2}{2 m_b^2}\right)+\frac{1 }{16}\left(\frac{r}{r_{V_a}}\right)^{2} \left(1-\frac{|{\bf p}|^2}{2 m_b^2}\right) \nonumber \\
		&+\frac{2 A g^2 m_a  \sqrt{r} }{\sqrt{\pi } \mpl^2 r_{V_a}^{3/2}}\left( D_2-\left(
		D_2-\frac{D_1^2}{2}\right)\frac{|{\bf p}|^2}{2 m_b^2}\right)+\cdots\Bigg)+\cdots \ .
\end{align}
Note that the mass and kinetic terms have been shifted by a small constant factor of order $\epsilon_{GV}$. This is just a dressing of the mass which can be absorbed in the definitions of the masses if desired. From the above, we see that the potential is given by
\begin{align}
	V(|{\bf p}|,r)=&-\frac{D_1^2 g^2 m_b m_a}{\mpl^2  r_{V_a}} \Bigg(\frac{ 1 }{2 \sqrt{\pi } }\left(\frac{r}{r_{V_a}}\right)^{1/2}\left(1-\frac{|{\bf p}|^2}{2 m_b^2}\right)-\frac{1 }{16}\left(\frac{r}{r_{V_a}}\right)^{2} \left(1-\frac{|{\bf p}|^2}{2 m_b^2}\right)+\cdots\Bigg)  \nonumber \\
	&+\frac{2 A g^2 m_a  \sqrt{r} }{\sqrt{\pi } \mpl^2 r_{V_a}^{3/2}}\left( D_2-\left(
	D_2-\frac{D_1^2}{2}\right)\frac{|{\bf p}|^2}{2 m_b^2}\right)+\cdots\Bigg)+\cdots \ .
\end{align}
Note that, when we include terms of order $\epsilon_{GV}^2$ or higher, we also include the backreaction to the equations of motion from the $\pi(0)$ terms since they start to contribute at this order. This effect can also be seen as a redressing of the mass, since it only shifts it by a constant factor of order $\epsilon_{GV}$, and can be absorbed in the definition of $m_a$.
\section{Scattering Angle and Phase Shift} \label{sec:scatAngle}
Given the results in the previous sections, we are now in a position to compute the scattering angle in a collision of two massive spinless particles conformally coupled to the Galileon field. The scattering angle is given as usual by
\begin{equation}
\chi=-2\int_{r_\text{min}}^{\infty} d r \frac{\partial p_{r}}{\partial L}-\pi \ , \label{chi}
\end{equation}
where $r_\text{min}$ is the distance of closest approach between the two particles, $p_r$ is their radial momentum, and $L$ the angular momentum. Since the energy $E$ is conserved, we set $E=\mathcal{H}$. By doing so, we can solve for the CM momentum as
\begin{equation}
	|{\bf{p}}|^2=	|{\bf{p}}_r|^2+\frac{L^2}{r^2}=	|{\bf{p}}_\infty|^2-V_\text{eff}(E,r) \ , \label{PVeff}
\end{equation}
where
\begin{equation}
|{\bf{p}}_\infty|^2=\frac{\left(E^{2}-m_{a}^{2}-m_{b}^{2}\right)^{2}-4 m_{a}^{2} m_{b}^{2}}{4 E^{2}}
\end{equation}
is the momentum at $r=\infty$, that is, when the interactions are switched off. Here, $|{\bf{p}}_\infty|^2$ can formally be thought of as an effective non-relativistic Hamiltonian \cite{Damour:2017zjx,Bjerrum-Bohr:2019kec,Kalin:2019rwq} where $V_\text{eff}$ represents a small perturbation. Since the integral in Eq.~\eqref{chi} requires the evaluation at $r_\text{min}$ which is a complicated function which sometimes has no analytic expression, it is useful to rewrite the expression for the scattering angle in a different way. By performing a change of variables to $u=\sqrt{r^2+r_\text{min}^2}$ and performing some algebraic manipulations, the scattering angle can be written as \cite{doi:10.1063/1.1705426,Wallace:1973iu,Bjerrum-Bohr:2019kec}
\begin{equation}
	\chi=\sum_{k=1}^{\infty} \widetilde{\chi}_{k}(b), \quad \widetilde{\chi}_{k}(b) \equiv \frac{2 b}{k !} \int_{0}^{\infty} \rd u\left(\frac{\rd}{\rd b^{2}}\right)^{k} \frac{V_\text{eff}^{k}(E,\sqrt{u^2+b^2}) (u^2+b^2)^{(k-1)}}{|{\bf{p}}_\infty|^{2 k}} \ , \label{chiSeries}
\end{equation}
where $b=L/|{\bf{p}}_\infty|$ is the impact parameter. This expression for the scattering angle is simply the Eikonal/WKB expansion. For example, the first term of this series corresponds to the standard {\it textbook} formula for the scattering angle when taking the non-relativistic limit \cite{BohmQT}. It is interesting to point out that some of the simplifications that appear in the computation of the scattering angle in gravity will not occur in our case. For the gravitational potential, it is known that the leading order term in Eq.~\eqref{chiSeries} in 4D is surprisingly valid up to second post-Minkowskian order, that is, up to order $G^2$. A detailed exploration which explains why this is the case has been performed in \cite{Bjerrum-Bohr:2019kec,Cristofoli:2020uzm}. For the Galileon potential, such special cancellations do not happen and we need to include the NLO in the series in Eq.~\eqref{chiSeries} to obtain all the $\mathcal{O}(G^2)$ terms. This is the case here since we have a double series expansion in both $\epsilon_{Gb}=g^2r_\text{Sch}/b$ and  $\epsilon_{Vb}=r_V/b$. While there is no contribution from the NLO term in Eq.~\eqref{chiSeries} at order $\epsilon_{Gb}^2\epsilon_{Vb}^0$, we should expect contributions for all $\epsilon_{Gb}^2\epsilon_{Vb}^{3n}$, with $n\neq 0$.

In the next subsections, we will find the analytic expressions for the scattering angle when the scattering occurs outside the screened region. Firstly, we want to analyze the parameter space for which  $ r_V<r_\text{min}< b$. This are the cases when the particles start outside their Vainshtein radius, but as they evolve they come inside it. Our results are not a good approximation for those cases. For non-relativistic scatterings, we can solve for the distance of closest approach by solving
\begin{equation}
	E\simeq m_a+m_b+\frac{L^2}{2m_a r^2}+\frac{L^2}{2m_b r^2}+V({\bf p},r)^\text{LO} \ ,
\end{equation}
with $V({\bf p},r)^\text{LO}$ given by the leading order of Eq.~\eqref{VGrv6}. At leading order we find
\begin{equation}
	r_\text{min}\simeq\frac{b \left(-\epsilon_{Gb}+\sqrt{4 \epsilon_{pm}+\epsilon_{Gb}^2}\right)}{2\epsilon_{pm}} \ ,
\end{equation}
where  we have defined $\epsilon_{pm}\equiv\left(E-m_a-m_b\right)/\mu$ with $\mu$ the reduced mass of the binary. One can notice that for $\epsilon_{pm}\gg\epsilon_{Gb}$, the minimum radius is of the order of the impact parameter. On the other hand, for $\epsilon_{pm}\ll\epsilon_{Gb}$ we have
\begin{equation}
	r_\text{min}\simeq\frac{b^2 \left(E-m_a-m_b\right)}{C_1^2 \mu \  r_{\text{Sch}}}=
	r_V\left(\frac{\epsilon_{pm}}{C_1^2 \ \epsilon_{Gb} \ \epsilon_{Vb}}\right) \ . \label{rmin}
\end{equation}
 Thus, we can see that when the parenthesis in the RHS is smaller than one, that is, when we are in the highly non-relativistic limit and the impact parameter is not too far from $r_V$, we can have the particle coming inside the Vainshtein radius even if $b$ was outside of it. This approximation is not valid for $r_\text{min}\ll r_V$ which occurs when the Galileon term dominates over the angular momentum contribution, but is sufficient for our purposes of understanding when the distance of closest approach will be in the screened region.

\subsection{Scattering Angle directly from Scattering Amplitudes for  $r_V< r_\text{min}< b$}
When the particles stay always outside the Vainshtein radius of each other, the conservative potential is a series in $\epsilon_G$ and $\epsilon_{Vr}$ as seen in the Subsection \ref{subsec:pot}. In this subsection, we would like to compute the scattering angle arising from such an interaction. For simplicity, we will take advantage of recently discovered relations between the scattering angle and scattering amplitudes. In \cite{Damour:2017zjx,Bjerrum-Bohr:2019kec,Kalin:2019rwq}, it has been shown that Eq.\eqref{PVeff} can be written in terms of the classical scattering amplitude in GR as
\begin{equation} \label{pRelM}
|{\bf{p}}|^2=|{\bf{p}}_\infty|^2-\frac{1}{2E} \mathcal{M}_\text{cl.}(|{\bf{p}}_\infty|,r) \ ,
\end{equation}
where $E=E_a+E_b$. A straightforward generalization of the proof in \cite{Bjerrum-Bohr:2019kec} using the implicit function theorem for functions $F:\mathbb R^3\rightarrow\mathbb R$ shows that this formula is valid for potentials with a double expansion such as ours, see Appendix \ref{ap:Proof} for details. Note that this relation is only valid in the conservative sector. At higher orders, radiative corrections become relevant for the classical dynamics and the relationship above becomes non-linear. We will neglect those effects here. Similarly, non-linear terms appear in $D>4$ \cite{Cristofoli:2020uzm}. We can systematically compute the scattering angle by using the results from previous sections together with Eq.~\eqref{chiSeries}, which after using Eq.~\eqref{pRelM}, can be written as
\begin{equation}
	\chi=\sum_{k=1}^{\infty} \widetilde{\chi}_{k}(b), \quad \widetilde{\chi}_{k}(b) \equiv \frac{2 b}{k !} \int_{0}^{\infty} \rd u\left(\frac{\rd}{\rd b^{2}}\right)^{k} \frac{1}{(2E)^k} \frac{\mathcal{M}_\text{cl.}^k(|{\bf{p}}_\infty|,\sqrt{u^2+b^2}) (u^2+b^2)^{(k-1)}}{|{\bf{p}}_\infty|^{2 k}} \ . \label{chiM}
\end{equation}
At next-to-leading order we find
\begin{subequations}
\label{chiU}
\begin{align}
	\chi_{\epsilon_{Gb}}&=\frac{4 D_1^2 g^2 G m^4 \nu^2 }{ E b |{\bf{p}}_\infty|^2} \Bigg(1-\frac{3}{16} \left(\frac{r_{V_a}^3}{b^3}+\frac{r_{V_b}^3}{b^3}\right) \nonumber \\
	&+\frac{32}{35} \left(\frac{r_{V_a}^6}{b^6}+\frac{39r_{V_a}^3r_{V_b}^3}{32 b^6}\left(1-\frac{2}{21\nu}\left(\frac{24}{\nu}-23\right)\frac{|{\bf{p}}_\infty|^2}{m^2}+\mathcal{O}(||{\bf{p}}_\infty|^4)\right)+\frac{r_{V_b}^6}{b^6}\right)\!\!\Bigg),\\
	\chi_{\epsilon_{Gb}^2}&=-\frac{16 D_1^4 g^4 G^2 m^8 \nu^4 }{\pi b^2 \mpl^2 |{\bf{p}}_\infty|^4 E^2} \left(\frac{r_{V_a}^3}{b^3}+\frac{r_{V_b}^3}{b^3}\right)-\frac{2\pi D_1^2(D_1^2+2D_2) g^4G^2 m^3 \nu^2}{ E  b |{\bf{p}}_\infty|^2} \ ,
\end{align}
\end{subequations}
where $m$ and $\nu$ are defined in Eq.~\eqref{eq:masses}. Here we have included the contribution at order $G^2r_V^3$ that arises from the next-to-leading order term in Eq.~\eqref{chiM}. Note that we have a non-trivial dependence on the energy of the particles in the series expansion. As in the case of the conservative potential, we do not have a fully PM expansion for $\chi_{\epsilon_{Gb}}$ since it requires resumation of higher momentum contributions in the H-type graphs. Instead, we have used the PN expansion for the scattering amplitudes at order $r_{V_a}^3r_{V_b}^3$ in the result above. Once we have the scattering angle, it is straightforward to obtain the phase shift, $\delta$, since
\begin{equation}
\chi=-\frac{1}{|{\bf{p}}_\infty|}\frac{\partial \delta}{\partial b} \ .
\end{equation}
The phase shift at next-to-leading order reads
\begin{subequations}
	\label{eq:PhaseShift}
	
	\begin{align}
		\delta_{\epsilon_{Gb}}&=-\frac{4 D_1^2 g^2 G m^4 \nu^2}{ E |{\bf{p}}_\infty|} \Bigg(\log(b)+\frac{1}{16} \left(\frac{r_{V_a}^3}{b^3}+\frac{r_{V_b}^3}{b^3}\right) \nonumber \\
		&-\frac{16}{105} \left(\frac{r_{V_a}^6}{b^6}+\frac{399r_{V_a}^3r_{V_b}^3}{32 b^6}\left(1-\frac{2}{21\nu}\left(\frac{24}{\nu}-23\right)\frac{|{\bf{p}}_\infty|^2}{m^2}+\mathcal{O}(||{\bf{p}}_\infty|^4)\right)+\frac{r_{V_b}^6}{b^6}\right)\!\!\Bigg)\\
		\delta_{\epsilon_{Gb}^2}&=-\frac{4 D_1^4 g^4 G^2m^8 \nu^4 }{\pi b \mpl^2 |{\bf{p}}_\infty|^3 E^2} \left(\frac{r_{V_a}^3}{b^3}+\frac{r_{V_b}^3}{b^3}\right)-\frac{2\pi D_1^2(D_1^2+2D_2) g^4G^2 m^3\nu^2 }{32\pi E \mpl^4 b^2 |{\bf{p}}_\infty|^2} \ .
	\end{align}
\end{subequations}
\subsection{Probe Particle Limit}
\begin{figure}[!t]
	\includegraphics[scale=1]{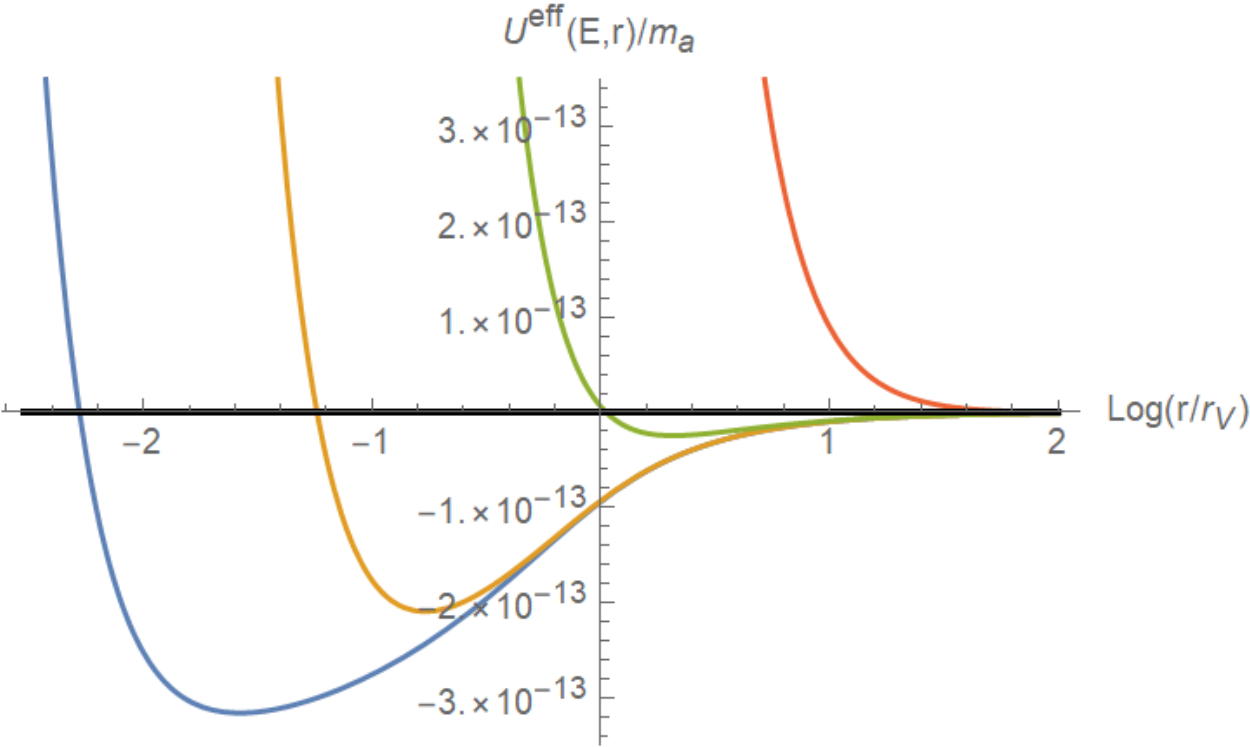}
		\caption{Plot of the effective potential for $\epsilon_{GV}=10^{-11}$ (which corresponds to $\Lambda=10^{-12}eV$ and a mass $m=10^6 M_\odot$), $\epsilon_{Vb}=10^{-2}$, and $m_b/m_a=10^{-2}$. The blue line correspond to $\epsilon_{pm}\epsilon_M=10^{-21}$. For the other color lines, $\epsilon_{pm}\epsilon_M$ increases by two orders of magnitude from left to right. The thick black line is $\epsilon_{pm}\epsilon_M$, thus the intersection of the color lines with the black line gives $r_\text{min}$. Note that these potentials were computed using the exact solution Eq.~\eqref{dGalileon}, so they are valid both inside and outside Vainshtein radius.}
		\label{Ueffplot}
\end{figure}

 In this subsection, we will compute the scattering angle analytically in the probe particle limit. We can analyze the effective potential to understand the expected orbits of the scattering particles. We define the effective potential as
\begin{equation}
U^\text{eff}=\mathcal{H}\left(p=\frac{L}{r}\right)-m_a-m_b \ ,
\end{equation}
where the Hamiltonian reads
\begin{equation}
\mathcal{H}=\sqrt{|{\bf p}|^2+m_a^2}+\sqrt{|{\bf p}|^2+m_b^2 A^2(\pi)} \ ,
\end{equation}
and we use the exact solution for the Galileon field, Eq.~\eqref{dGalileon}. In Fig.\ref{Ueffplot}, we can see a plot of the effective potential for different $\epsilon_{pm}$ and fixed $\epsilon_{GV}$, $\epsilon_{Vb}$, and $\epsilon_M\equiv m_b/m_a$. The distance of closest approach corresponds to the intersection of $U^\text{eff}/m_a$ and $\epsilon_{pm}\epsilon_M$ (black line). As expected from Eq.~\eqref{rmin}, we see that only in the highly non-relativistic limit does the particle enter the screened region. For cases where the Galileon potential dominates over the angular momentum contribution in a large region such as the blue graph in Fig.~\ref{Ueffplot}, the probe particle will come inside the Vainshtein radius. On the other hand, in cases where the Galileon term does not heavily dominate such as the red graph in Fig.~\ref{Ueffplot}, we always stay outside the screened region.\\

Now, we proceed to find the scattering angle when the particles are always outside each others Vainshtein radius by using Eq.~\eqref{chiSeries}; this reads
\begin{subequations}
\begin{align}
\chi_{\epsilon_{Gb}}&=\frac{\epsilon _{\text{Gb}} \epsilon _M  }{\epsilon _{\text{pm}}}\left(1-\frac{3}{16} \ \epsilon _{\text{Vb}}^3+\frac{32}{35\pi^2} \ \epsilon _{\text{Vb}}^6 + \cdots \right) \label{eq:chi1}\\
\chi_{\epsilon_{Gb}^2}&= -\frac{ \epsilon_{\text{Gb}}^2 \epsilon _M^2 \epsilon _{\text{Vb}}^3}{ \pi   \epsilon _{\text{pm}}^2}
-\frac{\epsilon_{\text{Gb}}^2 \epsilon _M\left(1+2D_2/D_1^2\right) \pi \left(1-\frac{8}{3\pi^2} \epsilon _{\text{Vb}}^3
\right)}{4  \epsilon _{\text{pm}}} \ . \label{eq:chi2}
\end{align}
\label{chiS}
\end{subequations}
We have check that these approximations are accurate in the corresponding situations by performing the integrals in Eq.~\eqref{chiSeries} numerically. Note that the case when the impact parameter is inside the Vainshtein radius requires the full series in $\epsilon_{Vb}$, while we do not give an analytic expression which might be possible to obtain via resummation, we note that the scattering angle can be obtained numerically as long as $\epsilon_{GV}\ll1$, which is expected in physically relevant cases. As a consistency check, the results in Eq.~\eqref{chiU} match those of Eq.~\eqref{chiS} in the probe particle, non-relativistic limit.

To get more insight into the expression for the scattering angle above, is worth understanding the order of magnitude of each different term. For example, for a supermassive black hole of mass $m=10^6 M_\odot$ and for $\Lambda=10^{-21}eV$ which is of the order of the largest value allowed by lunar laser ranging observations~\cite{Murphy_2012,Dvali:2002vf}, we have $\epsilon_{GV}=10^{-11}$. For smaller black holes, $\epsilon_{GV}$ will be much smaller. Let us further assume a not so large difference in masses $\epsilon_M=10^{-2}$, an impact parameter not so far from the Vainshtein radius $\epsilon_{Vb}=10^{-2}$, and $\epsilon_{pm}=10^{-10}$ so that the distance of closest approach is outside the Vainshtein radius. Then we have the following orders of magnitude for the different contribution to the scattering angle
\begin{equation}
	\chi_{\epsilon_{Gb}}\simeq10^{-5} \ , \quad \chi_{\epsilon_{Gb}\epsilon_{Vb}^3}\simeq10^{-12} \ , \quad \chi_{\epsilon_{Gb}\epsilon_{Vb}^6}\simeq10^{-18} \ , \quad \chi_{\epsilon_{Gb}^2}\simeq10^{-18}  \ , \quad \chi_{\epsilon_{Gb}^2\epsilon_{Vb}^3}\simeq10^{-16} \ .
\end{equation}
In this example, the order $G^2$ term is of the order of the $Gr_V^6$ contribution. In fact, if $\epsilon_{GV}>10^{-1}\epsilon_{Vb}^5$, then the $G^2$ correction would be larger. Furthermore, we can notice that the $G^2r_V^3$ contribution is larger than the  $Gr_V^6$  and $G^2$ ones. This can be surprising since we expect higher $r_V$ contributions to be suppressed. In fact, that is the case for the last term of Eq.\eqref{eq:chi2}, but not for the first one. The first term arises from the next-to-leading order Eikonal in Eq.~\eqref{chiSeries} while the last one comes from the leading order Eikonal. This means that the series expansion for the potential is correct, but when computing the scattering angle one should be careful and include higher order corrections which depend on lower order potential terms. Furthermore, this relevant correction can be computed away from the probe particle limit, as seen in Eq.~\eqref{chiU}, since it only requires lower order amplitudes which we have computed here. On the other hand, the contribution from scattering amplitudes of order $G^2r_V^3$ will always be subleading.

\section{Final remarks} \label{sec:final}
Starting from the scattering amplitudes, we have computed the conservative potential and scattering angle (phase shift) for the binary system of compact spinless objects coupled to a  gravitational spin-0 mode given by Galileons with cubic self-interactions. We have checked that in the point particle limit this reduces to well-known results. This shows that scattering amplitude methods which have been recently use to compute post-Minkowskian calculations are applicable beyond General Relativity and minimal couplings. When including non-minimal couplings, one should be careful in matching the correct scattering states between different EFTs. While we focused on a conformal coupling, we expect that the treatment of a disformal coupling should follow in a similar manner.

Most of the results in this paper are applicable when each object in the binary is outside the Vainshtein radius of the other. While this will not be the case for generic scenarios for which solar system tests impose an upper bound on the interaction scale $\Lambda$, this can be valid in more general scenarios or in special situations. Indeed, it is worth pointing out  that the coupling strengths can be redressed on specific backgrounds in which the binary is embedded. For adequate backgrounds, the redressed Vainshtein radius can end up being much smaller than the bare one. In these cases, the objects in the binary will always be outside the redressed Vainshtein radius of each other and perturbation theory will be valid. A special situation where such interesting dynamics can happen was analyzed in \cite{Brax:2020ujo}, where a binary of small black holes is considered inside the Vainshtein radius of  super-massive black hole. Our calculations are applicable for the dynamics of the small black hole binary and thus can help with tests of such three-body systems.

These new results open a window of opportunity for applying these techniques for more general theories of modified gravity. Here, we focused on the new physics arising in the decoupling limit, but extending the results to include interactions with the helicity-2 modes is straightforward. Similarly, one can extend these results to more general scalar-tensor theories and modified gravity theories which generically include extra degrees of freedom. It would also be interesting to apply these techniques to gravitational theories involving helicity-1 modes.  Another interesting case that could be analyzed with these methods is the case of quartic and quintic Galileons and other operators that manifest a screening mechanism. Standard methods to compute radiation from binary systems considering classical perturbations around a spherical background break down for the quartic Galileon \cite{deRham:2012fg} and it would be interesting to understand if using scattering amplitudes computations can give rise to any insights on how to correctly tackle these problems.

\section{Acknowledgments}

The authors would like to thank the Galileo Galilei Institute for Theoretical Physics for hosting the workshop on Gravitational scattering, inspiral, and radiation which fostered interesting conversations within the framework on this work.
The research of MCG, AJT and CdR is supported by STFC grants ST/P000762/1 and ST/T000791/1.  MCR and CdR are also supported by the European Union's Horizon 2020 Research Council grant 724659 MassiveCosmo ERC--2016--COG. CdR is supported by a Simons Foundation award ID 555326 under the Simons Foundation's Origins of the Universe initiative, `\textit{Cosmology Beyond Einstein's Theory}' and by a Simons Investigator award 690508.  AJT and CdR both thank the Royal Society for support at ICL through a Wolfson Research Merit Award.

\appendix
\section{Graphs Contributing in the Classical Limit.} \label{ap:Classical}
\begin{figure}[!h]
	\includegraphics[scale=0.8]{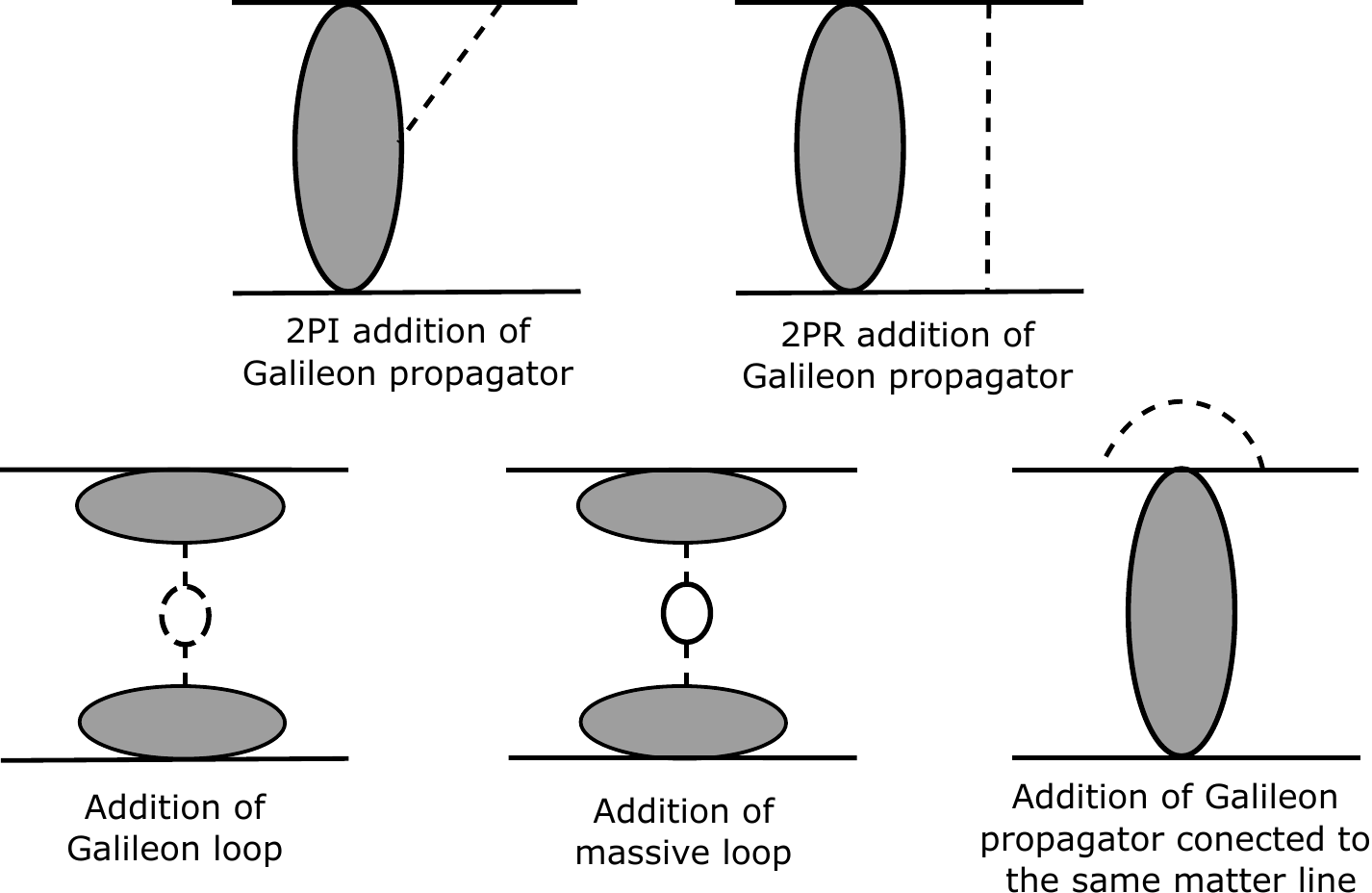}
	\caption{Different possibilities of loop contributions by adding a new propagator or loop to a graph with classical contributions. The grey blobs represent classical contributions. When we have two blobs, these were classical without the new insertion between them. The graphs in the top row can contribute in the classical limit. The ones in the bottom row are always quantum in nature. }
	\label{loopsInt}
\end{figure}

In this Appendix we analyze the types of graphs that can contribute in the classical limit. There is one noticeable difference with respect to the standard GR case. Here, not every graph gives an expansion in $r_V$. In GR, we only have one scale controlling all the interactions, namely $G_N$, so the result is a series in $r_\text{Schw}$. Here, we have two scales: $\mpl$ and $\Lambda^3$. Since the matter couplings scale in a different way than the self-interactions, insertions of loops scale differently than insertions of propagators connected to external massive scalars. Note that the classical contributions arise from the non-analytic structure $\sqrt{m^2/q^2}$, which can only appear in the presence of propagation of massless particles. We now turn to analyze each of the operations that can generate graphs with classical contributions. Here, we only discuss explicitly the case of a cubic an quartic matter couplings, but it is straightforward to understand interactions with more fields involved. Starting from an already classical graph, we can add extra interactions to build higher loop corrections. In the following, we analyze the different kind of extra interactions that can be added and which can contribute in the classical limit, examples of these graphs can be seen in Fig.~\ref{loopsInt}. For simplicity, we take all the external particles to have mass $m$ in this analysis.

\begin{description}
	\item[2PI addition of $\pi$ propagator] We consider two possible insertions of a new $\pi$ propagator. The first one corresponds to adding a propagator connected to an external $\phi$ and an internal $\pi$ to a graph. To give a classical contribution, it should add a power of  $M\lambda q/\Lambda^3$ to an already classical contribution. Adding this propagator to a graph that gives rise to a classical contribution, the new loop will lead to an amplitude of the form
	\begin{equation}
		\mathcal{M}\sim \int \frac{\rd^4 k}{(2\pi)^4} \frac{g m^2}{\Lambda^3 \mpl}\frac{N(q,p,k)}{(-(p+k)^2-m^2+i\epsilon)(-k^2+i \epsilon)(P(k,q))}A_\text{c}(q,p,k)  \ ,
	\end{equation}
	where $A_\text{c}(q,p,k)$ has classical scaling and $P(k,q)$ is the new propagator in the classical blob generated by the insertion of the $\pi$ line. The numerator $N(q,p,k)$ has 4 powers of momenta since it arises from the cubic Galileon vertex.  Performing the power counting from Section \ref{sec:ClassA} and realizing that this is a triangle integral, we see that after integration a factor of $\sqrt{m^2/q^2}$ will arise. Since $N$ has 4 factors of momenta from the 3-point vertex in the new loop, the total scaling of the graph at leading order in q will be
	\begin{equation}
		A_\text{c} \frac{g m^2}{\Lambda^3 \mpl} \frac{q^4}{m^2}\sqrt{\frac{m^2}{q^2}} \sim  A_\text{c} \ (r_V q)^3 \ ,
	\end{equation}
	which is a classical correction. Note that this integral will also contain higher orders in $q/m$ which we have ignore since they are quantum contributions.
	The second option is to change a cubic vertex $\phi^2\pi$ to a quartic one $\phi^2\pi^2$ and attach the other side of the $\pi$ propagator to a different matter line. The case when the Galileon propagator has both ends in the same matter line will be discussed below. In this case, no new Galileon vertex arises but the power counting in $1/\mpl$ is increased. The new amplitude will now scale as
	\begin{equation}
		\mathcal{M}\sim \int \frac{\rd^4 k}{(2\pi)^4} \frac{g^2 m^2}{\mpl^2}\frac{A_\text{c}(q,p,k)}{(-(p+k)^2-m^2+i\epsilon)(-k^2+i \epsilon)}  \ .
	\end{equation}
	Following the same procedure as before, we see that this will again gives rise to the necessary non-analyticity and the overall scaling of the graph will be
	\begin{equation}
		A_\text{c} \frac{g^2 m^2}{\mpl^2} \frac{q^2}{m^2}\sqrt{\frac{m^2}{q^2}} \sim  A_\text{c} \ (r_S q) \ ,
	\end{equation}
	which indeed gives a classical contribution.
	
	\item[2PR addition of $\pi$ propagator] This corresponds to the addition of a $\pi$ propagator connected to two external massive scalars and gives rise to a two massive particle irreducible graph that is approximated by
	\begin{equation}
		\hspace{-0.5cm}\mathcal{M}\sim \int \frac{\rd^4 k}{(2\pi)^4} {\left(\frac{g m^2}{\mpl}\right)^2}\frac{A_\text{c}(q,p,k)
		}{((p+k)^2+m^2-i\epsilon)((p'-k)^2+m^2-i\epsilon)((q-k)^2-i\epsilon)} \ . \label{2PRloop}
	\end{equation}
	This graph gives a classical contribution if it scales like as $A_\text{c}g^2m  q/\mpl^2$. Note that $ A_\text{c}(q,p,k) $ should contain a $k^2$ in the denominator. Thus, the scalar integral corresponds to a box integral and will not give rise to a classical contribution at leading order. Since we are performing an expansion in $q/M$, higher orders in the expansion could contribute classically and one should check in a case by case basis. More explicitly, this graphs scale as
	\begin{equation}
		A_\text{c} \frac{g^2 m^2}{\mpl^2}\left(1+\sqrt{\frac{q^2}{m^2}}+\cdots\right) \sim 	A_\text{c} (r_S m)+A_\text{c} (r_S q)\ ,
	\end{equation}
	where the second term is a classical contribution. These terms can vanish for certain graphs so it should be checked case by case.

	\item[Addition of a $\pi$ loop] The addition of a loop can performed in two ways. One is replacing a three point vertex by a loop and the second way we can add a loop is by inserting it in a $\pi$ propagator. In both cases we have
	\begin{equation}
		\mathcal{M}\sim \int \frac{\rd^4 k}{(2\pi)^4} \frac{1}{\Lambda^6}\frac{N^2(q,p,k)}{(k^2-i \epsilon)P(k,q)P'(k,q)}A_\text{c}(q,p,k)  \ .
	\end{equation}
	In order for these integrals to give classical contributions, they need to scale as $A_\text{c} Mq^5/\Lambda^6 $. Since we are adding a loop that does not involve massive particles, the non-analytic structure $\sqrt{m^2/q^2}$ that is require to give a classical correction cannot arise here.
	
	\item[Addition of a massive $\phi$ loop] The addition of a $\phi$ loop arises from inserting it in a $\pi$ propagator. The amplitude is now
	\begin{equation}
		\mathcal{M}\sim \int \frac{\rd^4 k}{(2\pi)^4} {\left(\frac{g m^2}{\mpl^2}\right)^2} \frac{A_\text{c}(q,p,k)}{(k^2+m^2-i \epsilon)P(k,q,m)P'(q,m)}  \ .
	\end{equation}
	The non-analytic contribution needed for the classical limit cannot arise from a massive loop, since it does not contribute to the long range propagation where the non-analyticity is generated.
	
	\item[Addition of $\pi$ propagator connected to the same matter line]
	We now consider the case of a $\pi$ propagator whose ends are attached to the same matter line. This occurs when we attach one end of the $\pi$ line to the matter line and the second one right after one or several interchanges of other Galileon fields. This is just a vertex correction or the so-called mushroom graphs. Here, the momentum of the $\pi$ propagator does not scale like $q$, and thus we can not obtain the required power counting for a classical contribution.
	
\end{description}
Intuitively, it is clear why the last three cases correspond to quantum processes. They are graphs that either contribute to renormalizing the wave-function, mass, or coupling strength; or contribute to the Galileon form factors for the scalars $\phi$.

\pagebreak

\section{Proof of the Relation between Momentum and Scattering Amplitude} \label{ap:Proof}
In this appendix we will show that Eq.~\eqref{pRelM} is satisfied for a potential of the form
\begin{equation}
V(|{\bf p}|,r)=\sum_{n,m}c_{n,m}(|{\bf p}|)\left(\frac{r_1}{r}\right)^n\left(\frac{r_2}{r}\right)^{m} \ . \label{eq:Vdouble}
\end{equation}
In our case $r_1$ and $r_2$ denote the Schwarzschild and Vainshtein radius respectively, but we keep the discussion general by taking the ratios $r_i/r$ to be arbitrary expansion parameters. The proof is a straightforward generalization of that in \cite{Bjerrum-Bohr:2019kec}. In this case, we use the implicit function theorem for $\mathbb{R}^3$ instead of $\mathbb{R}^2$. The implicit function theorem reads
\begin{theorem}
Let $F$ be a continuously differentiable function $F : \mathbb{R}^{n+m}\rightarrow\mathbb{R}^m$, where $(x,y)$ are coordinates of $\mathbb{R}^{n+m}$ and $(x_0, y_0) \in \mathbb{R}^{n+m}$ such that $F(x_0, y_0) = 0\in\mathbb{R}^m$. If the Jacobian matrix is invertible, i.e., if 
\begin{equation}
	\frac{\partial F_i}{\partial y_j}\neq 0 \ ,
\end{equation}
then there is an open set $U\subset \mathbb{R}^n$ containing $x_0$ such that there is a unique continuously differentiable function $g(x)$ that satisfies
\begin{equation}
 F(x, g(x)) = 0 \ , \quad\forall x\in U \ . \label{eq:ImplFTh}
\end{equation}
\end{theorem}
The derivatives of $g$ can be found by differentiating Eq.~\eqref{eq:ImplFTh}. Here we will focus on the case where $n=2$ and $m=1$. In such case the first derivatives of the function $g$ at $x_0=\left(x^1_{0},x^2_0\right)$ are
\begin{subequations}
\begin{align}
\partial_{x_i} g\left(x_0\right)&=-\left.\frac{\partial_{x_i} F}{\partial_{g} F}\right|_{x=x_{0}} \ , \label{eq:d1}\\
\partial_{x_i}\partial_{x_j} g\left(x_{0}\right)&=-\left.\frac{\partial_{x_i}\partial_{ x_j} F+ (\partial_{x_i} g) \partial_{x_j}\partial_{ g} F+ (\partial_{x_j} g) \partial_{x_i}\partial_{ g} F+(\partial_{x_i} g)(\partial_{x_j} g)\partial_{g}\partial_{ g} F }{\partial_{g} F}\right|_{x=x_{0}}  \ , \label{eq:d2}
\end{align}
\end{subequations}
while higher order ones are computed by differentiating Eq.~\eqref{eq:ImplFTh} to higher orders. To apply this theorem for our purposes, we choose $x=\{r_1,r_2\}$, $x_0=\{0,0\}$, $y_0=|{\bf p_\infty}|^{2}$ $g=|{\bf p}|^{2}(r_1,r_2)$, and
\begin{equation}
F(\{r_1,r_2\},y)=\sum_{i=1}^{2} \sqrt{|{\bf p}|^{2}+m_{i}^{2}}+V(|{\bf p}|, r)-E \ .
\end{equation}
Then the implicit theorem tells us that there exists a unique $|{\bf p}|^{2}$ such that
\begin{equation}
|{\bf p}|^{2}=|{\bf p_\infty}|^{2}+\sum_{n,m} \left.\frac{r_1^n r_2^m}{n! m!}\left(\partial_{r_1}^n\partial_{r_2}^m|{\bf p}|^{2}\right)\right|_{\{r_1,r_2\}=\{0,0\}} \ , \label{eq:impulse}
\end{equation}
where
\begin{subequations}
	\label{eq:dp2}
\begin{align}
\left.\partial_{r_i}|{\bf p}|^{2}\right|_{\{r_1,r_2\}=\{0,0\}} =&-\left.\frac{\partial_{r_i} V}{\frac{E}{2 E_1E_2}+\partial_{|{\bf p}|^{2}} V}\right|_{\{r_1,r_2\}=\{0,0\}} \ ,\\
\left.\partial_{r_i}\partial_{r_j}|{\bf p}|^{2}\right|_{\{r_1,r_2\}=\{0,0\}}  =&-\left.\frac{\partial_{r_i}\partial_{ r_j} V+ (\partial_{r_i} |{\bf p}|^{2}) \partial_{r_j}\partial_{ |{\bf p}|^{2}} V+ (\partial_{r_j} |{\bf p}|^{2}) \partial_{r_i}\partial_{ |{\bf p}|^{2}} V }{\frac{E}{2 E_1E_2}+\partial_{|{\bf p}|^{2}} V}\right|_{\{r_1,r_2\}=\{0,0\}} \nonumber \\
 &-\left.\frac{ (\partial_{r_i} |{\bf p}|^{2})(\partial_{r_j} |{\bf p}|^{2})(-\frac{1}{E_a^3}-\frac{1}{E_b^3}+\partial_{|{\bf p}|^{2}}\partial_{ |{\bf p}|^{2}} V) }{\frac{E}{2 E_1E_2}+\partial_{|{\bf p}|^{2}} V}\right|_{\{r_1,r_2\}=\{0,0\}} \ .
\end{align}
\end{subequations}
On the other hand, from the Lippmann-Schwinger equation one can write the classical amplitude as
\begin{equation}
\frac{\mathcal{M}_{c l .}({\bf p},{\bf p+q})}{4E_1E_2}=\sum_{n=0}^{\infty} \int_{k_{1}, k_{2}, \ldots, k_{n}} \frac{{V}\left({\bf p},{\bf k}_{1}\right) {V}\left({\bf k}_1,{\bf k}_{2}\right) \cdots {V}\left({\bf k}_n,{\bf p+q}\right)}{\left(E_{p}-E_{k_{1}}\right)\left(E_{k_{1}}-E_{k_{2}}\right) \cdots\left(E_{k_{n-1}}-E_{k_{n}}\right)} \ ,
\end{equation}
where ${V}\left({\bf k}_i,{\bf k}_j\right)$ is the Fourier transform of Eq.~\eqref{eq:Vdouble} and reads
\begin{equation}
{V}\left({\bf k}_i,{\bf k}_j\right)=\sum_{n,m}r_1^nr_2^m\frac{(4 \pi)^{\frac{3}{2}}}{2^{n+m}} \frac{\Gamma\left(\frac{3-n-m}{2}\right)}{\Gamma\left(\frac{n+m}{2}\right)} \frac{c_{n,m}\left(|{\bf p}|\right)}{\left|k_{i}-k_{j}\right|^{3-n-m}} \ , \quad|{\bf p}|=\frac{{\bf k}_i^2+{\bf k}_j^2}{2} \ .
\end{equation}
By setting $N=n+m$, it is easy to see that the analysis of \cite{Bjerrum-Bohr:2019kec} is applicable for the present case, and thus we can write the classical amplitude as
\begin{equation}
\frac{\mathcal{M}_{c l .}(|{\bf p}|, r)}{4E_1E_2}=V(|{\bf p}|,r)-2 \frac{E_1E_2}{E} V(|{\bf p}|,r) \partial_{|{\bf p}|^{2}} V(|{\bf p}|,r)-\frac{E_1^2-E_1E_2+E_2^2}{2 (E_1^2E_2+E_1E_2^2)} V(|{\bf p}|,r)^{2}+\ldots \label{eq:MclasV}
\end{equation}
Using the explicit expression for the potential in Eq.~\eqref{eq:Vdouble}, and comparing Eqs.~\eqref{eq:impulse},\eqref{eq:dp2} with Eq.~\eqref{eq:MclasV} one can see that
\begin{equation}
|{\bf{p}}|^2=|{\bf{p}}_\infty|^2-\frac{1}{2E} \mathcal{M}_\text{cl.}(|{\bf{p}}_\infty|,r) \ .
\end{equation}
\bibliography{bibl}

\begin{thebibliography}{110}%
\makeatletter
\providecommand \@ifxundefined [1]{%
 \@ifx{#1\undefined}
}%
\providecommand \@ifnum [1]{%
 \ifnum #1\expandafter \@firstoftwo
 \else \expandafter \@secondoftwo
 \fi
}%
\providecommand \@ifx [1]{%
 \ifx #1\expandafter \@firstoftwo
 \else \expandafter \@secondoftwo
 \fi
}%
\providecommand \natexlab [1]{#1}%
\providecommand \enquote  [1]{``#1''}%
\providecommand \bibnamefont  [1]{#1}%
\providecommand \bibfnamefont [1]{#1}%
\providecommand \citenamefont [1]{#1}%
\providecommand \href@noop [0]{\@secondoftwo}%
\providecommand \href [0]{\begingroup \@sanitize@url \@href}%
\providecommand \@href[1]{\@@startlink{#1}\@@href}%
\providecommand \@@href[1]{\endgroup#1\@@endlink}%
\providecommand \@sanitize@url [0]{\catcode `\\12\catcode `\$12\catcode
  `\&12\catcode `\#12\catcode `\^12\catcode `\_12\catcode `\%12\relax}%
\providecommand \@@startlink[1]{}%
\providecommand \@@endlink[0]{}%
\providecommand \url  [0]{\begingroup\@sanitize@url \@url }%
\providecommand \@url [1]{\endgroup\@href {#1}{\urlprefix }}%
\providecommand \urlprefix  [0]{URL }%
\providecommand \Eprint [0]{\href }%
\providecommand \doibase [0]{http://dx.doi.org/}%
\providecommand \selectlanguage [0]{\@gobble}%
\providecommand \bibinfo  [0]{\@secondoftwo}%
\providecommand \bibfield  [0]{\@secondoftwo}%
\providecommand \translation [1]{[#1]}%
\providecommand \BibitemOpen [0]{}%
\providecommand \bibitemStop [0]{}%
\providecommand \bibitemNoStop [0]{.\EOS\space}%
\providecommand \EOS [0]{\spacefactor3000\relax}%
\providecommand \BibitemShut  [1]{\csname bibitem#1\endcsname}%
\let\auto@bib@innerbib\@empty
\bibitem [{\citenamefont {Abbott}\ \emph {et~al.}(2016)\citenamefont {Abbott}
  \emph {et~al.}}]{Abbott:2016blz}%
  \BibitemOpen
  \bibfield  {author} {\bibinfo {author} {\bibfnamefont {B.P.}\ \bibnamefont
  {Abbott}} \emph {et~al.} (\bibinfo {collaboration} {LIGO Scientific,
  Virgo}),\ }\bibfield  {title} {\enquote {\bibinfo {title} {{Observation of
  Gravitational Waves from a Binary Black Hole Merger}},}\ }\href {\doibase
  10.1103/PhysRevLett.116.061102} {\bibfield  {journal} {\bibinfo  {journal}
  {Phys. Rev. Lett.}\ }\textbf {\bibinfo {volume} {116}},\ \bibinfo {pages}
  {061102} (\bibinfo {year} {2016})},\ \Eprint
  {http://arxiv.org/abs/1602.03837} {arXiv:1602.03837 [gr-qc]} \BibitemShut
  {NoStop}%
\bibitem [{\citenamefont {Abbott}\ \emph {et~al.}(2017)\citenamefont {Abbott}
  \emph {et~al.}}]{TheLIGOScientific:2017qsa}%
  \BibitemOpen
  \bibfield  {author} {\bibinfo {author} {\bibfnamefont {B.P.}\ \bibnamefont
  {Abbott}} \emph {et~al.} (\bibinfo {collaboration} {LIGO Scientific,
  Virgo}),\ }\bibfield  {title} {\enquote {\bibinfo {title} {{GW170817:
  Observation of Gravitational Waves from a Binary Neutron Star Inspiral}},}\
  }\href {\doibase 10.1103/PhysRevLett.119.161101} {\bibfield  {journal}
  {\bibinfo  {journal} {Phys. Rev. Lett.}\ }\textbf {\bibinfo {volume} {119}},\
  \bibinfo {pages} {161101} (\bibinfo {year} {2017})},\ \Eprint
  {http://arxiv.org/abs/1710.05832} {arXiv:1710.05832 [gr-qc]} \BibitemShut
  {NoStop}%
\bibitem [{\citenamefont {Buonanno}\ and\ \citenamefont
  {Damour}(1999)}]{Buonanno:1998gg}%
  \BibitemOpen
  \bibfield  {author} {\bibinfo {author} {\bibfnamefont {A.}~\bibnamefont
  {Buonanno}}\ and\ \bibinfo {author} {\bibfnamefont {T.}~\bibnamefont
  {Damour}},\ }\bibfield  {title} {\enquote {\bibinfo {title} {{Effective
  one-body approach to general relativistic two-body dynamics}},}\ }\href
  {\doibase 10.1103/PhysRevD.59.084006} {\bibfield  {journal} {\bibinfo
  {journal} {Phys. Rev. D}\ }\textbf {\bibinfo {volume} {59}},\ \bibinfo
  {pages} {084006} (\bibinfo {year} {1999})},\ \Eprint
  {http://arxiv.org/abs/gr-qc/9811091} {arXiv:gr-qc/9811091} \BibitemShut
  {NoStop}%
\bibitem [{\citenamefont {Buonanno}\ and\ \citenamefont
  {Damour}(2000)}]{Buonanno:2000ef}%
  \BibitemOpen
  \bibfield  {author} {\bibinfo {author} {\bibfnamefont {Alessandra}\
  \bibnamefont {Buonanno}}\ and\ \bibinfo {author} {\bibfnamefont {Thibault}\
  \bibnamefont {Damour}},\ }\bibfield  {title} {\enquote {\bibinfo {title}
  {{Transition from inspiral to plunge in binary black hole coalescences}},}\
  }\href {\doibase 10.1103/PhysRevD.62.064015} {\bibfield  {journal} {\bibinfo
  {journal} {Phys. Rev. D}\ }\textbf {\bibinfo {volume} {62}},\ \bibinfo
  {pages} {064015} (\bibinfo {year} {2000})},\ \Eprint
  {http://arxiv.org/abs/gr-qc/0001013} {arXiv:gr-qc/0001013} \BibitemShut
  {NoStop}%
\bibitem [{\citenamefont {Campanelli}\ \emph {et~al.}(2006)\citenamefont
  {Campanelli}, \citenamefont {Lousto}, \citenamefont {Marronetti},\ and\
  \citenamefont {Zlochower}}]{Campanelli:2005dd}%
  \BibitemOpen
  \bibfield  {author} {\bibinfo {author} {\bibfnamefont {Manuela}\ \bibnamefont
  {Campanelli}}, \bibinfo {author} {\bibfnamefont {C.~O.}\ \bibnamefont
  {Lousto}}, \bibinfo {author} {\bibfnamefont {P.}~\bibnamefont {Marronetti}},
  \ and\ \bibinfo {author} {\bibfnamefont {Y.}~\bibnamefont {Zlochower}},\
  }\bibfield  {title} {\enquote {\bibinfo {title} {{Accurate evolutions of
  orbiting black-hole binaries without excision}},}\ }\href {\doibase
  10.1103/PhysRevLett.96.111101} {\bibfield  {journal} {\bibinfo  {journal}
  {Phys. Rev. Lett.}\ }\textbf {\bibinfo {volume} {96}},\ \bibinfo {pages}
  {111101} (\bibinfo {year} {2006})},\ \Eprint
  {http://arxiv.org/abs/gr-qc/0511048} {arXiv:gr-qc/0511048} \BibitemShut
  {NoStop}%
\bibitem [{\citenamefont {Baker}\ \emph {et~al.}(2006)\citenamefont {Baker},
  \citenamefont {Centrella}, \citenamefont {Choi}, \citenamefont {Koppitz},\
  and\ \citenamefont {van Meter}}]{Baker:2005vv}%
  \BibitemOpen
  \bibfield  {author} {\bibinfo {author} {\bibfnamefont {John~G.}\ \bibnamefont
  {Baker}}, \bibinfo {author} {\bibfnamefont {Joan}\ \bibnamefont {Centrella}},
  \bibinfo {author} {\bibfnamefont {Dae-Il}\ \bibnamefont {Choi}}, \bibinfo
  {author} {\bibfnamefont {Michael}\ \bibnamefont {Koppitz}}, \ and\ \bibinfo
  {author} {\bibfnamefont {James}\ \bibnamefont {van Meter}},\ }\bibfield
  {title} {\enquote {\bibinfo {title} {{Gravitational wave extraction from an
  inspiraling configuration of merging black holes}},}\ }\href {\doibase
  10.1103/PhysRevLett.96.111102} {\bibfield  {journal} {\bibinfo  {journal}
  {Phys. Rev. Lett.}\ }\textbf {\bibinfo {volume} {96}},\ \bibinfo {pages}
  {111102} (\bibinfo {year} {2006})},\ \Eprint
  {http://arxiv.org/abs/gr-qc/0511103} {arXiv:gr-qc/0511103} \BibitemShut
  {NoStop}%
\bibitem [{\citenamefont {Pretorius}(2005)}]{Pretorius:2005gq}%
  \BibitemOpen
  \bibfield  {author} {\bibinfo {author} {\bibfnamefont {Frans}\ \bibnamefont
  {Pretorius}},\ }\bibfield  {title} {\enquote {\bibinfo {title} {{Evolution of
  binary black hole spacetimes}},}\ }\href {\doibase
  10.1103/PhysRevLett.95.121101} {\bibfield  {journal} {\bibinfo  {journal}
  {Phys. Rev. Lett.}\ }\textbf {\bibinfo {volume} {95}},\ \bibinfo {pages}
  {121101} (\bibinfo {year} {2005})},\ \Eprint
  {http://arxiv.org/abs/gr-qc/0507014} {arXiv:gr-qc/0507014} \BibitemShut
  {NoStop}%
\bibitem [{\citenamefont {Mino}\ \emph {et~al.}(1997)\citenamefont {Mino},
  \citenamefont {Sasaki},\ and\ \citenamefont {Tanaka}}]{Mino:1996nk}%
  \BibitemOpen
  \bibfield  {author} {\bibinfo {author} {\bibfnamefont {Yasushi}\ \bibnamefont
  {Mino}}, \bibinfo {author} {\bibfnamefont {Misao}\ \bibnamefont {Sasaki}}, \
  and\ \bibinfo {author} {\bibfnamefont {Takahiro}\ \bibnamefont {Tanaka}},\
  }\bibfield  {title} {\enquote {\bibinfo {title} {{Gravitational radiation
  reaction to a particle motion}},}\ }\href {\doibase 10.1103/PhysRevD.55.3457}
  {\bibfield  {journal} {\bibinfo  {journal} {Phys. Rev. D}\ }\textbf {\bibinfo
  {volume} {55}},\ \bibinfo {pages} {3457--3476} (\bibinfo {year} {1997})},\
  \Eprint {http://arxiv.org/abs/gr-qc/9606018} {arXiv:gr-qc/9606018}
  \BibitemShut {NoStop}%
\bibitem [{\citenamefont {Quinn}\ and\ \citenamefont
  {Wald}(1997)}]{Quinn:1996am}%
  \BibitemOpen
  \bibfield  {author} {\bibinfo {author} {\bibfnamefont {Theodore~C.}\
  \bibnamefont {Quinn}}\ and\ \bibinfo {author} {\bibfnamefont {Robert~M.}\
  \bibnamefont {Wald}},\ }\bibfield  {title} {\enquote {\bibinfo {title} {{An
  Axiomatic approach to electromagnetic and gravitational radiation reaction of
  particles in curved space-time}},}\ }\href {\doibase
  10.1103/PhysRevD.56.3381} {\bibfield  {journal} {\bibinfo  {journal} {Phys.
  Rev. D}\ }\textbf {\bibinfo {volume} {56}},\ \bibinfo {pages} {3381--3394}
  (\bibinfo {year} {1997})},\ \Eprint {http://arxiv.org/abs/gr-qc/9610053}
  {arXiv:gr-qc/9610053} \BibitemShut {NoStop}%
\bibitem [{\citenamefont {Blanchet}(2014)}]{Blanchet:2013haa}%
  \BibitemOpen
  \bibfield  {author} {\bibinfo {author} {\bibfnamefont {Luc}\ \bibnamefont
  {Blanchet}},\ }\bibfield  {title} {\enquote {\bibinfo {title} {{Gravitational
  Radiation from Post-Newtonian Sources and Inspiralling Compact Binaries}},}\
  }\href {\doibase 10.12942/lrr-2014-2} {\bibfield  {journal} {\bibinfo
  {journal} {Living Rev. Rel.}\ }\textbf {\bibinfo {volume} {17}},\ \bibinfo
  {pages} {2} (\bibinfo {year} {2014})},\ \Eprint
  {http://arxiv.org/abs/1310.1528} {arXiv:1310.1528 [gr-qc]} \BibitemShut
  {NoStop}%
\bibitem [{\citenamefont {Sch\"afer}\ and\ \citenamefont
  {Jaranowski}(2018)}]{Schafer:2018kuf}%
  \BibitemOpen
  \bibfield  {author} {\bibinfo {author} {\bibfnamefont {Gerhard}\ \bibnamefont
  {Sch\"afer}}\ and\ \bibinfo {author} {\bibfnamefont {Piotr}\ \bibnamefont
  {Jaranowski}},\ }\bibfield  {title} {\enquote {\bibinfo {title} {{Hamiltonian
  formulation of general relativity and post-Newtonian dynamics of compact
  binaries}},}\ }\href {\doibase 10.1007/s41114-018-0016-5} {\bibfield
  {journal} {\bibinfo  {journal} {Living Rev. Rel.}\ }\textbf {\bibinfo
  {volume} {21}},\ \bibinfo {pages} {7} (\bibinfo {year} {2018})},\ \Eprint
  {http://arxiv.org/abs/1805.07240} {arXiv:1805.07240 [gr-qc]} \BibitemShut
  {NoStop}%
\bibitem [{\citenamefont {Barack}\ and\ \citenamefont
  {Pound}(2019)}]{Barack:2018yvs}%
  \BibitemOpen
  \bibfield  {author} {\bibinfo {author} {\bibfnamefont {Leor}\ \bibnamefont
  {Barack}}\ and\ \bibinfo {author} {\bibfnamefont {Adam}\ \bibnamefont
  {Pound}},\ }\bibfield  {title} {\enquote {\bibinfo {title} {{Self-force and
  radiation reaction in general relativity}},}\ }\href {\doibase
  10.1088/1361-6633/aae552} {\bibfield  {journal} {\bibinfo  {journal} {Rept.
  Prog. Phys.}\ }\textbf {\bibinfo {volume} {82}},\ \bibinfo {pages} {016904}
  (\bibinfo {year} {2019})},\ \Eprint {http://arxiv.org/abs/1805.10385}
  {arXiv:1805.10385 [gr-qc]} \BibitemShut {NoStop}%
\bibitem [{\citenamefont {Barack}\ \emph {et~al.}(2019)\citenamefont {Barack}
  \emph {et~al.}}]{Barack:2018yly}%
  \BibitemOpen
  \bibfield  {author} {\bibinfo {author} {\bibfnamefont {Leor}\ \bibnamefont
  {Barack}} \emph {et~al.},\ }\bibfield  {title} {\enquote {\bibinfo {title}
  {{Black holes, gravitational waves and fundamental physics: a roadmap}},}\
  }\href {\doibase 10.1088/1361-6382/ab0587} {\bibfield  {journal} {\bibinfo
  {journal} {Class. Quant. Grav.}\ }\textbf {\bibinfo {volume} {36}},\ \bibinfo
  {pages} {143001} (\bibinfo {year} {2019})},\ \Eprint
  {http://arxiv.org/abs/1806.05195} {arXiv:1806.05195 [gr-qc]} \BibitemShut
  {NoStop}%
\bibitem [{\citenamefont {Porto}(2016)}]{Porto:2016pyg}%
  \BibitemOpen
  \bibfield  {author} {\bibinfo {author} {\bibfnamefont {Rafael~A.}\
  \bibnamefont {Porto}},\ }\bibfield  {title} {\enquote {\bibinfo {title} {{The
  effective field theorist\textquoteright{}s approach to gravitational
  dynamics}},}\ }\href {\doibase 10.1016/j.physrep.2016.04.003} {\bibfield
  {journal} {\bibinfo  {journal} {Phys. Rept.}\ }\textbf {\bibinfo {volume}
  {633}},\ \bibinfo {pages} {1--104} (\bibinfo {year} {2016})},\ \Eprint
  {http://arxiv.org/abs/1601.04914} {arXiv:1601.04914 [hep-th]} \BibitemShut
  {NoStop}%
\bibitem [{\citenamefont {Levi}(2020)}]{Levi:2018nxp}%
  \BibitemOpen
  \bibfield  {author} {\bibinfo {author} {\bibfnamefont {Mich\`ele}\
  \bibnamefont {Levi}},\ }\bibfield  {title} {\enquote {\bibinfo {title}
  {{Effective Field Theories of Post-Newtonian Gravity: A comprehensive
  review}},}\ }\href {\doibase 10.1088/1361-6633/ab12bc} {\bibfield  {journal}
  {\bibinfo  {journal} {Rept. Prog. Phys.}\ }\textbf {\bibinfo {volume} {83}},\
  \bibinfo {pages} {075901} (\bibinfo {year} {2020})},\ \Eprint
  {http://arxiv.org/abs/1807.01699} {arXiv:1807.01699 [hep-th]} \BibitemShut
  {NoStop}%
\bibitem [{\citenamefont {Cheung}\ \emph {et~al.}(2018)\citenamefont {Cheung},
  \citenamefont {Rothstein},\ and\ \citenamefont {Solon}}]{Cheung:2018wkq}%
  \BibitemOpen
  \bibfield  {author} {\bibinfo {author} {\bibfnamefont {Clifford}\
  \bibnamefont {Cheung}}, \bibinfo {author} {\bibfnamefont {Ira~Z.}\
  \bibnamefont {Rothstein}}, \ and\ \bibinfo {author} {\bibfnamefont
  {Mikhail~P.}\ \bibnamefont {Solon}},\ }\bibfield  {title} {\enquote {\bibinfo
  {title} {{From Scattering Amplitudes to Classical Potentials in the
  Post-Minkowskian Expansion}},}\ }\href {\doibase
  10.1103/PhysRevLett.121.251101} {\bibfield  {journal} {\bibinfo  {journal}
  {Phys. Rev. Lett.}\ }\textbf {\bibinfo {volume} {121}},\ \bibinfo {pages}
  {251101} (\bibinfo {year} {2018})},\ \Eprint
  {http://arxiv.org/abs/1808.02489} {arXiv:1808.02489 [hep-th]} \BibitemShut
  {NoStop}%
\bibitem [{\citenamefont {Bern}\ \emph
  {et~al.}(2019{\natexlab{a}})\citenamefont {Bern}, \citenamefont {Cheung},
  \citenamefont {Roiban}, \citenamefont {Shen}, \citenamefont {Solon},\ and\
  \citenamefont {Zeng}}]{Bern:2019nnu}%
  \BibitemOpen
  \bibfield  {author} {\bibinfo {author} {\bibfnamefont {Zvi}\ \bibnamefont
  {Bern}}, \bibinfo {author} {\bibfnamefont {Clifford}\ \bibnamefont {Cheung}},
  \bibinfo {author} {\bibfnamefont {Radu}\ \bibnamefont {Roiban}}, \bibinfo
  {author} {\bibfnamefont {Chia-Hsien}\ \bibnamefont {Shen}}, \bibinfo {author}
  {\bibfnamefont {Mikhail~P.}\ \bibnamefont {Solon}}, \ and\ \bibinfo {author}
  {\bibfnamefont {Mao}\ \bibnamefont {Zeng}},\ }\bibfield  {title} {\enquote
  {\bibinfo {title} {{Scattering Amplitudes and the Conservative Hamiltonian
  for Binary Systems at Third Post-Minkowskian Order}},}\ }\href {\doibase
  10.1103/PhysRevLett.122.201603} {\bibfield  {journal} {\bibinfo  {journal}
  {Phys. Rev. Lett.}\ }\textbf {\bibinfo {volume} {122}},\ \bibinfo {pages}
  {201603} (\bibinfo {year} {2019}{\natexlab{a}})},\ \Eprint
  {http://arxiv.org/abs/1901.04424} {arXiv:1901.04424 [hep-th]} \BibitemShut
  {NoStop}%
\bibitem [{\citenamefont {Bern}\ \emph
  {et~al.}(2019{\natexlab{b}})\citenamefont {Bern}, \citenamefont {Cheung},
  \citenamefont {Roiban}, \citenamefont {Shen}, \citenamefont {Solon},\ and\
  \citenamefont {Zeng}}]{Bern:2019crd}%
  \BibitemOpen
  \bibfield  {author} {\bibinfo {author} {\bibfnamefont {Zvi}\ \bibnamefont
  {Bern}}, \bibinfo {author} {\bibfnamefont {Clifford}\ \bibnamefont {Cheung}},
  \bibinfo {author} {\bibfnamefont {Radu}\ \bibnamefont {Roiban}}, \bibinfo
  {author} {\bibfnamefont {Chia-Hsien}\ \bibnamefont {Shen}}, \bibinfo {author}
  {\bibfnamefont {Mikhail~P.}\ \bibnamefont {Solon}}, \ and\ \bibinfo {author}
  {\bibfnamefont {Mao}\ \bibnamefont {Zeng}},\ }\bibfield  {title} {\enquote
  {\bibinfo {title} {{Black Hole Binary Dynamics from the Double Copy and
  Effective Theory}},}\ }\href {\doibase 10.1007/JHEP10(2019)206} {\bibfield
  {journal} {\bibinfo  {journal} {JHEP}\ }\textbf {\bibinfo {volume} {10}},\
  \bibinfo {pages} {206} (\bibinfo {year} {2019}{\natexlab{b}})},\ \Eprint
  {http://arxiv.org/abs/1908.01493} {arXiv:1908.01493 [hep-th]} \BibitemShut
  {NoStop}%
\bibitem [{\citenamefont {Bjerrum-Bohr}\ \emph {et~al.}(2018)\citenamefont
  {Bjerrum-Bohr}, \citenamefont {Damgaard}, \citenamefont {Festuccia},
  \citenamefont {Plant\'e},\ and\ \citenamefont
  {Vanhove}}]{Bjerrum-Bohr:2018xdl}%
  \BibitemOpen
  \bibfield  {author} {\bibinfo {author} {\bibfnamefont {N.~E.~J.}\
  \bibnamefont {Bjerrum-Bohr}}, \bibinfo {author} {\bibfnamefont {Poul~H.}\
  \bibnamefont {Damgaard}}, \bibinfo {author} {\bibfnamefont {Guido}\
  \bibnamefont {Festuccia}}, \bibinfo {author} {\bibfnamefont {Ludovic}\
  \bibnamefont {Plant\'e}}, \ and\ \bibinfo {author} {\bibfnamefont {Pierre}\
  \bibnamefont {Vanhove}},\ }\bibfield  {title} {\enquote {\bibinfo {title}
  {{General Relativity from Scattering Amplitudes}},}\ }\href {\doibase
  10.1103/PhysRevLett.121.171601} {\bibfield  {journal} {\bibinfo  {journal}
  {Phys. Rev. Lett.}\ }\textbf {\bibinfo {volume} {121}},\ \bibinfo {pages}
  {171601} (\bibinfo {year} {2018})},\ \Eprint
  {http://arxiv.org/abs/1806.04920} {arXiv:1806.04920 [hep-th]} \BibitemShut
  {NoStop}%
\bibitem [{\citenamefont {Ciafaloni}\ \emph {et~al.}(2019)\citenamefont
  {Ciafaloni}, \citenamefont {Colferai},\ and\ \citenamefont
  {Veneziano}}]{Ciafaloni:2018uwe}%
  \BibitemOpen
  \bibfield  {author} {\bibinfo {author} {\bibfnamefont {Marcello}\
  \bibnamefont {Ciafaloni}}, \bibinfo {author} {\bibfnamefont {Dimitri}\
  \bibnamefont {Colferai}}, \ and\ \bibinfo {author} {\bibfnamefont {Gabriele}\
  \bibnamefont {Veneziano}},\ }\bibfield  {title} {\enquote {\bibinfo {title}
  {{Infrared features of gravitational scattering and radiation in the eikonal
  approach}},}\ }\href {\doibase 10.1103/PhysRevD.99.066008} {\bibfield
  {journal} {\bibinfo  {journal} {Phys. Rev. D}\ }\textbf {\bibinfo {volume}
  {99}},\ \bibinfo {pages} {066008} (\bibinfo {year} {2019})},\ \Eprint
  {http://arxiv.org/abs/1812.08137} {arXiv:1812.08137 [hep-th]} \BibitemShut
  {NoStop}%
\bibitem [{\citenamefont {Bjerrum-Bohr}\ \emph {et~al.}(2020)\citenamefont
  {Bjerrum-Bohr}, \citenamefont {Cristofoli},\ and\ \citenamefont
  {Damgaard}}]{Bjerrum-Bohr:2019kec}%
  \BibitemOpen
  \bibfield  {author} {\bibinfo {author} {\bibfnamefont {N.~E.~J.}\
  \bibnamefont {Bjerrum-Bohr}}, \bibinfo {author} {\bibfnamefont {Andrea}\
  \bibnamefont {Cristofoli}}, \ and\ \bibinfo {author} {\bibfnamefont
  {Poul~H.}\ \bibnamefont {Damgaard}},\ }\bibfield  {title} {\enquote {\bibinfo
  {title} {{Post-Minkowskian Scattering Angle in Einstein Gravity}},}\ }\href
  {\doibase 10.1007/JHEP08(2020)038} {\bibfield  {journal} {\bibinfo  {journal}
  {JHEP}\ }\textbf {\bibinfo {volume} {08}},\ \bibinfo {pages} {038} (\bibinfo
  {year} {2020})},\ \Eprint {http://arxiv.org/abs/1910.09366} {arXiv:1910.09366
  [hep-th]} \BibitemShut {NoStop}%
\bibitem [{\citenamefont {Cachazo}\ and\ \citenamefont
  {Guevara}(2020)}]{Cachazo:2017jef}%
  \BibitemOpen
  \bibfield  {author} {\bibinfo {author} {\bibfnamefont {Freddy}\ \bibnamefont
  {Cachazo}}\ and\ \bibinfo {author} {\bibfnamefont {Alfredo}\ \bibnamefont
  {Guevara}},\ }\bibfield  {title} {\enquote {\bibinfo {title} {{Leading
  Singularities and Classical Gravitational Scattering}},}\ }\href {\doibase
  10.1007/JHEP02(2020)181} {\bibfield  {journal} {\bibinfo  {journal} {JHEP}\
  }\textbf {\bibinfo {volume} {02}},\ \bibinfo {pages} {181} (\bibinfo {year}
  {2020})},\ \Eprint {http://arxiv.org/abs/1705.10262} {arXiv:1705.10262
  [hep-th]} \BibitemShut {NoStop}%
\bibitem [{\citenamefont {Cristofoli}\ \emph {et~al.}(2019)\citenamefont
  {Cristofoli}, \citenamefont {Bjerrum-Bohr}, \citenamefont {Damgaard},\ and\
  \citenamefont {Vanhove}}]{Cristofoli:2019neg}%
  \BibitemOpen
  \bibfield  {author} {\bibinfo {author} {\bibfnamefont {Andrea}\ \bibnamefont
  {Cristofoli}}, \bibinfo {author} {\bibfnamefont {N.E.J.}\ \bibnamefont
  {Bjerrum-Bohr}}, \bibinfo {author} {\bibfnamefont {Poul~H.}\ \bibnamefont
  {Damgaard}}, \ and\ \bibinfo {author} {\bibfnamefont {Pierre}\ \bibnamefont
  {Vanhove}},\ }\bibfield  {title} {\enquote {\bibinfo {title}
  {{Post-Minkowskian Hamiltonians in general relativity}},}\ }\href {\doibase
  10.1103/PhysRevD.100.084040} {\bibfield  {journal} {\bibinfo  {journal}
  {Phys. Rev. D}\ }\textbf {\bibinfo {volume} {100}},\ \bibinfo {pages}
  {084040} (\bibinfo {year} {2019})},\ \Eprint
  {http://arxiv.org/abs/1906.01579} {arXiv:1906.01579 [hep-th]} \BibitemShut
  {NoStop}%
\bibitem [{\citenamefont {Damgaard}\ \emph {et~al.}(2019)\citenamefont
  {Damgaard}, \citenamefont {Haddad},\ and\ \citenamefont
  {Helset}}]{Damgaard:2019lfh}%
  \BibitemOpen
  \bibfield  {author} {\bibinfo {author} {\bibfnamefont {Poul~H.}\ \bibnamefont
  {Damgaard}}, \bibinfo {author} {\bibfnamefont {Kays}\ \bibnamefont {Haddad}},
  \ and\ \bibinfo {author} {\bibfnamefont {Andreas}\ \bibnamefont {Helset}},\
  }\bibfield  {title} {\enquote {\bibinfo {title} {{Heavy Black Hole Effective
  Theory}},}\ }\href {\doibase 10.1007/JHEP11(2019)070} {\bibfield  {journal}
  {\bibinfo  {journal} {JHEP}\ }\textbf {\bibinfo {volume} {11}},\ \bibinfo
  {pages} {070} (\bibinfo {year} {2019})},\ \Eprint
  {http://arxiv.org/abs/1908.10308} {arXiv:1908.10308 [hep-ph]} \BibitemShut
  {NoStop}%
\bibitem [{\citenamefont {Cristofoli}\ \emph {et~al.}(2020)\citenamefont
  {Cristofoli}, \citenamefont {Damgaard}, \citenamefont {Di~Vecchia},\ and\
  \citenamefont {Heissenberg}}]{Cristofoli:2020uzm}%
  \BibitemOpen
  \bibfield  {author} {\bibinfo {author} {\bibfnamefont {Andrea}\ \bibnamefont
  {Cristofoli}}, \bibinfo {author} {\bibfnamefont {Poul~H.}\ \bibnamefont
  {Damgaard}}, \bibinfo {author} {\bibfnamefont {Paolo}\ \bibnamefont
  {Di~Vecchia}}, \ and\ \bibinfo {author} {\bibfnamefont {Carlo}\ \bibnamefont
  {Heissenberg}},\ }\bibfield  {title} {\enquote {\bibinfo {title}
  {{Second-order Post-Minkowskian scattering in arbitrary dimensions}},}\
  }\href {\doibase 10.1007/JHEP07(2020)122} {\bibfield  {journal} {\bibinfo
  {journal} {JHEP}\ }\textbf {\bibinfo {volume} {07}},\ \bibinfo {pages} {122}
  (\bibinfo {year} {2020})},\ \Eprint {http://arxiv.org/abs/2003.10274}
  {arXiv:2003.10274 [hep-th]} \BibitemShut {NoStop}%
\bibitem [{\citenamefont {Kosower}\ \emph {et~al.}(2019)\citenamefont
  {Kosower}, \citenamefont {Maybee},\ and\ \citenamefont
  {O'Connell}}]{Kosower:2018adc}%
  \BibitemOpen
  \bibfield  {author} {\bibinfo {author} {\bibfnamefont {David~A.}\
  \bibnamefont {Kosower}}, \bibinfo {author} {\bibfnamefont {Ben}\ \bibnamefont
  {Maybee}}, \ and\ \bibinfo {author} {\bibfnamefont {Donal}\ \bibnamefont
  {O'Connell}},\ }\bibfield  {title} {\enquote {\bibinfo {title} {{Amplitudes,
  Observables, and Classical Scattering}},}\ }\href {\doibase
  10.1007/JHEP02(2019)137} {\bibfield  {journal} {\bibinfo  {journal} {JHEP}\
  }\textbf {\bibinfo {volume} {02}},\ \bibinfo {pages} {137} (\bibinfo {year}
  {2019})},\ \Eprint {http://arxiv.org/abs/1811.10950} {arXiv:1811.10950
  [hep-th]} \BibitemShut {NoStop}%
\bibitem [{\citenamefont {Maybee}\ \emph {et~al.}(2019)\citenamefont {Maybee},
  \citenamefont {O'Connell},\ and\ \citenamefont {Vines}}]{Maybee:2019jus}%
  \BibitemOpen
  \bibfield  {author} {\bibinfo {author} {\bibfnamefont {Ben}\ \bibnamefont
  {Maybee}}, \bibinfo {author} {\bibfnamefont {Donal}\ \bibnamefont
  {O'Connell}}, \ and\ \bibinfo {author} {\bibfnamefont {Justin}\ \bibnamefont
  {Vines}},\ }\bibfield  {title} {\enquote {\bibinfo {title} {{Observables and
  amplitudes for spinning particles and black holes}},}\ }\href {\doibase
  10.1007/JHEP12(2019)156} {\bibfield  {journal} {\bibinfo  {journal} {JHEP}\
  }\textbf {\bibinfo {volume} {12}},\ \bibinfo {pages} {156} (\bibinfo {year}
  {2019})},\ \Eprint {http://arxiv.org/abs/1906.09260} {arXiv:1906.09260
  [hep-th]} \BibitemShut {NoStop}%
\bibitem [{\citenamefont {Koemans~Collado}\ \emph {et~al.}(2019)\citenamefont
  {Koemans~Collado}, \citenamefont {Di~Vecchia},\ and\ \citenamefont
  {Russo}}]{KoemansCollado:2019ggb}%
  \BibitemOpen
  \bibfield  {author} {\bibinfo {author} {\bibfnamefont {Arnau}\ \bibnamefont
  {Koemans~Collado}}, \bibinfo {author} {\bibfnamefont {Paolo}\ \bibnamefont
  {Di~Vecchia}}, \ and\ \bibinfo {author} {\bibfnamefont {Rodolfo}\
  \bibnamefont {Russo}},\ }\bibfield  {title} {\enquote {\bibinfo {title}
  {{Revisiting the second post-Minkowskian eikonal and the dynamics of binary
  black holes}},}\ }\href {\doibase 10.1103/PhysRevD.100.066028} {\bibfield
  {journal} {\bibinfo  {journal} {Phys. Rev. D}\ }\textbf {\bibinfo {volume}
  {100}},\ \bibinfo {pages} {066028} (\bibinfo {year} {2019})},\ \Eprint
  {http://arxiv.org/abs/1904.02667} {arXiv:1904.02667 [hep-th]} \BibitemShut
  {NoStop}%
\bibitem [{\citenamefont {Mougiakakos}\ and\ \citenamefont
  {Vanhove}(2021)}]{Mougiakakos:2020laz}%
  \BibitemOpen
  \bibfield  {author} {\bibinfo {author} {\bibfnamefont {Stavros}\ \bibnamefont
  {Mougiakakos}}\ and\ \bibinfo {author} {\bibfnamefont {Pierre}\ \bibnamefont
  {Vanhove}},\ }\bibfield  {title} {\enquote {\bibinfo {title}
  {{Schwarzschild-Tangherlini metric from scattering amplitudes in various
  dimensions}},}\ }\href {\doibase 10.1103/PhysRevD.103.026001} {\bibfield
  {journal} {\bibinfo  {journal} {Phys. Rev. D}\ }\textbf {\bibinfo {volume}
  {103}},\ \bibinfo {pages} {026001} (\bibinfo {year} {2021})},\ \Eprint
  {http://arxiv.org/abs/2010.08882} {arXiv:2010.08882 [hep-th]} \BibitemShut
  {NoStop}%
\bibitem [{\citenamefont {Parra-Martinez}\ \emph {et~al.}(2020)\citenamefont
  {Parra-Martinez}, \citenamefont {Ruf},\ and\ \citenamefont
  {Zeng}}]{Parra-Martinez:2020dzs}%
  \BibitemOpen
  \bibfield  {author} {\bibinfo {author} {\bibfnamefont {Julio}\ \bibnamefont
  {Parra-Martinez}}, \bibinfo {author} {\bibfnamefont {Michael~S.}\
  \bibnamefont {Ruf}}, \ and\ \bibinfo {author} {\bibfnamefont {Mao}\
  \bibnamefont {Zeng}},\ }\bibfield  {title} {\enquote {\bibinfo {title}
  {{Extremal black hole scattering at $\mathcal{O}(G^3)$: graviton dominance,
  eikonal exponentiation, and differential equations}},}\ }\href {\doibase
  10.1007/JHEP11(2020)023} {\bibfield  {journal} {\bibinfo  {journal} {JHEP}\
  }\textbf {\bibinfo {volume} {11}},\ \bibinfo {pages} {023} (\bibinfo {year}
  {2020})},\ \Eprint {http://arxiv.org/abs/2005.04236} {arXiv:2005.04236
  [hep-th]} \BibitemShut {NoStop}%
\bibitem [{\citenamefont {Bern}\ \emph {et~al.}(2020)\citenamefont {Bern},
  \citenamefont {Luna}, \citenamefont {Roiban}, \citenamefont {Shen},\ and\
  \citenamefont {Zeng}}]{Bern:2020buy}%
  \BibitemOpen
  \bibfield  {author} {\bibinfo {author} {\bibfnamefont {Zvi}\ \bibnamefont
  {Bern}}, \bibinfo {author} {\bibfnamefont {Andres}\ \bibnamefont {Luna}},
  \bibinfo {author} {\bibfnamefont {Radu}\ \bibnamefont {Roiban}}, \bibinfo
  {author} {\bibfnamefont {Chia-Hsien}\ \bibnamefont {Shen}}, \ and\ \bibinfo
  {author} {\bibfnamefont {Mao}\ \bibnamefont {Zeng}},\ }\bibfield  {title}
  {\enquote {\bibinfo {title} {{Spinning Black Hole Binary Dynamics, Scattering
  Amplitudes and Effective Field Theory}},}\ }\href@noop {} {\  (\bibinfo
  {year} {2020})},\ \Eprint {http://arxiv.org/abs/2005.03071} {arXiv:2005.03071
  [hep-th]} \BibitemShut {NoStop}%
\bibitem [{\citenamefont {Bern}\ \emph {et~al.}(2021)\citenamefont {Bern},
  \citenamefont {Parra-Martinez}, \citenamefont {Roiban}, \citenamefont {Ruf},
  \citenamefont {Shen}, \citenamefont {Solon},\ and\ \citenamefont
  {Zeng}}]{Bern:2021dqo}%
  \BibitemOpen
  \bibfield  {author} {\bibinfo {author} {\bibfnamefont {Zvi}\ \bibnamefont
  {Bern}}, \bibinfo {author} {\bibfnamefont {Julio}\ \bibnamefont
  {Parra-Martinez}}, \bibinfo {author} {\bibfnamefont {Radu}\ \bibnamefont
  {Roiban}}, \bibinfo {author} {\bibfnamefont {Michael~S.}\ \bibnamefont
  {Ruf}}, \bibinfo {author} {\bibfnamefont {Chia-Hsien}\ \bibnamefont {Shen}},
  \bibinfo {author} {\bibfnamefont {Mikhail~P.}\ \bibnamefont {Solon}}, \ and\
  \bibinfo {author} {\bibfnamefont {Mao}\ \bibnamefont {Zeng}},\ }\bibfield
  {title} {\enquote {\bibinfo {title} {{Scattering Amplitudes and Conservative
  Binary Dynamics at ${\cal O}(G^4)$}},}\ }\href@noop {} {\  (\bibinfo {year}
  {2021})},\ \Eprint {http://arxiv.org/abs/2101.07254} {arXiv:2101.07254
  [hep-th]} \BibitemShut {NoStop}%
\bibitem [{\citenamefont {Herrmann}\ \emph {et~al.}(2021)\citenamefont
  {Herrmann}, \citenamefont {Parra-Martinez}, \citenamefont {Ruf},\ and\
  \citenamefont {Zeng}}]{Herrmann:2021tct}%
  \BibitemOpen
  \bibfield  {author} {\bibinfo {author} {\bibfnamefont {Enrico}\ \bibnamefont
  {Herrmann}}, \bibinfo {author} {\bibfnamefont {Julio}\ \bibnamefont
  {Parra-Martinez}}, \bibinfo {author} {\bibfnamefont {Michael~S.}\
  \bibnamefont {Ruf}}, \ and\ \bibinfo {author} {\bibfnamefont {Mao}\
  \bibnamefont {Zeng}},\ }\bibfield  {title} {\enquote {\bibinfo {title}
  {{Radiative Classical Gravitational Observables at $\mathcal O(G^3)$ from
  Scattering Amplitudes}},}\ }\href@noop {} {\  (\bibinfo {year} {2021})},\
  \Eprint {http://arxiv.org/abs/2104.03957} {arXiv:2104.03957 [hep-th]}
  \BibitemShut {NoStop}%
\bibitem [{\citenamefont {Di~Vecchia}\ \emph {et~al.}(2020)\citenamefont
  {Di~Vecchia}, \citenamefont {Heissenberg}, \citenamefont {Russo},\ and\
  \citenamefont {Veneziano}}]{DiVecchia:2020ymx}%
  \BibitemOpen
  \bibfield  {author} {\bibinfo {author} {\bibfnamefont {Paolo}\ \bibnamefont
  {Di~Vecchia}}, \bibinfo {author} {\bibfnamefont {Carlo}\ \bibnamefont
  {Heissenberg}}, \bibinfo {author} {\bibfnamefont {Rodolfo}\ \bibnamefont
  {Russo}}, \ and\ \bibinfo {author} {\bibfnamefont {Gabriele}\ \bibnamefont
  {Veneziano}},\ }\bibfield  {title} {\enquote {\bibinfo {title} {{Universality
  of ultra-relativistic gravitational scattering}},}\ }\href {\doibase
  10.1016/j.physletb.2020.135924} {\bibfield  {journal} {\bibinfo  {journal}
  {Phys. Lett. B}\ }\textbf {\bibinfo {volume} {811}},\ \bibinfo {pages}
  {135924} (\bibinfo {year} {2020})},\ \Eprint
  {http://arxiv.org/abs/2008.12743} {arXiv:2008.12743 [hep-th]} \BibitemShut
  {NoStop}%
\bibitem [{\citenamefont {K\"alin}\ and\ \citenamefont
  {Porto}(2020{\natexlab{a}})}]{Kalin:2019rwq}%
  \BibitemOpen
  \bibfield  {author} {\bibinfo {author} {\bibfnamefont {Gregor}\ \bibnamefont
  {K\"alin}}\ and\ \bibinfo {author} {\bibfnamefont {Rafael~A.}\ \bibnamefont
  {Porto}},\ }\bibfield  {title} {\enquote {\bibinfo {title} {{From Boundary
  Data to Bound States}},}\ }\href {\doibase 10.1007/JHEP01(2020)072}
  {\bibfield  {journal} {\bibinfo  {journal} {JHEP}\ }\textbf {\bibinfo
  {volume} {01}},\ \bibinfo {pages} {072} (\bibinfo {year}
  {2020}{\natexlab{a}})},\ \Eprint {http://arxiv.org/abs/1910.03008}
  {arXiv:1910.03008 [hep-th]} \BibitemShut {NoStop}%
\bibitem [{\citenamefont {K\"alin}\ and\ \citenamefont
  {Porto}(2020{\natexlab{b}})}]{Kalin:2019inp}%
  \BibitemOpen
  \bibfield  {author} {\bibinfo {author} {\bibfnamefont {Gregor}\ \bibnamefont
  {K\"alin}}\ and\ \bibinfo {author} {\bibfnamefont {Rafael~A.}\ \bibnamefont
  {Porto}},\ }\bibfield  {title} {\enquote {\bibinfo {title} {{From boundary
  data to bound states. Part II. Scattering angle to dynamical invariants (with
  twist)}},}\ }\href {\doibase 10.1007/JHEP02(2020)120} {\bibfield  {journal}
  {\bibinfo  {journal} {JHEP}\ }\textbf {\bibinfo {volume} {02}},\ \bibinfo
  {pages} {120} (\bibinfo {year} {2020}{\natexlab{b}})},\ \Eprint
  {http://arxiv.org/abs/1911.09130} {arXiv:1911.09130 [hep-th]} \BibitemShut
  {NoStop}%
\bibitem [{\citenamefont {Bjerrum-Bohr}\ \emph
  {et~al.}(2021{\natexlab{a}})\citenamefont {Bjerrum-Bohr}, \citenamefont
  {Damgaard}, \citenamefont {Plant\'e},\ and\ \citenamefont
  {Vanhove}}]{Bjerrum-Bohr:2021vuf}%
  \BibitemOpen
  \bibfield  {author} {\bibinfo {author} {\bibfnamefont {N.~Emil~J.}\
  \bibnamefont {Bjerrum-Bohr}}, \bibinfo {author} {\bibfnamefont {Poul~H.}\
  \bibnamefont {Damgaard}}, \bibinfo {author} {\bibfnamefont {Ludovic}\
  \bibnamefont {Plant\'e}}, \ and\ \bibinfo {author} {\bibfnamefont {Pierre}\
  \bibnamefont {Vanhove}},\ }\bibfield  {title} {\enquote {\bibinfo {title}
  {{Classical Gravity from Loop Amplitudes}},}\ }\href@noop {} {\  (\bibinfo
  {year} {2021}{\natexlab{a}})},\ \Eprint {http://arxiv.org/abs/2104.04510}
  {arXiv:2104.04510 [hep-th]} \BibitemShut {NoStop}%
\bibitem [{\citenamefont {Cheung}\ and\ \citenamefont
  {Solon}(2020)}]{Cheung:2020gyp}%
  \BibitemOpen
  \bibfield  {author} {\bibinfo {author} {\bibfnamefont {Clifford}\
  \bibnamefont {Cheung}}\ and\ \bibinfo {author} {\bibfnamefont {Mikhail~P.}\
  \bibnamefont {Solon}},\ }\bibfield  {title} {\enquote {\bibinfo {title}
  {{Classical gravitational scattering at $ \mathcal{O} $(G$^{3}$) from Feynman
  diagrams}},}\ }\href {\doibase 10.1007/JHEP06(2020)144} {\bibfield  {journal}
  {\bibinfo  {journal} {JHEP}\ }\textbf {\bibinfo {volume} {06}},\ \bibinfo
  {pages} {144} (\bibinfo {year} {2020})},\ \Eprint
  {http://arxiv.org/abs/2003.08351} {arXiv:2003.08351 [hep-th]} \BibitemShut
  {NoStop}%
\bibitem [{\citenamefont {Damour}(2020)}]{Damour:2020tta}%
  \BibitemOpen
  \bibfield  {author} {\bibinfo {author} {\bibfnamefont {Thibault}\
  \bibnamefont {Damour}},\ }\bibfield  {title} {\enquote {\bibinfo {title}
  {{Radiative contribution to classical gravitational scattering at the third
  order in $G$}},}\ }\href {\doibase 10.1103/PhysRevD.102.124008} {\bibfield
  {journal} {\bibinfo  {journal} {Phys. Rev. D}\ }\textbf {\bibinfo {volume}
  {102}},\ \bibinfo {pages} {124008} (\bibinfo {year} {2020})},\ \Eprint
  {http://arxiv.org/abs/2010.01641} {arXiv:2010.01641 [gr-qc]} \BibitemShut
  {NoStop}%
\bibitem [{\citenamefont {K\"alin}\ \emph {et~al.}(2020)\citenamefont
  {K\"alin}, \citenamefont {Liu},\ and\ \citenamefont {Porto}}]{Kalin:2020lmz}%
  \BibitemOpen
  \bibfield  {author} {\bibinfo {author} {\bibfnamefont {Gregor}\ \bibnamefont
  {K\"alin}}, \bibinfo {author} {\bibfnamefont {Zhengwen}\ \bibnamefont {Liu}},
  \ and\ \bibinfo {author} {\bibfnamefont {Rafael~A.}\ \bibnamefont {Porto}},\
  }\bibfield  {title} {\enquote {\bibinfo {title} {{Conservative Tidal Effects
  in Compact Binary Systems to Next-to-Leading Post-Minkowskian Order}},}\
  }\href {\doibase 10.1103/PhysRevD.102.124025} {\bibfield  {journal} {\bibinfo
   {journal} {Phys. Rev. D}\ }\textbf {\bibinfo {volume} {102}},\ \bibinfo
  {pages} {124025} (\bibinfo {year} {2020})},\ \Eprint
  {http://arxiv.org/abs/2008.06047} {arXiv:2008.06047 [hep-th]} \BibitemShut
  {NoStop}%
\bibitem [{\citenamefont {Di~Vecchia}\ \emph
  {et~al.}(2021{\natexlab{a}})\citenamefont {Di~Vecchia}, \citenamefont
  {Heissenberg}, \citenamefont {Russo},\ and\ \citenamefont
  {Veneziano}}]{DiVecchia:2021bdo}%
  \BibitemOpen
  \bibfield  {author} {\bibinfo {author} {\bibfnamefont {Paolo}\ \bibnamefont
  {Di~Vecchia}}, \bibinfo {author} {\bibfnamefont {Carlo}\ \bibnamefont
  {Heissenberg}}, \bibinfo {author} {\bibfnamefont {Rodolfo}\ \bibnamefont
  {Russo}}, \ and\ \bibinfo {author} {\bibfnamefont {Gabriele}\ \bibnamefont
  {Veneziano}},\ }\bibfield  {title} {\enquote {\bibinfo {title} {{The Eikonal
  Approach to Gravitational Scattering and Radiation at $\mathcal O(G^3)$}},}\
  }\href@noop {} {\  (\bibinfo {year} {2021}{\natexlab{a}})},\ \Eprint
  {http://arxiv.org/abs/2104.03256} {arXiv:2104.03256 [hep-th]} \BibitemShut
  {NoStop}%
\bibitem [{\citenamefont {Liu}\ \emph {et~al.}(2021)\citenamefont {Liu},
  \citenamefont {Porto},\ and\ \citenamefont {Yang}}]{Liu:2021zxr}%
  \BibitemOpen
  \bibfield  {author} {\bibinfo {author} {\bibfnamefont {Zhengwen}\
  \bibnamefont {Liu}}, \bibinfo {author} {\bibfnamefont {Rafael~A.}\
  \bibnamefont {Porto}}, \ and\ \bibinfo {author} {\bibfnamefont {Zixin}\
  \bibnamefont {Yang}},\ }\bibfield  {title} {\enquote {\bibinfo {title} {{Spin
  Effects in the Effective Field Theory Approach to Post-Minkowskian
  Conservative Dynamics}},}\ }\href {\doibase 10.1007/JHEP06(2021)012}
  {\bibfield  {journal} {\bibinfo  {journal} {JHEP}\ }\textbf {\bibinfo
  {volume} {06}},\ \bibinfo {pages} {012} (\bibinfo {year} {2021})},\ \Eprint
  {http://arxiv.org/abs/2102.10059} {arXiv:2102.10059 [hep-th]} \BibitemShut
  {NoStop}%
\bibitem [{\citenamefont {Di~Vecchia}\ \emph
  {et~al.}(2021{\natexlab{b}})\citenamefont {Di~Vecchia}, \citenamefont
  {Heissenberg}, \citenamefont {Russo},\ and\ \citenamefont
  {Veneziano}}]{DiVecchia:2021ndb}%
  \BibitemOpen
  \bibfield  {author} {\bibinfo {author} {\bibfnamefont {Paolo}\ \bibnamefont
  {Di~Vecchia}}, \bibinfo {author} {\bibfnamefont {Carlo}\ \bibnamefont
  {Heissenberg}}, \bibinfo {author} {\bibfnamefont {Rodolfo}\ \bibnamefont
  {Russo}}, \ and\ \bibinfo {author} {\bibfnamefont {Gabriele}\ \bibnamefont
  {Veneziano}},\ }\bibfield  {title} {\enquote {\bibinfo {title} {{Radiation
  Reaction from Soft Theorems}},}\ }\href {\doibase
  10.1016/j.physletb.2021.136379} {\bibfield  {journal} {\bibinfo  {journal}
  {Phys. Lett. B}\ }\textbf {\bibinfo {volume} {818}},\ \bibinfo {pages}
  {136379} (\bibinfo {year} {2021}{\natexlab{b}})},\ \Eprint
  {http://arxiv.org/abs/2101.05772} {arXiv:2101.05772 [hep-th]} \BibitemShut
  {NoStop}%
\bibitem [{\citenamefont {Cho}\ \emph {et~al.}(2021)\citenamefont {Cho},
  \citenamefont {Pardo},\ and\ \citenamefont {Porto}}]{Cho:2021mqw}%
  \BibitemOpen
  \bibfield  {author} {\bibinfo {author} {\bibfnamefont {Gihyuk}\ \bibnamefont
  {Cho}}, \bibinfo {author} {\bibfnamefont {Brian}\ \bibnamefont {Pardo}}, \
  and\ \bibinfo {author} {\bibfnamefont {Rafael~A.}\ \bibnamefont {Porto}},\
  }\bibfield  {title} {\enquote {\bibinfo {title} {{Gravitational radiation
  from inspiralling compact objects: Spin-spin effects completed at the
  next-to-leading post-Newtonian order}},}\ }\href@noop {} {\  (\bibinfo {year}
  {2021})},\ \Eprint {http://arxiv.org/abs/2103.14612} {arXiv:2103.14612
  [gr-qc]} \BibitemShut {NoStop}%
\bibitem [{\citenamefont {Bjerrum-Bohr}\ \emph
  {et~al.}(2021{\natexlab{b}})\citenamefont {Bjerrum-Bohr}, \citenamefont
  {Damgaard}, \citenamefont {Plant\'e},\ and\ \citenamefont
  {Vanhove}}]{Bjerrum-Bohr:2021din}%
  \BibitemOpen
  \bibfield  {author} {\bibinfo {author} {\bibfnamefont {N.~E.~J.}\
  \bibnamefont {Bjerrum-Bohr}}, \bibinfo {author} {\bibfnamefont {P.~H.}\
  \bibnamefont {Damgaard}}, \bibinfo {author} {\bibfnamefont {L.}~\bibnamefont
  {Plant\'e}}, \ and\ \bibinfo {author} {\bibfnamefont {P.}~\bibnamefont
  {Vanhove}},\ }\bibfield  {title} {\enquote {\bibinfo {title} {{The Amplitude
  for Classical Gravitational Scattering at Third Post-Minkowskian Order}},}\
  }\href@noop {} {\  (\bibinfo {year} {2021}{\natexlab{b}})},\ \Eprint
  {http://arxiv.org/abs/2105.05218} {arXiv:2105.05218 [hep-th]} \BibitemShut
  {NoStop}%
\bibitem [{\citenamefont {Dlapa}\ \emph {et~al.}(2021)\citenamefont {Dlapa},
  \citenamefont {K\"alin}, \citenamefont {Liu},\ and\ \citenamefont
  {Porto}}]{Dlapa:2021npj}%
  \BibitemOpen
  \bibfield  {author} {\bibinfo {author} {\bibfnamefont {Christoph}\
  \bibnamefont {Dlapa}}, \bibinfo {author} {\bibfnamefont {Gregor}\
  \bibnamefont {K\"alin}}, \bibinfo {author} {\bibfnamefont {Zhengwen}\
  \bibnamefont {Liu}}, \ and\ \bibinfo {author} {\bibfnamefont {Rafael~A.}\
  \bibnamefont {Porto}},\ }\bibfield  {title} {\enquote {\bibinfo {title}
  {{Dynamics of Binary Systems to Fourth Post-Minkowskian Order from the
  Effective Field Theory Approach}},}\ }\href@noop {} {\  (\bibinfo {year}
  {2021})},\ \Eprint {http://arxiv.org/abs/2106.08276} {arXiv:2106.08276
  [hep-th]} \BibitemShut {NoStop}%
\bibitem [{\citenamefont {Cristofoli}\ \emph {et~al.}(2021)\citenamefont
  {Cristofoli}, \citenamefont {Gonzo}, \citenamefont {Kosower},\ and\
  \citenamefont {O'Connell}}]{Cristofoli:2021vyo}%
  \BibitemOpen
  \bibfield  {author} {\bibinfo {author} {\bibfnamefont {Andrea}\ \bibnamefont
  {Cristofoli}}, \bibinfo {author} {\bibfnamefont {Riccardo}\ \bibnamefont
  {Gonzo}}, \bibinfo {author} {\bibfnamefont {David~A.}\ \bibnamefont
  {Kosower}}, \ and\ \bibinfo {author} {\bibfnamefont {Donal}\ \bibnamefont
  {O'Connell}},\ }\bibfield  {title} {\enquote {\bibinfo {title} {{Waveforms
  from Amplitudes}},}\ }\href@noop {} {\  (\bibinfo {year} {2021})},\ \Eprint
  {http://arxiv.org/abs/2107.10193} {arXiv:2107.10193 [hep-th]} \BibitemShut
  {NoStop}%
\bibitem [{\citenamefont {Bautista}\ \emph {et~al.}(2021)\citenamefont
  {Bautista}, \citenamefont {Guevara}, \citenamefont {Kavanagh},\ and\
  \citenamefont {Vines}}]{Bautista:2021wfy}%
  \BibitemOpen
  \bibfield  {author} {\bibinfo {author} {\bibfnamefont {Yilber~Fabian}\
  \bibnamefont {Bautista}}, \bibinfo {author} {\bibfnamefont {Alfredo}\
  \bibnamefont {Guevara}}, \bibinfo {author} {\bibfnamefont {Chris}\
  \bibnamefont {Kavanagh}}, \ and\ \bibinfo {author} {\bibfnamefont {Justin}\
  \bibnamefont {Vines}},\ }\bibfield  {title} {\enquote {\bibinfo {title}
  {{From Scattering in Black Hole Backgrounds to Higher-Spin Amplitudes: Part
  I}},}\ }\href@noop {} {\  (\bibinfo {year} {2021})},\ \Eprint
  {http://arxiv.org/abs/2107.10179} {arXiv:2107.10179 [hep-th]} \BibitemShut
  {NoStop}%
\bibitem [{\citenamefont {Kosmopoulos}\ and\ \citenamefont
  {Luna}(2021)}]{Kosmopoulos:2021zoq}%
  \BibitemOpen
  \bibfield  {author} {\bibinfo {author} {\bibfnamefont {Dimitrios}\
  \bibnamefont {Kosmopoulos}}\ and\ \bibinfo {author} {\bibfnamefont {Andres}\
  \bibnamefont {Luna}},\ }\bibfield  {title} {\enquote {\bibinfo {title}
  {{Quadratic-in-spin Hamiltonian at $ \mathcal{O} $(G$^{2}$) from scattering
  amplitudes}},}\ }\href {\doibase 10.1007/JHEP07(2021)037} {\bibfield
  {journal} {\bibinfo  {journal} {JHEP}\ }\textbf {\bibinfo {volume} {07}},\
  \bibinfo {pages} {037} (\bibinfo {year} {2021})},\ \Eprint
  {http://arxiv.org/abs/2102.10137} {arXiv:2102.10137 [hep-th]} \BibitemShut
  {NoStop}%
\bibitem [{\citenamefont {de~la Cruz}\ \emph {et~al.}(2020)\citenamefont {de~la
  Cruz}, \citenamefont {Maybee}, \citenamefont {O'Connell},\ and\ \citenamefont
  {Ross}}]{delaCruz:2020bbn}%
  \BibitemOpen
  \bibfield  {author} {\bibinfo {author} {\bibfnamefont {Leonardo}\
  \bibnamefont {de~la Cruz}}, \bibinfo {author} {\bibfnamefont {Ben}\
  \bibnamefont {Maybee}}, \bibinfo {author} {\bibfnamefont {Donal}\
  \bibnamefont {O'Connell}}, \ and\ \bibinfo {author} {\bibfnamefont
  {Alasdair}\ \bibnamefont {Ross}},\ }\bibfield  {title} {\enquote {\bibinfo
  {title} {{Classical Yang-Mills observables from amplitudes}},}\ }\href
  {\doibase 10.1007/JHEP12(2020)076} {\bibfield  {journal} {\bibinfo  {journal}
  {JHEP}\ }\textbf {\bibinfo {volume} {12}},\ \bibinfo {pages} {076} (\bibinfo
  {year} {2020})},\ \Eprint {http://arxiv.org/abs/2009.03842} {arXiv:2009.03842
  [hep-th]} \BibitemShut {NoStop}%
\bibitem [{\citenamefont {de~la Cruz}\ \emph {et~al.}(2021)\citenamefont {de~la
  Cruz}, \citenamefont {Luna},\ and\ \citenamefont
  {Scheopner}}]{delaCruz:2021gjp}%
  \BibitemOpen
  \bibfield  {author} {\bibinfo {author} {\bibfnamefont {Leonardo}\
  \bibnamefont {de~la Cruz}}, \bibinfo {author} {\bibfnamefont {Andres}\
  \bibnamefont {Luna}}, \ and\ \bibinfo {author} {\bibfnamefont {Trevor}\
  \bibnamefont {Scheopner}},\ }\bibfield  {title} {\enquote {\bibinfo {title}
  {{Yang-Mills observables: from KMOC to eikonal through EFT}},}\ }\href@noop
  {} {\  (\bibinfo {year} {2021})},\ \Eprint {http://arxiv.org/abs/2108.02178}
  {arXiv:2108.02178 [hep-th]} \BibitemShut {NoStop}%
\bibitem [{\citenamefont {Brandhuber}\ and\ \citenamefont
  {Travaglini}(2020)}]{Brandhuber:2019qpg}%
  \BibitemOpen
  \bibfield  {author} {\bibinfo {author} {\bibfnamefont {Andreas}\ \bibnamefont
  {Brandhuber}}\ and\ \bibinfo {author} {\bibfnamefont {Gabriele}\ \bibnamefont
  {Travaglini}},\ }\bibfield  {title} {\enquote {\bibinfo {title} {{On
  higher-derivative effects on the gravitational potential and particle
  bending}},}\ }\href {\doibase 10.1007/JHEP01(2020)010} {\bibfield  {journal}
  {\bibinfo  {journal} {JHEP}\ }\textbf {\bibinfo {volume} {01}},\ \bibinfo
  {pages} {010} (\bibinfo {year} {2020})},\ \Eprint
  {http://arxiv.org/abs/1905.05657} {arXiv:1905.05657 [hep-th]} \BibitemShut
  {NoStop}%
\bibitem [{\citenamefont {Accettulli~Huber}\ \emph
  {et~al.}(2020{\natexlab{a}})\citenamefont {Accettulli~Huber}, \citenamefont
  {Brandhuber}, \citenamefont {De~Angelis},\ and\ \citenamefont
  {Travaglini}}]{AccettulliHuber:2019jqo}%
  \BibitemOpen
  \bibfield  {author} {\bibinfo {author} {\bibfnamefont {Manuel}\ \bibnamefont
  {Accettulli~Huber}}, \bibinfo {author} {\bibfnamefont {Andreas}\ \bibnamefont
  {Brandhuber}}, \bibinfo {author} {\bibfnamefont {Stefano}\ \bibnamefont
  {De~Angelis}}, \ and\ \bibinfo {author} {\bibfnamefont {Gabriele}\
  \bibnamefont {Travaglini}},\ }\bibfield  {title} {\enquote {\bibinfo {title}
  {{Note on the absence of $R^2$ corrections to Newton's potential}},}\ }\href
  {\doibase 10.1103/PhysRevD.101.046011} {\bibfield  {journal} {\bibinfo
  {journal} {Phys. Rev. D}\ }\textbf {\bibinfo {volume} {101}},\ \bibinfo
  {pages} {046011} (\bibinfo {year} {2020}{\natexlab{a}})},\ \Eprint
  {http://arxiv.org/abs/1911.10108} {arXiv:1911.10108 [hep-th]} \BibitemShut
  {NoStop}%
\bibitem [{\citenamefont {Emond}\ and\ \citenamefont
  {Moynihan}(2019)}]{Emond:2019crr}%
  \BibitemOpen
  \bibfield  {author} {\bibinfo {author} {\bibfnamefont {William~T.}\
  \bibnamefont {Emond}}\ and\ \bibinfo {author} {\bibfnamefont {Nathan}\
  \bibnamefont {Moynihan}},\ }\bibfield  {title} {\enquote {\bibinfo {title}
  {{Scattering Amplitudes, Black Holes and Leading Singularities in Cubic
  Theories of Gravity}},}\ }\href {\doibase 10.1007/JHEP12(2019)019} {\bibfield
   {journal} {\bibinfo  {journal} {JHEP}\ }\textbf {\bibinfo {volume} {12}},\
  \bibinfo {pages} {019} (\bibinfo {year} {2019})},\ \Eprint
  {http://arxiv.org/abs/1905.08213} {arXiv:1905.08213 [hep-th]} \BibitemShut
  {NoStop}%
\bibitem [{\citenamefont {Accettulli~Huber}\ \emph
  {et~al.}(2020{\natexlab{b}})\citenamefont {Accettulli~Huber}, \citenamefont
  {Brandhuber}, \citenamefont {De~Angelis},\ and\ \citenamefont
  {Travaglini}}]{AccettulliHuber:2020oou}%
  \BibitemOpen
  \bibfield  {author} {\bibinfo {author} {\bibfnamefont {Manuel}\ \bibnamefont
  {Accettulli~Huber}}, \bibinfo {author} {\bibfnamefont {Andreas}\ \bibnamefont
  {Brandhuber}}, \bibinfo {author} {\bibfnamefont {Stefano}\ \bibnamefont
  {De~Angelis}}, \ and\ \bibinfo {author} {\bibfnamefont {Gabriele}\
  \bibnamefont {Travaglini}},\ }\bibfield  {title} {\enquote {\bibinfo {title}
  {{Eikonal phase matrix, deflection angle and time delay in effective field
  theories of gravity}},}\ }\href {\doibase 10.1103/PhysRevD.102.046014}
  {\bibfield  {journal} {\bibinfo  {journal} {Phys. Rev. D}\ }\textbf {\bibinfo
  {volume} {102}},\ \bibinfo {pages} {046014} (\bibinfo {year}
  {2020}{\natexlab{b}})},\ \Eprint {http://arxiv.org/abs/2006.02375}
  {arXiv:2006.02375 [hep-th]} \BibitemShut {NoStop}%
\bibitem [{\citenamefont {Accettulli~Huber}\ \emph {et~al.}(2021)\citenamefont
  {Accettulli~Huber}, \citenamefont {Brandhuber}, \citenamefont {De~Angelis},\
  and\ \citenamefont {Travaglini}}]{AccettulliHuber:2020dal}%
  \BibitemOpen
  \bibfield  {author} {\bibinfo {author} {\bibfnamefont {Manuel}\ \bibnamefont
  {Accettulli~Huber}}, \bibinfo {author} {\bibfnamefont {Andreas}\ \bibnamefont
  {Brandhuber}}, \bibinfo {author} {\bibfnamefont {Stefano}\ \bibnamefont
  {De~Angelis}}, \ and\ \bibinfo {author} {\bibfnamefont {Gabriele}\
  \bibnamefont {Travaglini}},\ }\bibfield  {title} {\enquote {\bibinfo {title}
  {{From amplitudes to gravitational radiation with cubic interactions and
  tidal effects}},}\ }\href {\doibase 10.1103/PhysRevD.103.045015} {\bibfield
  {journal} {\bibinfo  {journal} {Phys. Rev. D}\ }\textbf {\bibinfo {volume}
  {103}},\ \bibinfo {pages} {045015} (\bibinfo {year} {2021})},\ \Eprint
  {http://arxiv.org/abs/2012.06548} {arXiv:2012.06548 [hep-th]} \BibitemShut
  {NoStop}%
\bibitem [{\citenamefont {Deffayet}\ \emph {et~al.}(2010)\citenamefont
  {Deffayet}, \citenamefont {Deser},\ and\ \citenamefont
  {Esposito-Farese}}]{Deffayet:2010zh}%
  \BibitemOpen
  \bibfield  {author} {\bibinfo {author} {\bibfnamefont {C.}~\bibnamefont
  {Deffayet}}, \bibinfo {author} {\bibfnamefont {S.}~\bibnamefont {Deser}}, \
  and\ \bibinfo {author} {\bibfnamefont {G.}~\bibnamefont {Esposito-Farese}},\
  }\bibfield  {title} {\enquote {\bibinfo {title} {{Arbitrary $p$-form
  Galileons}},}\ }\href {\doibase 10.1103/PhysRevD.82.061501} {\bibfield
  {journal} {\bibinfo  {journal} {Phys. Rev.}\ }\textbf {\bibinfo {volume}
  {D82}},\ \bibinfo {pages} {061501} (\bibinfo {year} {2010})},\ \Eprint
  {http://arxiv.org/abs/1007.5278} {arXiv:1007.5278 [gr-qc]} \BibitemShut
  {NoStop}%
\bibitem [{\citenamefont {Heisenberg}(2014)}]{Heisenberg:2014rta}%
  \BibitemOpen
  \bibfield  {author} {\bibinfo {author} {\bibfnamefont {Lavinia}\ \bibnamefont
  {Heisenberg}},\ }\bibfield  {title} {\enquote {\bibinfo {title}
  {{Generalization of the Proca Action}},}\ }\href {\doibase
  10.1088/1475-7516/2014/05/015} {\bibfield  {journal} {\bibinfo  {journal}
  {JCAP}\ }\textbf {\bibinfo {volume} {1405}},\ \bibinfo {pages} {015}
  (\bibinfo {year} {2014})},\ \Eprint {http://arxiv.org/abs/1402.7026}
  {arXiv:1402.7026 [hep-th]} \BibitemShut {NoStop}%
\bibitem [{\citenamefont {Tasinato}(2014)}]{Tasinato:2014eka}%
  \BibitemOpen
  \bibfield  {author} {\bibinfo {author} {\bibfnamefont {Gianmassimo}\
  \bibnamefont {Tasinato}},\ }\bibfield  {title} {\enquote {\bibinfo {title}
  {{Cosmic Acceleration from Abelian Symmetry Breaking}},}\ }\href {\doibase
  10.1007/JHEP04(2014)067} {\bibfield  {journal} {\bibinfo  {journal} {JHEP}\
  }\textbf {\bibinfo {volume} {04}},\ \bibinfo {pages} {067} (\bibinfo {year}
  {2014})},\ \Eprint {http://arxiv.org/abs/1402.6450} {arXiv:1402.6450
  [hep-th]} \BibitemShut {NoStop}%
\bibitem [{\citenamefont {Allys}\ \emph
  {et~al.}(2016{\natexlab{a}})\citenamefont {Allys}, \citenamefont {Peter},\
  and\ \citenamefont {Rodriguez}}]{Allys:2015sht}%
  \BibitemOpen
  \bibfield  {author} {\bibinfo {author} {\bibfnamefont {Erwan}\ \bibnamefont
  {Allys}}, \bibinfo {author} {\bibfnamefont {Patrick}\ \bibnamefont {Peter}},
  \ and\ \bibinfo {author} {\bibfnamefont {Yeinzon}\ \bibnamefont
  {Rodriguez}},\ }\bibfield  {title} {\enquote {\bibinfo {title} {{Generalized
  Proca action for an Abelian vector field}},}\ }\href {\doibase
  10.1088/1475-7516/2016/02/004} {\bibfield  {journal} {\bibinfo  {journal}
  {JCAP}\ }\textbf {\bibinfo {volume} {1602}},\ \bibinfo {pages} {004}
  (\bibinfo {year} {2016}{\natexlab{a}})},\ \Eprint
  {http://arxiv.org/abs/1511.03101} {arXiv:1511.03101 [hep-th]} \BibitemShut
  {NoStop}%
\bibitem [{\citenamefont {Hull}\ \emph {et~al.}(2016)\citenamefont {Hull},
  \citenamefont {Koyama},\ and\ \citenamefont {Tasinato}}]{Hull:2015uwa}%
  \BibitemOpen
  \bibfield  {author} {\bibinfo {author} {\bibfnamefont {Matthew}\ \bibnamefont
  {Hull}}, \bibinfo {author} {\bibfnamefont {Kazuya}\ \bibnamefont {Koyama}}, \
  and\ \bibinfo {author} {\bibfnamefont {Gianmassimo}\ \bibnamefont
  {Tasinato}},\ }\bibfield  {title} {\enquote {\bibinfo {title} {{Covariantized
  vector Galileons}},}\ }\href {\doibase 10.1103/PhysRevD.93.064012} {\bibfield
   {journal} {\bibinfo  {journal} {Phys. Rev.}\ }\textbf {\bibinfo {volume}
  {D93}},\ \bibinfo {pages} {064012} (\bibinfo {year} {2016})},\ \Eprint
  {http://arxiv.org/abs/1510.07029} {arXiv:1510.07029 [hep-th]} \BibitemShut
  {NoStop}%
\bibitem [{\citenamefont {Allys}\ \emph
  {et~al.}(2016{\natexlab{b}})\citenamefont {Allys}, \citenamefont
  {Beltran~Almeida}, \citenamefont {Peter},\ and\ \citenamefont
  {Rodr\'\i{}guez}}]{Allys:2016jaq}%
  \BibitemOpen
  \bibfield  {author} {\bibinfo {author} {\bibfnamefont {Erwan}\ \bibnamefont
  {Allys}}, \bibinfo {author} {\bibfnamefont {Juan~P.}\ \bibnamefont
  {Beltran~Almeida}}, \bibinfo {author} {\bibfnamefont {Patrick}\ \bibnamefont
  {Peter}}, \ and\ \bibinfo {author} {\bibfnamefont {Yeinzon}\ \bibnamefont
  {Rodr\'\i{}guez}},\ }\bibfield  {title} {\enquote {\bibinfo {title} {{On the
  4D generalized Proca action for an Abelian vector field}},}\ }\href {\doibase
  10.1088/1475-7516/2016/09/026} {\bibfield  {journal} {\bibinfo  {journal}
  {JCAP}\ }\textbf {\bibinfo {volume} {09}},\ \bibinfo {pages} {026} (\bibinfo
  {year} {2016}{\natexlab{b}})},\ \Eprint {http://arxiv.org/abs/1605.08355}
  {arXiv:1605.08355 [hep-th]} \BibitemShut {NoStop}%
\bibitem [{\citenamefont {Heisenberg}\ \emph {et~al.}(2016)\citenamefont
  {Heisenberg}, \citenamefont {Kase},\ and\ \citenamefont
  {Tsujikawa}}]{Heisenberg:2016eld}%
  \BibitemOpen
  \bibfield  {author} {\bibinfo {author} {\bibfnamefont {Lavinia}\ \bibnamefont
  {Heisenberg}}, \bibinfo {author} {\bibfnamefont {Ryotaro}\ \bibnamefont
  {Kase}}, \ and\ \bibinfo {author} {\bibfnamefont {Shinji}\ \bibnamefont
  {Tsujikawa}},\ }\bibfield  {title} {\enquote {\bibinfo {title} {{Beyond
  generalized Proca theories}},}\ }\href {\doibase
  10.1016/j.physletb.2016.07.052} {\bibfield  {journal} {\bibinfo  {journal}
  {Phys. Lett.}\ }\textbf {\bibinfo {volume} {B760}},\ \bibinfo {pages}
  {617--626} (\bibinfo {year} {2016})},\ \Eprint
  {http://arxiv.org/abs/1605.05565} {arXiv:1605.05565 [hep-th]} \BibitemShut
  {NoStop}%
\bibitem [{\citenamefont {Beltran~Jimenez}\ and\ \citenamefont
  {Heisenberg}(2016)}]{Jimenez:2016isa}%
  \BibitemOpen
  \bibfield  {author} {\bibinfo {author} {\bibfnamefont {Jose}\ \bibnamefont
  {Beltran~Jimenez}}\ and\ \bibinfo {author} {\bibfnamefont {Lavinia}\
  \bibnamefont {Heisenberg}},\ }\bibfield  {title} {\enquote {\bibinfo {title}
  {{Derivative self-interactions for a massive vector field}},}\ }\href
  {\doibase 10.1016/j.physletb.2016.04.017} {\bibfield  {journal} {\bibinfo
  {journal} {Phys. Lett.}\ }\textbf {\bibinfo {volume} {B757}},\ \bibinfo
  {pages} {405--411} (\bibinfo {year} {2016})},\ \Eprint
  {http://arxiv.org/abs/1602.03410} {arXiv:1602.03410 [hep-th]} \BibitemShut
  {NoStop}%
\bibitem [{\citenamefont {De~Felice}\ \emph {et~al.}(2016)\citenamefont
  {De~Felice}, \citenamefont {Heisenberg}, \citenamefont {Kase}, \citenamefont
  {Mukohyama}, \citenamefont {Tsujikawa},\ and\ \citenamefont
  {Zhang}}]{DeFelice:2016yws}%
  \BibitemOpen
  \bibfield  {author} {\bibinfo {author} {\bibfnamefont {Antonio}\ \bibnamefont
  {De~Felice}}, \bibinfo {author} {\bibfnamefont {Lavinia}\ \bibnamefont
  {Heisenberg}}, \bibinfo {author} {\bibfnamefont {Ryotaro}\ \bibnamefont
  {Kase}}, \bibinfo {author} {\bibfnamefont {Shinji}\ \bibnamefont
  {Mukohyama}}, \bibinfo {author} {\bibfnamefont {Shinji}\ \bibnamefont
  {Tsujikawa}}, \ and\ \bibinfo {author} {\bibfnamefont {Ying-li}\ \bibnamefont
  {Zhang}},\ }\bibfield  {title} {\enquote {\bibinfo {title} {{Cosmology in
  generalized Proca theories}},}\ }\href {\doibase
  10.1088/1475-7516/2016/06/048} {\bibfield  {journal} {\bibinfo  {journal}
  {JCAP}\ }\textbf {\bibinfo {volume} {1606}},\ \bibinfo {pages} {048}
  (\bibinfo {year} {2016})},\ \Eprint {http://arxiv.org/abs/1603.05806}
  {arXiv:1603.05806 [gr-qc]} \BibitemShut {NoStop}%
\bibitem [{\citenamefont {Allys}(2017)}]{Allys:2017map}%
  \BibitemOpen
  \bibfield  {author} {\bibinfo {author} {\bibfnamefont {Erwan}\ \bibnamefont
  {Allys}},\ }\emph {\bibinfo {title} {{Au-delà des modèles standards en
  cosmologie}}},\ \href@noop {} {Ph.D. thesis},\ \bibinfo  {school} {UPMC,
  Paris (main)} (\bibinfo {year} {2017}),\ \Eprint
  {http://arxiv.org/abs/1710.02143} {arXiv:1710.02143 [astro-ph.CO]}
  \BibitemShut {NoStop}%
\bibitem [{\citenamefont {de~Rham}\ and\ \citenamefont
  {Pozsgay}(2020)}]{deRham:2020yet}%
  \BibitemOpen
  \bibfield  {author} {\bibinfo {author} {\bibfnamefont {Claudia}\ \bibnamefont
  {de~Rham}}\ and\ \bibinfo {author} {\bibfnamefont {Victor}\ \bibnamefont
  {Pozsgay}},\ }\bibfield  {title} {\enquote {\bibinfo {title} {{New class of
  Proca interactions}},}\ }\href {\doibase 10.1103/PhysRevD.102.083508}
  {\bibfield  {journal} {\bibinfo  {journal} {Phys. Rev. D}\ }\textbf {\bibinfo
  {volume} {102}},\ \bibinfo {pages} {083508} (\bibinfo {year} {2020})},\
  \Eprint {http://arxiv.org/abs/2003.13773} {arXiv:2003.13773 [hep-th]}
  \BibitemShut {NoStop}%
\bibitem [{\citenamefont {Will}(2014)}]{Will:2014kxa}%
  \BibitemOpen
  \bibfield  {author} {\bibinfo {author} {\bibfnamefont {Clifford~M.}\
  \bibnamefont {Will}},\ }\bibfield  {title} {\enquote {\bibinfo {title} {{The
  Confrontation between General Relativity and Experiment}},}\ }\href {\doibase
  10.12942/lrr-2014-4} {\bibfield  {journal} {\bibinfo  {journal} {Living Rev.
  Rel.}\ }\textbf {\bibinfo {volume} {17}},\ \bibinfo {pages} {4} (\bibinfo
  {year} {2014})},\ \Eprint {http://arxiv.org/abs/1403.7377} {arXiv:1403.7377
  [gr-qc]} \BibitemShut {NoStop}%
\bibitem [{\citenamefont {Bertotti}\ \emph {et~al.}(2003)\citenamefont
  {Bertotti}, \citenamefont {Iess},\ and\ \citenamefont
  {Tortora}}]{Bertotti:2003rm}%
  \BibitemOpen
  \bibfield  {author} {\bibinfo {author} {\bibfnamefont {B.}~\bibnamefont
  {Bertotti}}, \bibinfo {author} {\bibfnamefont {L.}~\bibnamefont {Iess}}, \
  and\ \bibinfo {author} {\bibfnamefont {P.}~\bibnamefont {Tortora}},\
  }\bibfield  {title} {\enquote {\bibinfo {title} {{A test of general
  relativity using radio links with the Cassini spacecraft}},}\ }\href
  {\doibase 10.1038/nature01997} {\bibfield  {journal} {\bibinfo  {journal}
  {Nature}\ }\textbf {\bibinfo {volume} {425}},\ \bibinfo {pages} {374--376}
  (\bibinfo {year} {2003})}\BibitemShut {NoStop}%
\bibitem [{\citenamefont {Vainshtein}(1972)}]{Vainshtein:1972sx}%
  \BibitemOpen
  \bibfield  {author} {\bibinfo {author} {\bibfnamefont {A.~I.}\ \bibnamefont
  {Vainshtein}},\ }\bibfield  {title} {\enquote {\bibinfo {title} {{To the
  problem of nonvanishing gravitation mass}},}\ }\href {\doibase
  10.1016/0370-2693(72)90147-5} {\bibfield  {journal} {\bibinfo  {journal}
  {Phys. Lett.}\ }\textbf {\bibinfo {volume} {39B}},\ \bibinfo {pages}
  {393--394} (\bibinfo {year} {1972})}\BibitemShut {NoStop}%
\bibitem [{\citenamefont {Babichev}\ and\ \citenamefont
  {Deffayet}(2013)}]{Babichev:2013usa}%
  \BibitemOpen
  \bibfield  {author} {\bibinfo {author} {\bibfnamefont {Eugeny}\ \bibnamefont
  {Babichev}}\ and\ \bibinfo {author} {\bibfnamefont {C{\'e}dric}\ \bibnamefont
  {Deffayet}},\ }\bibfield  {title} {\enquote {\bibinfo {title} {{An
  introduction to the Vainshtein mechanism}},}\ }\href {\doibase
  10.1088/0264-9381/30/18/184001} {\bibfield  {journal} {\bibinfo  {journal}
  {Class. Quant. Grav.}\ }\textbf {\bibinfo {volume} {30}},\ \bibinfo {pages}
  {184001} (\bibinfo {year} {2013})},\ \Eprint {http://arxiv.org/abs/1304.7240}
  {arXiv:1304.7240 [gr-qc]} \BibitemShut {NoStop}%
\bibitem [{\citenamefont {Dvali}\ \emph {et~al.}(2000)\citenamefont {Dvali},
  \citenamefont {Gabadadze},\ and\ \citenamefont {Porrati}}]{Dvali:2000hr}%
  \BibitemOpen
  \bibfield  {author} {\bibinfo {author} {\bibfnamefont {G.~R.}\ \bibnamefont
  {Dvali}}, \bibinfo {author} {\bibfnamefont {Gregory}\ \bibnamefont
  {Gabadadze}}, \ and\ \bibinfo {author} {\bibfnamefont {Massimo}\ \bibnamefont
  {Porrati}},\ }\bibfield  {title} {\enquote {\bibinfo {title} {{4-D gravity on
  a brane in 5-D Minkowski space}},}\ }\href {\doibase
  10.1016/S0370-2693(00)00669-9} {\bibfield  {journal} {\bibinfo  {journal}
  {Phys. Lett.}\ }\textbf {\bibinfo {volume} {B485}},\ \bibinfo {pages}
  {208--214} (\bibinfo {year} {2000})},\ \Eprint
  {http://arxiv.org/abs/hep-th/0005016} {arXiv:hep-th/0005016 [hep-th]}
  \BibitemShut {NoStop}%
\bibitem [{\citenamefont {de~Rham}\ \emph
  {et~al.}(2008{\natexlab{a}})\citenamefont {de~Rham}, \citenamefont {Dvali},
  \citenamefont {Hofmann}, \citenamefont {Khoury}, \citenamefont {Pujolas},
  \citenamefont {Redi},\ and\ \citenamefont {Tolley}}]{deRham:2007xp}%
  \BibitemOpen
  \bibfield  {author} {\bibinfo {author} {\bibfnamefont {Claudia}\ \bibnamefont
  {de~Rham}}, \bibinfo {author} {\bibfnamefont {Gia}\ \bibnamefont {Dvali}},
  \bibinfo {author} {\bibfnamefont {Stefan}\ \bibnamefont {Hofmann}}, \bibinfo
  {author} {\bibfnamefont {Justin}\ \bibnamefont {Khoury}}, \bibinfo {author}
  {\bibfnamefont {Oriol}\ \bibnamefont {Pujolas}}, \bibinfo {author}
  {\bibfnamefont {Michele}\ \bibnamefont {Redi}}, \ and\ \bibinfo {author}
  {\bibfnamefont {Andrew~J.}\ \bibnamefont {Tolley}},\ }\bibfield  {title}
  {\enquote {\bibinfo {title} {{Cascading gravity: Extending the
  Dvali-Gabadadze-Porrati model to higher dimension}},}\ }\href {\doibase
  10.1103/PhysRevLett.100.251603} {\bibfield  {journal} {\bibinfo  {journal}
  {Phys. Rev. Lett.}\ }\textbf {\bibinfo {volume} {100}},\ \bibinfo {pages}
  {251603} (\bibinfo {year} {2008}{\natexlab{a}})},\ \Eprint
  {http://arxiv.org/abs/0711.2072} {arXiv:0711.2072 [hep-th]} \BibitemShut
  {NoStop}%
\bibitem [{\citenamefont {de~Rham}\ \emph
  {et~al.}(2008{\natexlab{b}})\citenamefont {de~Rham}, \citenamefont {Hofmann},
  \citenamefont {Khoury},\ and\ \citenamefont {Tolley}}]{deRham:2007rw}%
  \BibitemOpen
  \bibfield  {author} {\bibinfo {author} {\bibfnamefont {Claudia}\ \bibnamefont
  {de~Rham}}, \bibinfo {author} {\bibfnamefont {Stefan}\ \bibnamefont
  {Hofmann}}, \bibinfo {author} {\bibfnamefont {Justin}\ \bibnamefont
  {Khoury}}, \ and\ \bibinfo {author} {\bibfnamefont {Andrew~J.}\ \bibnamefont
  {Tolley}},\ }\bibfield  {title} {\enquote {\bibinfo {title} {{Cascading
  Gravity and Degravitation}},}\ }\href {\doibase
  10.1088/1475-7516/2008/02/011} {\bibfield  {journal} {\bibinfo  {journal}
  {JCAP}\ }\textbf {\bibinfo {volume} {02}},\ \bibinfo {pages} {011} (\bibinfo
  {year} {2008}{\natexlab{b}})},\ \Eprint {http://arxiv.org/abs/0712.2821}
  {arXiv:0712.2821 [hep-th]} \BibitemShut {NoStop}%
\bibitem [{\citenamefont {de~Rham}\ \emph {et~al.}(2009)\citenamefont
  {de~Rham}, \citenamefont {Khoury},\ and\ \citenamefont
  {Tolley}}]{deRham:2009wb}%
  \BibitemOpen
  \bibfield  {author} {\bibinfo {author} {\bibfnamefont {Claudia}\ \bibnamefont
  {de~Rham}}, \bibinfo {author} {\bibfnamefont {Justin}\ \bibnamefont
  {Khoury}}, \ and\ \bibinfo {author} {\bibfnamefont {Andrew~J.}\ \bibnamefont
  {Tolley}},\ }\bibfield  {title} {\enquote {\bibinfo {title} {{Flat 3-Brane
  with Tension in Cascading Gravity}},}\ }\href {\doibase
  10.1103/PhysRevLett.103.161601} {\bibfield  {journal} {\bibinfo  {journal}
  {Phys. Rev. Lett.}\ }\textbf {\bibinfo {volume} {103}},\ \bibinfo {pages}
  {161601} (\bibinfo {year} {2009})},\ \Eprint {http://arxiv.org/abs/0907.0473}
  {arXiv:0907.0473 [hep-th]} \BibitemShut {NoStop}%
\bibitem [{\citenamefont {de~Rham}\ and\ \citenamefont
  {Gabadadze}(2010{\natexlab{a}})}]{deRham:2010ik}%
  \BibitemOpen
  \bibfield  {author} {\bibinfo {author} {\bibfnamefont {Claudia}\ \bibnamefont
  {de~Rham}}\ and\ \bibinfo {author} {\bibfnamefont {Gregory}\ \bibnamefont
  {Gabadadze}},\ }\bibfield  {title} {\enquote {\bibinfo {title}
  {{Generalization of the Fierz-Pauli Action}},}\ }\href {\doibase
  10.1103/PhysRevD.82.044020} {\bibfield  {journal} {\bibinfo  {journal} {Phys.
  Rev. D}\ }\textbf {\bibinfo {volume} {82}},\ \bibinfo {pages} {044020}
  (\bibinfo {year} {2010}{\natexlab{a}})},\ \Eprint
  {http://arxiv.org/abs/1007.0443} {arXiv:1007.0443 [hep-th]} \BibitemShut
  {NoStop}%
\bibitem [{\citenamefont {de~Rham}\ \emph {et~al.}(2011)\citenamefont
  {de~Rham}, \citenamefont {Gabadadze},\ and\ \citenamefont
  {Tolley}}]{deRham:2010kj}%
  \BibitemOpen
  \bibfield  {author} {\bibinfo {author} {\bibfnamefont {Claudia}\ \bibnamefont
  {de~Rham}}, \bibinfo {author} {\bibfnamefont {Gregory}\ \bibnamefont
  {Gabadadze}}, \ and\ \bibinfo {author} {\bibfnamefont {Andrew~J.}\
  \bibnamefont {Tolley}},\ }\bibfield  {title} {\enquote {\bibinfo {title}
  {{Resummation of Massive Gravity}},}\ }\href {\doibase
  10.1103/PhysRevLett.106.231101} {\bibfield  {journal} {\bibinfo  {journal}
  {Phys. Rev. Lett.}\ }\textbf {\bibinfo {volume} {106}},\ \bibinfo {pages}
  {231101} (\bibinfo {year} {2011})},\ \Eprint {http://arxiv.org/abs/1011.1232}
  {arXiv:1011.1232 [hep-th]} \BibitemShut {NoStop}%
\bibitem [{\citenamefont {Deffayet}\ \emph {et~al.}(2002)\citenamefont
  {Deffayet}, \citenamefont {Dvali}, \citenamefont {Gabadadze},\ and\
  \citenamefont {Vainshtein}}]{Deffayet:2001uk}%
  \BibitemOpen
  \bibfield  {author} {\bibinfo {author} {\bibfnamefont {Cedric}\ \bibnamefont
  {Deffayet}}, \bibinfo {author} {\bibfnamefont {G.~R.}\ \bibnamefont {Dvali}},
  \bibinfo {author} {\bibfnamefont {Gregory}\ \bibnamefont {Gabadadze}}, \ and\
  \bibinfo {author} {\bibfnamefont {Arkady~I.}\ \bibnamefont {Vainshtein}},\
  }\bibfield  {title} {\enquote {\bibinfo {title} {{Nonperturbative continuity
  in graviton mass versus perturbative discontinuity}},}\ }\href {\doibase
  10.1103/PhysRevD.65.044026} {\bibfield  {journal} {\bibinfo  {journal} {Phys.
  Rev. D}\ }\textbf {\bibinfo {volume} {65}},\ \bibinfo {pages} {044026}
  (\bibinfo {year} {2002})},\ \Eprint {http://arxiv.org/abs/hep-th/0106001}
  {arXiv:hep-th/0106001} \BibitemShut {NoStop}%
\bibitem [{\citenamefont {de~Rham}(2014)}]{deRham:2014zqa}%
  \BibitemOpen
  \bibfield  {author} {\bibinfo {author} {\bibfnamefont {Claudia}\ \bibnamefont
  {de~Rham}},\ }\bibfield  {title} {\enquote {\bibinfo {title} {{Massive
  Gravity}},}\ }\href {\doibase 10.12942/lrr-2014-7} {\bibfield  {journal}
  {\bibinfo  {journal} {Living Rev. Rel.}\ }\textbf {\bibinfo {volume} {17}},\
  \bibinfo {pages} {7} (\bibinfo {year} {2014})},\ \Eprint
  {http://arxiv.org/abs/1401.4173} {arXiv:1401.4173 [hep-th]} \BibitemShut
  {NoStop}%
\bibitem [{\citenamefont {de~Rham}\ and\ \citenamefont
  {Ribeiro}(2014)}]{deRham:2014wfa}%
  \BibitemOpen
  \bibfield  {author} {\bibinfo {author} {\bibfnamefont {Claudia}\ \bibnamefont
  {de~Rham}}\ and\ \bibinfo {author} {\bibfnamefont {Raquel~H.}\ \bibnamefont
  {Ribeiro}},\ }\bibfield  {title} {\enquote {\bibinfo {title} {{Riding on
  irrelevant operators}},}\ }\href {\doibase 10.1088/1475-7516/2014/11/016}
  {\bibfield  {journal} {\bibinfo  {journal} {JCAP}\ }\textbf {\bibinfo
  {volume} {1411}},\ \bibinfo {pages} {016} (\bibinfo {year} {2014})},\ \Eprint
  {http://arxiv.org/abs/1405.5213} {arXiv:1405.5213 [hep-th]} \BibitemShut
  {NoStop}%
\bibitem [{\citenamefont {Joyce}\ \emph {et~al.}(2015)\citenamefont {Joyce},
  \citenamefont {Jain}, \citenamefont {Khoury},\ and\ \citenamefont
  {Trodden}}]{Joyce:2014kja}%
  \BibitemOpen
  \bibfield  {author} {\bibinfo {author} {\bibfnamefont {Austin}\ \bibnamefont
  {Joyce}}, \bibinfo {author} {\bibfnamefont {Bhuvnesh}\ \bibnamefont {Jain}},
  \bibinfo {author} {\bibfnamefont {Justin}\ \bibnamefont {Khoury}}, \ and\
  \bibinfo {author} {\bibfnamefont {Mark}\ \bibnamefont {Trodden}},\ }\bibfield
   {title} {\enquote {\bibinfo {title} {{Beyond the Cosmological Standard
  Model}},}\ }\href {\doibase 10.1016/j.physrep.2014.12.002} {\bibfield
  {journal} {\bibinfo  {journal} {Phys. Rept.}\ }\textbf {\bibinfo {volume}
  {568}},\ \bibinfo {pages} {1--98} (\bibinfo {year} {2015})},\ \Eprint
  {http://arxiv.org/abs/1407.0059} {arXiv:1407.0059 [astro-ph.CO]} \BibitemShut
  {NoStop}%
\bibitem [{\citenamefont {de~Rham}\ \emph
  {et~al.}(2013{\natexlab{a}})\citenamefont {de~Rham}, \citenamefont {Tolley},\
  and\ \citenamefont {Wesley}}]{deRham:2012fw}%
  \BibitemOpen
  \bibfield  {author} {\bibinfo {author} {\bibfnamefont {Claudia}\ \bibnamefont
  {de~Rham}}, \bibinfo {author} {\bibfnamefont {Andrew~J.}\ \bibnamefont
  {Tolley}}, \ and\ \bibinfo {author} {\bibfnamefont {Daniel~H.}\ \bibnamefont
  {Wesley}},\ }\bibfield  {title} {\enquote {\bibinfo {title} {{Vainshtein
  Mechanism in Binary Pulsars}},}\ }\href {\doibase 10.1103/PhysRevD.87.044025}
  {\bibfield  {journal} {\bibinfo  {journal} {Phys. Rev. D}\ }\textbf {\bibinfo
  {volume} {87}},\ \bibinfo {pages} {044025} (\bibinfo {year}
  {2013}{\natexlab{a}})},\ \Eprint {http://arxiv.org/abs/1208.0580}
  {arXiv:1208.0580 [gr-qc]} \BibitemShut {NoStop}%
\bibitem [{\citenamefont {Chu}\ and\ \citenamefont
  {Trodden}(2013)}]{Chu:2012kz}%
  \BibitemOpen
  \bibfield  {author} {\bibinfo {author} {\bibfnamefont {Yi-Zen}\ \bibnamefont
  {Chu}}\ and\ \bibinfo {author} {\bibfnamefont {Mark}\ \bibnamefont
  {Trodden}},\ }\bibfield  {title} {\enquote {\bibinfo {title} {{Retarded
  Green\textquoteright{}s function of a Vainshtein system and Galileon
  waves}},}\ }\href {\doibase 10.1103/PhysRevD.87.024011} {\bibfield  {journal}
  {\bibinfo  {journal} {Phys. Rev. D}\ }\textbf {\bibinfo {volume} {87}},\
  \bibinfo {pages} {024011} (\bibinfo {year} {2013})},\ \Eprint
  {http://arxiv.org/abs/1210.6651} {arXiv:1210.6651 [astro-ph.CO]} \BibitemShut
  {NoStop}%
\bibitem [{\citenamefont {de~Rham}\ \emph
  {et~al.}(2013{\natexlab{b}})\citenamefont {de~Rham}, \citenamefont {Matas},\
  and\ \citenamefont {Tolley}}]{deRham:2012fg}%
  \BibitemOpen
  \bibfield  {author} {\bibinfo {author} {\bibfnamefont {Claudia}\ \bibnamefont
  {de~Rham}}, \bibinfo {author} {\bibfnamefont {Andrew}\ \bibnamefont {Matas}},
  \ and\ \bibinfo {author} {\bibfnamefont {Andrew~J.}\ \bibnamefont {Tolley}},\
  }\bibfield  {title} {\enquote {\bibinfo {title} {{Galileon Radiation from
  Binary Systems}},}\ }\href {\doibase 10.1103/PhysRevD.87.064024} {\bibfield
  {journal} {\bibinfo  {journal} {Phys. Rev. D}\ }\textbf {\bibinfo {volume}
  {87}},\ \bibinfo {pages} {064024} (\bibinfo {year} {2013}{\natexlab{b}})},\
  \Eprint {http://arxiv.org/abs/1212.5212} {arXiv:1212.5212 [hep-th]}
  \BibitemShut {NoStop}%
\bibitem [{\citenamefont {Dar}\ \emph {et~al.}(2019)\citenamefont {Dar},
  \citenamefont {De~Rham}, \citenamefont {Deskins}, \citenamefont {Giblin},\
  and\ \citenamefont {Tolley}}]{Dar:2018dra}%
  \BibitemOpen
  \bibfield  {author} {\bibinfo {author} {\bibfnamefont {Furqan}\ \bibnamefont
  {Dar}}, \bibinfo {author} {\bibfnamefont {Claudia}\ \bibnamefont {De~Rham}},
  \bibinfo {author} {\bibfnamefont {J.~Tate}\ \bibnamefont {Deskins}}, \bibinfo
  {author} {\bibfnamefont {John~T.}\ \bibnamefont {Giblin}}, \ and\ \bibinfo
  {author} {\bibfnamefont {Andrew~J.}\ \bibnamefont {Tolley}},\ }\bibfield
  {title} {\enquote {\bibinfo {title} {{Scalar Gravitational Radiation from
  Binaries: Vainshtein Mechanism in Time-dependent Systems}},}\ }\href
  {\doibase 10.1088/1361-6382/aaf5e8} {\bibfield  {journal} {\bibinfo
  {journal} {Class. Quant. Grav.}\ }\textbf {\bibinfo {volume} {36}},\ \bibinfo
  {pages} {025008} (\bibinfo {year} {2019})},\ \Eprint
  {http://arxiv.org/abs/1808.02165} {arXiv:1808.02165 [hep-th]} \BibitemShut
  {NoStop}%
\bibitem [{\citenamefont {Brax}\ \emph {et~al.}(2017)\citenamefont {Brax},
  \citenamefont {Davis},\ and\ \citenamefont {Jha}}]{Brax:2017wcj}%
  \BibitemOpen
  \bibfield  {author} {\bibinfo {author} {\bibfnamefont {Philippe}\
  \bibnamefont {Brax}}, \bibinfo {author} {\bibfnamefont {Anne-Christine}\
  \bibnamefont {Davis}}, \ and\ \bibinfo {author} {\bibfnamefont {Rahul}\
  \bibnamefont {Jha}},\ }\bibfield  {title} {\enquote {\bibinfo {title}
  {{Neutron Stars in Screened Modified Gravity: Chameleon vs Dilaton}},}\
  }\href {\doibase 10.1103/PhysRevD.95.083514} {\bibfield  {journal} {\bibinfo
  {journal} {Phys. Rev. D}\ }\textbf {\bibinfo {volume} {95}},\ \bibinfo
  {pages} {083514} (\bibinfo {year} {2017})},\ \Eprint
  {http://arxiv.org/abs/1702.02983} {arXiv:1702.02983 [gr-qc]} \BibitemShut
  {NoStop}%
\bibitem [{\citenamefont {Kuntz}(2019)}]{Kuntz:2019plo}%
  \BibitemOpen
  \bibfield  {author} {\bibinfo {author} {\bibfnamefont {Adrien}\ \bibnamefont
  {Kuntz}},\ }\bibfield  {title} {\enquote {\bibinfo {title} {{Two-body
  potential of Vainshtein screened theories}},}\ }\href {\doibase
  10.1103/PhysRevD.100.024024} {\bibfield  {journal} {\bibinfo  {journal}
  {Phys. Rev. D}\ }\textbf {\bibinfo {volume} {100}},\ \bibinfo {pages}
  {024024} (\bibinfo {year} {2019})},\ \Eprint
  {http://arxiv.org/abs/1905.07340} {arXiv:1905.07340 [gr-qc]} \BibitemShut
  {NoStop}%
\bibitem [{\citenamefont {de~Aguiar}\ and\ \citenamefont
  {Mendes}(2020)}]{deAguiar:2020urb}%
  \BibitemOpen
  \bibfield  {author} {\bibinfo {author} {\bibfnamefont {Bernardo~F.}\
  \bibnamefont {de~Aguiar}}\ and\ \bibinfo {author} {\bibfnamefont {Raissa
  F.~P.}\ \bibnamefont {Mendes}},\ }\bibfield  {title} {\enquote {\bibinfo
  {title} {{Highly compact neutron stars and screening mechanisms: Equilibrium
  and stability}},}\ }\href {\doibase 10.1103/PhysRevD.102.024064} {\bibfield
  {journal} {\bibinfo  {journal} {Phys. Rev. D}\ }\textbf {\bibinfo {volume}
  {102}},\ \bibinfo {pages} {024064} (\bibinfo {year} {2020})},\ \Eprint
  {http://arxiv.org/abs/2006.10080} {arXiv:2006.10080 [gr-qc]} \BibitemShut
  {NoStop}%
\bibitem [{\citenamefont {Bezares}\ \emph
  {et~al.}(2021{\natexlab{a}})\citenamefont {Bezares}, \citenamefont {ter
  Haar}, \citenamefont {Crisostomi}, \citenamefont {Barausse},\ and\
  \citenamefont {Palenzuela}}]{Bezares:2021yek}%
  \BibitemOpen
  \bibfield  {author} {\bibinfo {author} {\bibfnamefont {Miguel}\ \bibnamefont
  {Bezares}}, \bibinfo {author} {\bibfnamefont {Lotte}\ \bibnamefont {ter
  Haar}}, \bibinfo {author} {\bibfnamefont {Marco}\ \bibnamefont {Crisostomi}},
  \bibinfo {author} {\bibfnamefont {Enrico}\ \bibnamefont {Barausse}}, \ and\
  \bibinfo {author} {\bibfnamefont {Carlos}\ \bibnamefont {Palenzuela}},\
  }\bibfield  {title} {\enquote {\bibinfo {title} {{Kinetic screening in
  non-linear stellar oscillations and gravitational collapse}},}\ }\href@noop
  {} {\  (\bibinfo {year} {2021}{\natexlab{a}})},\ \Eprint
  {http://arxiv.org/abs/2105.13992} {arXiv:2105.13992 [gr-qc]} \BibitemShut
  {NoStop}%
\bibitem [{\citenamefont {Brax}\ \emph {et~al.}(2020)\citenamefont {Brax},
  \citenamefont {Heisenberg},\ and\ \citenamefont {Kuntz}}]{Brax:2020ujo}%
  \BibitemOpen
  \bibfield  {author} {\bibinfo {author} {\bibfnamefont {Philippe}\
  \bibnamefont {Brax}}, \bibinfo {author} {\bibfnamefont {Lavinia}\
  \bibnamefont {Heisenberg}}, \ and\ \bibinfo {author} {\bibfnamefont {Adrien}\
  \bibnamefont {Kuntz}},\ }\bibfield  {title} {\enquote {\bibinfo {title}
  {{Unveiling the Galileon in a three-body system : scalar and gravitational
  wave production}},}\ }\href {\doibase 10.1088/1475-7516/2020/05/012}
  {\bibfield  {journal} {\bibinfo  {journal} {JCAP}\ }\textbf {\bibinfo
  {volume} {05}},\ \bibinfo {pages} {012} (\bibinfo {year} {2020})},\ \Eprint
  {http://arxiv.org/abs/2002.12590} {arXiv:2002.12590 [gr-qc]} \BibitemShut
  {NoStop}%
\bibitem [{\citenamefont {Renevey}\ \emph {et~al.}(2021)\citenamefont
  {Renevey}, \citenamefont {McManus}, \citenamefont {Dalang},\ and\
  \citenamefont {Lombriser}}]{Renevey:2021tcz}%
  \BibitemOpen
  \bibfield  {author} {\bibinfo {author} {\bibfnamefont {Cyril}\ \bibnamefont
  {Renevey}}, \bibinfo {author} {\bibfnamefont {Ryan}\ \bibnamefont {McManus}},
  \bibinfo {author} {\bibfnamefont {Charles}\ \bibnamefont {Dalang}}, \ and\
  \bibinfo {author} {\bibfnamefont {Lucas}\ \bibnamefont {Lombriser}},\
  }\bibfield  {title} {\enquote {\bibinfo {title} {{The effect of screening
  mechanisms on black hole binary inspiral waveforms}},}\ }\href@noop {} {\
  (\bibinfo {year} {2021})},\ \Eprint {http://arxiv.org/abs/2106.05678}
  {arXiv:2106.05678 [gr-qc]} \BibitemShut {NoStop}%
\bibitem [{\citenamefont {Dima}\ \emph {et~al.}(2021)\citenamefont {Dima},
  \citenamefont {Bezares},\ and\ \citenamefont {Barausse}}]{Dima:2021pwx}%
  \BibitemOpen
  \bibfield  {author} {\bibinfo {author} {\bibfnamefont {Alexandru}\
  \bibnamefont {Dima}}, \bibinfo {author} {\bibfnamefont {Miguel}\ \bibnamefont
  {Bezares}}, \ and\ \bibinfo {author} {\bibfnamefont {Enrico}\ \bibnamefont
  {Barausse}},\ }\bibfield  {title} {\enquote {\bibinfo {title} {{Dynamical
  Chameleon Neutron Stars: stability, radial oscillations and scalar radiation
  in spherical symmetry}},}\ }\href@noop {} {\  (\bibinfo {year} {2021})},\
  \Eprint {http://arxiv.org/abs/2107.04359} {arXiv:2107.04359 [gr-qc]}
  \BibitemShut {NoStop}%
\bibitem [{\citenamefont {Bezares}\ \emph
  {et~al.}(2021{\natexlab{b}})\citenamefont {Bezares}, \citenamefont
  {Aguilera-Miret}, \citenamefont {ter Haar}, \citenamefont {Crisostomi},
  \citenamefont {Palenzuela},\ and\ \citenamefont
  {Barausse}}]{Bezares:2021dma}%
  \BibitemOpen
  \bibfield  {author} {\bibinfo {author} {\bibfnamefont {Miguel}\ \bibnamefont
  {Bezares}}, \bibinfo {author} {\bibfnamefont {Ricard}\ \bibnamefont
  {Aguilera-Miret}}, \bibinfo {author} {\bibfnamefont {Lotte}\ \bibnamefont
  {ter Haar}}, \bibinfo {author} {\bibfnamefont {Marco}\ \bibnamefont
  {Crisostomi}}, \bibinfo {author} {\bibfnamefont {Carlos}\ \bibnamefont
  {Palenzuela}}, \ and\ \bibinfo {author} {\bibfnamefont {Enrico}\ \bibnamefont
  {Barausse}},\ }\bibfield  {title} {\enquote {\bibinfo {title} {{No evidence
  of kinetic screening in merging binary neutron stars}},}\ }\href@noop {} {\
  (\bibinfo {year} {2021}{\natexlab{b}})},\ \Eprint
  {http://arxiv.org/abs/2107.05648} {arXiv:2107.05648 [gr-qc]} \BibitemShut
  {NoStop}%
\bibitem [{\citenamefont {Nicolis}\ \emph {et~al.}(2009)\citenamefont
  {Nicolis}, \citenamefont {Rattazzi},\ and\ \citenamefont
  {Trincherini}}]{Nicolis:2008in}%
  \BibitemOpen
  \bibfield  {author} {\bibinfo {author} {\bibfnamefont {Alberto}\ \bibnamefont
  {Nicolis}}, \bibinfo {author} {\bibfnamefont {Riccardo}\ \bibnamefont
  {Rattazzi}}, \ and\ \bibinfo {author} {\bibfnamefont {Enrico}\ \bibnamefont
  {Trincherini}},\ }\bibfield  {title} {\enquote {\bibinfo {title} {{The
  Galileon as a local modification of gravity}},}\ }\href {\doibase
  10.1103/PhysRevD.79.064036} {\bibfield  {journal} {\bibinfo  {journal} {Phys.
  Rev. D}\ }\textbf {\bibinfo {volume} {79}},\ \bibinfo {pages} {064036}
  (\bibinfo {year} {2009})},\ \Eprint {http://arxiv.org/abs/0811.2197}
  {arXiv:0811.2197 [hep-th]} \BibitemShut {NoStop}%
\bibitem [{\citenamefont {Luty}\ \emph {et~al.}(2003)\citenamefont {Luty},
  \citenamefont {Porrati},\ and\ \citenamefont {Rattazzi}}]{Luty:2003vm}%
  \BibitemOpen
  \bibfield  {author} {\bibinfo {author} {\bibfnamefont {Markus~A.}\
  \bibnamefont {Luty}}, \bibinfo {author} {\bibfnamefont {Massimo}\
  \bibnamefont {Porrati}}, \ and\ \bibinfo {author} {\bibfnamefont {Riccardo}\
  \bibnamefont {Rattazzi}},\ }\bibfield  {title} {\enquote {\bibinfo {title}
  {{Strong interactions and stability in the DGP model}},}\ }\href {\doibase
  10.1088/1126-6708/2003/09/029} {\bibfield  {journal} {\bibinfo  {journal}
  {JHEP}\ }\textbf {\bibinfo {volume} {09}},\ \bibinfo {pages} {029} (\bibinfo
  {year} {2003})},\ \Eprint {http://arxiv.org/abs/hep-th/0303116}
  {arXiv:hep-th/0303116} \BibitemShut {NoStop}%
\bibitem [{\citenamefont {Nicolis}\ and\ \citenamefont
  {Rattazzi}(2004)}]{Nicolis:2004qq}%
  \BibitemOpen
  \bibfield  {author} {\bibinfo {author} {\bibfnamefont {Alberto}\ \bibnamefont
  {Nicolis}}\ and\ \bibinfo {author} {\bibfnamefont {Riccardo}\ \bibnamefont
  {Rattazzi}},\ }\bibfield  {title} {\enquote {\bibinfo {title} {{Classical and
  quantum consistency of the DGP model}},}\ }\href {\doibase
  10.1088/1126-6708/2004/06/059} {\bibfield  {journal} {\bibinfo  {journal}
  {JHEP}\ }\textbf {\bibinfo {volume} {06}},\ \bibinfo {pages} {059} (\bibinfo
  {year} {2004})},\ \Eprint {http://arxiv.org/abs/hep-th/0404159}
  {arXiv:hep-th/0404159} \BibitemShut {NoStop}%
\bibitem [{\citenamefont {de~Rham}(2010)}]{deRham:2009rm}%
  \BibitemOpen
  \bibfield  {author} {\bibinfo {author} {\bibfnamefont {Claudia}\ \bibnamefont
  {de~Rham}},\ }\bibfield  {title} {\enquote {\bibinfo {title} {{Massive
  gravity from Dirichlet boundary conditions}},}\ }\href {\doibase
  10.1016/j.physletb.2010.04.005} {\bibfield  {journal} {\bibinfo  {journal}
  {Phys. Lett. B}\ }\textbf {\bibinfo {volume} {688}},\ \bibinfo {pages}
  {137--141} (\bibinfo {year} {2010})},\ \Eprint
  {http://arxiv.org/abs/0910.5474} {arXiv:0910.5474 [hep-th]} \BibitemShut
  {NoStop}%
\bibitem [{\citenamefont {de~Rham}\ and\ \citenamefont
  {Gabadadze}(2010{\natexlab{b}})}]{deRham:2010gu}%
  \BibitemOpen
  \bibfield  {author} {\bibinfo {author} {\bibfnamefont {Claudia}\ \bibnamefont
  {de~Rham}}\ and\ \bibinfo {author} {\bibfnamefont {Gregory}\ \bibnamefont
  {Gabadadze}},\ }\bibfield  {title} {\enquote {\bibinfo {title} {{Selftuned
  Massive Spin-2}},}\ }\href {\doibase 10.1016/j.physletb.2010.08.043}
  {\bibfield  {journal} {\bibinfo  {journal} {Phys. Lett. B}\ }\textbf
  {\bibinfo {volume} {693}},\ \bibinfo {pages} {334--338} (\bibinfo {year}
  {2010}{\natexlab{b}})},\ \Eprint {http://arxiv.org/abs/1006.4367}
  {arXiv:1006.4367 [hep-th]} \BibitemShut {NoStop}%
\bibitem [{\citenamefont {de~Rham}\ and\ \citenamefont
  {Tolley}(2010)}]{deRham:2010eu}%
  \BibitemOpen
  \bibfield  {author} {\bibinfo {author} {\bibfnamefont {Claudia}\ \bibnamefont
  {de~Rham}}\ and\ \bibinfo {author} {\bibfnamefont {Andrew~J.}\ \bibnamefont
  {Tolley}},\ }\bibfield  {title} {\enquote {\bibinfo {title} {{DBI and the
  Galileon reunited}},}\ }\href {\doibase 10.1088/1475-7516/2010/05/015}
  {\bibfield  {journal} {\bibinfo  {journal} {JCAP}\ }\textbf {\bibinfo
  {volume} {05}},\ \bibinfo {pages} {015} (\bibinfo {year} {2010})},\ \Eprint
  {http://arxiv.org/abs/1003.5917} {arXiv:1003.5917 [hep-th]} \BibitemShut
  {NoStop}%
\bibitem [{\citenamefont {Beneke}\ and\ \citenamefont
  {Smirnov}(1998)}]{Beneke:1997zp}%
  \BibitemOpen
  \bibfield  {author} {\bibinfo {author} {\bibfnamefont {M.}~\bibnamefont
  {Beneke}}\ and\ \bibinfo {author} {\bibfnamefont {Vladimir~A.}\ \bibnamefont
  {Smirnov}},\ }\bibfield  {title} {\enquote {\bibinfo {title} {{Asymptotic
  expansion of Feynman integrals near threshold}},}\ }\href {\doibase
  10.1016/S0550-3213(98)00138-2} {\bibfield  {journal} {\bibinfo  {journal}
  {Nucl. Phys. B}\ }\textbf {\bibinfo {volume} {522}},\ \bibinfo {pages}
  {321--344} (\bibinfo {year} {1998})},\ \Eprint
  {http://arxiv.org/abs/hep-ph/9711391} {arXiv:hep-ph/9711391} \BibitemShut
  {NoStop}%
\bibitem [{\citenamefont {Smirnov}(2004)}]{Smirnov:2004ym}%
  \BibitemOpen
  \bibfield  {author} {\bibinfo {author} {\bibfnamefont {Vladimir~A.}\
  \bibnamefont {Smirnov}},\ }\bibfield  {title} {\enquote {\bibinfo {title}
  {{Evaluating Feynman integrals}},}\ }\href@noop {} {\bibfield  {journal}
  {\bibinfo  {journal} {Springer Tracts Mod. Phys.}\ }\textbf {\bibinfo
  {volume} {211}},\ \bibinfo {pages} {1--244} (\bibinfo {year}
  {2004})}\BibitemShut {NoStop}%
\bibitem [{\citenamefont {Passarino}\ and\ \citenamefont
  {Veltman}(1979)}]{Passarino:1978jh}%
  \BibitemOpen
  \bibfield  {author} {\bibinfo {author} {\bibfnamefont {G.}~\bibnamefont
  {Passarino}}\ and\ \bibinfo {author} {\bibfnamefont {M.~J.~G.}\ \bibnamefont
  {Veltman}},\ }\bibfield  {title} {\enquote {\bibinfo {title} {{One Loop
  Corrections for e+ e- Annihilation Into mu+ mu- in the Weinberg Model}},}\
  }\href {\doibase 10.1016/0550-3213(79)90234-7} {\bibfield  {journal}
  {\bibinfo  {journal} {Nucl. Phys. B}\ }\textbf {\bibinfo {volume} {160}},\
  \bibinfo {pages} {151--207} (\bibinfo {year} {1979})}\BibitemShut {NoStop}%
\bibitem [{\citenamefont {Iwasaki}(1971)}]{Iwasaki:1971vb}%
  \BibitemOpen
  \bibfield  {author} {\bibinfo {author} {\bibfnamefont {Y.}~\bibnamefont
  {Iwasaki}},\ }\bibfield  {title} {\enquote {\bibinfo {title} {{Quantum theory
  of gravitation vs. classical theory. - fourth-order potential}},}\ }\href
  {\doibase 10.1143/PTP.46.1587} {\bibfield  {journal} {\bibinfo  {journal}
  {Prog. Theor. Phys.}\ }\textbf {\bibinfo {volume} {46}},\ \bibinfo {pages}
  {1587--1609} (\bibinfo {year} {1971})}\BibitemShut {NoStop}%
\bibitem [{\citenamefont {Neill}\ and\ \citenamefont
  {Rothstein}(2013)}]{Neill:2013wsa}%
  \BibitemOpen
  \bibfield  {author} {\bibinfo {author} {\bibfnamefont {Duff}\ \bibnamefont
  {Neill}}\ and\ \bibinfo {author} {\bibfnamefont {Ira~Z.}\ \bibnamefont
  {Rothstein}},\ }\bibfield  {title} {\enquote {\bibinfo {title} {{Classical
  Space-Times from the S Matrix}},}\ }\href {\doibase
  10.1016/j.nuclphysb.2013.09.007} {\bibfield  {journal} {\bibinfo  {journal}
  {Nucl. Phys. B}\ }\textbf {\bibinfo {volume} {877}},\ \bibinfo {pages}
  {177--189} (\bibinfo {year} {2013})},\ \Eprint
  {http://arxiv.org/abs/1304.7263} {arXiv:1304.7263 [hep-th]} \BibitemShut
  {NoStop}%
\bibitem [{\citenamefont {Damour}(2018)}]{Damour:2017zjx}%
  \BibitemOpen
  \bibfield  {author} {\bibinfo {author} {\bibfnamefont {Thibault}\
  \bibnamefont {Damour}},\ }\bibfield  {title} {\enquote {\bibinfo {title}
  {{High-energy gravitational scattering and the general relativistic two-body
  problem}},}\ }\href {\doibase 10.1103/PhysRevD.97.044038} {\bibfield
  {journal} {\bibinfo  {journal} {Phys. Rev. D}\ }\textbf {\bibinfo {volume}
  {97}},\ \bibinfo {pages} {044038} (\bibinfo {year} {2018})},\ \Eprint
  {http://arxiv.org/abs/1710.10599} {arXiv:1710.10599 [gr-qc]} \BibitemShut
  {NoStop}%
\bibitem [{\citenamefont {Paliov}\ and\ \citenamefont
  {Rosendorff}(1967)}]{doi:10.1063/1.1705426}%
  \BibitemOpen
  \bibfield  {author} {\bibinfo {author} {\bibfnamefont {A.}~\bibnamefont
  {Paliov}}\ and\ \bibinfo {author} {\bibfnamefont {S.}~\bibnamefont
  {Rosendorff}},\ }\bibfield  {title} {\enquote {\bibinfo {title}
  {High‐energy phase shifts produced by repulsive singular potentials},}\
  }\href {\doibase 10.1063/1.1705426} {\bibfield  {journal} {\bibinfo
  {journal} {Journal of Mathematical Physics}\ }\textbf {\bibinfo {volume}
  {8}},\ \bibinfo {pages} {1829--1837} (\bibinfo {year} {1967})}\BibitemShut
  {NoStop}%
\bibitem [{\citenamefont {Wallace}(1973)}]{Wallace:1973iu}%
  \BibitemOpen
  \bibfield  {author} {\bibinfo {author} {\bibfnamefont {S.~J.}\ \bibnamefont
  {Wallace}},\ }\bibfield  {title} {\enquote {\bibinfo {title} {{Eikonal
  expansion}},}\ }\href {\doibase 10.1016/0003-4916(73)90008-0} {\bibfield
  {journal} {\bibinfo  {journal} {Annals Phys.}\ }\textbf {\bibinfo {volume}
  {78}},\ \bibinfo {pages} {190--257} (\bibinfo {year} {1973})}\BibitemShut
  {NoStop}%
\bibitem [{\citenamefont {Bohm}(1989)}]{BohmQT}%
  \BibitemOpen
  \bibfield  {author} {\bibinfo {author} {\bibfnamefont {David}\ \bibnamefont
  {Bohm}},\ }\href@noop {} {\emph {\bibinfo {title} {Quantum Theory}}}\
  (\bibinfo  {publisher} {Dover Publications},\ \bibinfo {year}
  {1989})\BibitemShut {NoStop}%
\bibitem [{\citenamefont {Murphy}\ \emph {et~al.}(2012)\citenamefont {Murphy},
  \citenamefont {Adelberger}, \citenamefont {Battat}, \citenamefont {Hoyle},
  \citenamefont {Johnson}, \citenamefont {McMillan}, \citenamefont {Stubbs},\
  and\ \citenamefont {Swanson}}]{Murphy_2012}%
  \BibitemOpen
  \bibfield  {author} {\bibinfo {author} {\bibfnamefont {T~W}\ \bibnamefont
  {Murphy}}, \bibinfo {author} {\bibfnamefont {E~G}\ \bibnamefont
  {Adelberger}}, \bibinfo {author} {\bibfnamefont {J~B~R}\ \bibnamefont
  {Battat}}, \bibinfo {author} {\bibfnamefont {C~D}\ \bibnamefont {Hoyle}},
  \bibinfo {author} {\bibfnamefont {N~H}\ \bibnamefont {Johnson}}, \bibinfo
  {author} {\bibfnamefont {R~J}\ \bibnamefont {McMillan}}, \bibinfo {author}
  {\bibfnamefont {C~W}\ \bibnamefont {Stubbs}}, \ and\ \bibinfo {author}
  {\bibfnamefont {H~E}\ \bibnamefont {Swanson}},\ }\bibfield  {title} {\enquote
  {\bibinfo {title} {{APOLLO}: millimeter lunar laser ranging},}\ }\href
  {\doibase 10.1088/0264-9381/29/18/184005} {\bibfield  {journal} {\bibinfo
  {journal} {Classical and Quantum Gravity}\ }\textbf {\bibinfo {volume}
  {29}},\ \bibinfo {pages} {184005} (\bibinfo {year} {2012})}\BibitemShut
  {NoStop}%
\bibitem [{\citenamefont {Dvali}\ \emph {et~al.}(2003)\citenamefont {Dvali},
  \citenamefont {Gruzinov},\ and\ \citenamefont {Zaldarriaga}}]{Dvali:2002vf}%
  \BibitemOpen
  \bibfield  {author} {\bibinfo {author} {\bibfnamefont {Gia}\ \bibnamefont
  {Dvali}}, \bibinfo {author} {\bibfnamefont {Andrei}\ \bibnamefont
  {Gruzinov}}, \ and\ \bibinfo {author} {\bibfnamefont {Matias}\ \bibnamefont
  {Zaldarriaga}},\ }\bibfield  {title} {\enquote {\bibinfo {title} {{The
  Accelerated universe and the moon}},}\ }\href {\doibase
  10.1103/PhysRevD.68.024012} {\bibfield  {journal} {\bibinfo  {journal} {Phys.
  Rev. D}\ }\textbf {\bibinfo {volume} {68}},\ \bibinfo {pages} {024012}
  (\bibinfo {year} {2003})},\ \Eprint {http://arxiv.org/abs/hep-ph/0212069}
  {arXiv:hep-ph/0212069} \BibitemShut {NoStop}%
\end{thebibliography}%
\end{document}